\documentclass[useAMS,usenatbib,twocolumn]{mnras}

\usepackage{amsmath}
\usepackage{graphicx}
\usepackage{dcolumn}
\usepackage{bm}
\usepackage{epsfig}
\usepackage{amssymb,latexsym,mathrsfs}
\usepackage{graphicx}
\usepackage{enumitem}
\usepackage{color}
\usepackage{hyperref}
\usepackage{subcaption}
\usepackage{float}
\usepackage{natbib}
\usepackage{tikz}
\usepackage{ulem}
\usetikzlibrary{positioning}
\usepackage[utf8]{inputenc}
\usepackage[T1]{fontenc}

\hypersetup{
    colorlinks=true,
    linkcolor=red,
    citecolor=blue,
} 

\def\apj{{ ApJ}}
\def\apjl{{ ApJL}}
\def\apjs{{ ApJS}}

\def\aap{{ A\&A}}

\def\nat {{ Nature}}

\newcommand{\be}{\begin{equation}}
\newcommand{\ee}{\end{equation}}
\newcommand{\bs}{\begin{split}} 
\newcommand{\bea}{\begin{eqnarray}}
\newcommand{\eea}{\end{eqnarray}}

\title[Nucleosynthesis in protomagnetar outflows]
{Systematic exploration of heavy element nucleosynthesis in protomagnetar outflows}

\author[Ekanger et al.]{
Nick Ekanger$^{1}$\thanks{enick1@vt.edu}, Mukul Bhattacharya$^{1,2}$, Shunsaku Horiuchi$^{1,3}$\\ 
${}^1$Center for Neutrino Physics, Department of Physics, Virginia Tech, Blacksburg, VA 24061, USA\\
${}^2$Department of Physics, Department of Astronomy and Astrophysics, Pennsylvania State University, University Park, PA 16802, USA \\
${}^3$Kavli IPMU (WPI), UTIAS, The University of Tokyo, Kashiwa, Chiba 277-8583, Japan 
} 

\begin{document}

\date{Accepted 2022 March 28; Received 2022 March 23; in original form 2022 January 13}

\pagerange{\pageref{firstpage}--\pageref{lastpage}} \pubyear{2022}

\maketitle

\label{firstpage}

\begin{abstract} 
We study the nucleosynthesis products in neutrino-driven winds from rapidly rotating, highly magnetized and misaligned protomagnetars using the nuclear reaction network {\tt SkyNet}. We adopt a semi-analytic parametrized model for the protomagnetar and systematically study the capabilities of its neutrino-driven wind for synthesizing nuclei and eventually producing ultra-high energy cosmic rays (UHECRs). We find that for neutron-rich outflows ($Y_e<0.5$), synthesis of heavy elements ($\overline{A}\sim 20-65$) is possible during the first $\sim 10\, {\rm seconds}$ of the outflow, but these nuclei are subjected to composition-altering photodisintegration during the epoch of particle acceleration at the dissipation radii. However, after the first $\sim 10\, {\rm seconds}$ of the outflow, nucleosynthesis reaches lighter elements ($\overline{A}\sim 10-50$) that are not subjected to subsequent photodisintegration. For proton-rich ($Y_e \geq 0.5$) outflows, synthesis is more limited ($\overline{A}\sim 4-15$). These suggest that while protomagnetars typically do not synthesize nuclei heavier than second r-process peak elements, they are intriguing sources of intermediate/heavy mass UHECRs. For all configurations, the most rapidly rotating protomagnetars are more conducive for nucleosynthesis with a weaker dependence on the magnetic field strength.
\end{abstract}


\section{Introduction}

The origin of ultra-high energy cosmic rays (UHECRs) remains an important unresolved question in astrophysics (see \citealt{2005JPhG...31R..95H,Kotera_2011,Anchordoqui_2019,Alves_Batista_2019} for recent reviews). Broadly speaking, there are three key observational quantities which can be used to elucidate their sources. The first is the UHECR energy spectrum. Air shower observatories like Pierre Auger Observatory (PAO) \citep{PAO2015}, Telescope Array (TA) \citep{Abu_Zayyad_2013}, and HiRes \citep{Abbasi_2009} have measured a suppression in the UHECR flux at the highest energies, compatible with their interactions with the cosmic background radiation as predicted by \citet{Greisen:1966jv} and \citet{1966JETPL...4...78Z}; this is the so-called Greisen-Zatsepin-Kuzmin (GZK) cutoff, providing support of the extragalactic nature of UHECR sources. The second observational measure is the arrival directions of UHECRs. A number of studies have reported anisotropic distributions for UHECRs (see e.g. \citealt{Abreu_2010,Aab_2014c,Aab_2015}), including a correlation with starburst galaxies \citep{Aab_2018}. However, directional studies are subject to not just the acceleration, survival and propagation from the source models, but also to uncertainties of the nuclear composition of UHECRs and extragalactic magnetic fields. Directional information, such as the large-scale anisotropy and lack of anisotropy from the Galactic Centre \citep{2017aab}, is compatible with many source classes, e.g., active galactic nuclei (AGN), starburst winds, gamma-ray bursts (GRBs), and others (e.g. \citealt{2016arXiv161000944B,Zhang_2018,Aab_2018}). Therefore, additional observational features are warranted.

The third observational measure is the nuclear composition of UHECRs. The composition is primarily determined by the distribution of particle shower maxima, $X_{\textrm{max}}$, which is proportional to the logarithm of mass number $A$, and its second moment. Studies with PAO data suggest a population of UHECRs from nitrogen and silicon groups up to iron groups (\citealt{Aab_2014,Aab_2014b,Unger_2015,Batista_2019}), which is in statistical agreement with TA (\citealt{2015arXiv151102103T,Abbasi_2018,2019arXiv190909073T}). However, this method is not precise enough to confirm the composition of individual cosmic rays. Therefore, the precise mass distribution is still under debate. Nevertheless, the pure-proton/proton-helium compositions are generally disfavored, at least for the highest energies (e.g. \citealt{2019arXiv190909073T,Jiang:2020arb,Kuznetsov_2021}).

The composition of UHECRs is particularly informative to their source models. Various types of AGNs, GRBs and core-collapse supernovae (CCSNe) have been proposed as potential UHECR sites and have already been studied in detail. AGNs, for example, are not expected to produce a heavy-dominated composition (\citealt{Lemoine_2009,MGH11,Horiuchi_2012}) (although the low-luminosity AGN models of \citealt{Rodrigues_2021} seem to support the notion of heavy element dominated composition at ultra-high energies, based on the galactic cosmic ray composition). On the other hand, CCSNe and related phenomena are more conducive to generating higher fractions of heavy nuclei. 

In this paper, we study rapidly rotating and highly magnetized protoneutron stars (PNSs) (also known as `millisecond (ms) protomagnetars') that are formed upon the core collapse of massive stars (\citealt{1992ApJ...392L...9D,1992Natur.357..472U}). These newly-born magnetars have been discussed as sources of long-duration GRBs (\citealt{Wheeler_2000,Thompson_2004,Metzger_2011}), that can accelerate protons (\citealt{Arons_2003,Kotera_2011_uhecr}) and heavier particles (\citealt{MGH11,Horiuchi_2012,Zhang_2018}) to ultra-high energies, making these PNS central engines promising sources of heavier UHECRs. Magnetar theory more generally has also been linked more recently to GRB observations (e.g. \citealt{2013margutti,2014margutti}) and used to constrain magnetic field strength and spin period using X-ray afterglow plateau data (e.g. \citealt{2001zhang,2014lu}).

Here we consider relativistically expanding winds launched by PNS and study their composition as a function of the source magnetic field strength and spin period. Not all configurations of magnetic field and spin period may lead to successful GRBs. Alternatively, winds embedded inside of their progenitor/supernova are not freely expanding, and in the extreme case outflows can be choked \citep[e.g.,][]{Metzger_2011}. However, the parameter space where this happens is highly uncertain and also dependent on the progenitor model, and we defer this to a future investigation. Note that our focus also contrasts with more typical pulsars with lower field strengths, which represent a sufficiently different scenario, where nucleosynthesis and particle acceleration are coupled and the results may be significantly different from those found in this work (see \citealt{2012fang}).

A number of papers have analytically studied the composition of CCSNe outflows more broadly (e.g. \citealt{Qian_1996,Hoffman_1997,MGH11}) but there is a benefit of using numerical treatments that allow the study of these systems in extraordinary detail. With large networks of nuclides and reactions, one can study the time evolution of abundance patterns of thousands of nuclei. Nuclear reaction networks like {\tt SkyNet} \citep{Lippuner_2017}, {\tt WinNet} \citep{2014PhDT.......206W}, and {\tt PRISM} \citep{Mumpower_2017} have been applied to a number of astrophysical scenarios including CCSNe outflows (e.g. \citealt{Roberts_2010,Arcones_2011,Halevi_2018,M_sta_2018,Grimmett_2020,Reichert_2021}), NS-NS and neutron star-black hole (NS-BH) mergers (e.g. \citealt{Roberts_2016,Lippuner_2017,C_t__2018,2021chen}). In particular, although NS-NS mergers are considered the primary source (e.g. \citealt{Abbott_2017,Yong_2021}), CCSNe are an additional proposed source for r-process elements in the universe. These sources would have to agree with the solar abundance patterns and those of ultra-metal-poor (UMP) stars (\citealt{Arnould_2007,sneden2008,Yong_2021}). NS-NS merger events may also be inefficient accelerators of heavy nuclei, making alternative transient phenomena worthy of study \citep{2016kyutoku}. There is, in fact, some observational evidence for the association of r-process elements and hypernovae (\citealt{Yong_2021,Sk_lad_ttir_2021}).

For the scope of this study, we use the nuclear reaction network {\tt SkyNet} to understand the composition of protomagnetar outflows as a function of their physical parameters. We show that heavy nuclei ($A>82$, first r-process peak) are synthesized in lesser quantities compared to previous analytic studies (e.g. \citealt{MGH11,MB_2021}), but the results support an intermediate-mass composition ($\overline{A}\sim 10-50$) which can eventually become UHECRs. The paper is structure as follows. In $\S$\ref{modelandnetwork}, we describe our physical model, its tunable model parameters, and the nuclear network used. In $\S$\ref{results}, we discuss the nucleosynthesis results as a function of the model parameters. In $\S$\ref{discussion} we discuss the general trends in our results and extend this discussion to a broader context. Finally, we conclude in $\S$\ref{summary}. 

\section{Protomagnetar Model and Nuclear Network}\label{modelandnetwork}

\subsection{Protomagnetar outflow properties}\label{properties}

The model for the protomagnetar outflows comes primarily from the neutrino-driven winds prescription of \citet{Qian_1996} (hereafter QW96), the protomagnetar-GRB model of \citet{Metzger_2011}, and updates from \citet{MGH11}. \citet{MGH11} used the analytical models of primarily \citet{Qian_1996}, \citet{Hoffman_1997}, and \citet{Metzger_2011} to estimate heavy element nucleosynthesis in the context of UHECRs. We build on these models to numerically study the outflow composition, which is useful to confirm the key results of analytical studies and reveal more detailed information on the composition of magnetar outflows.

We first describe our modeling of the neutrino-driven outflows launched from rapidly rotating protomagnetars. The important quantities of interest are the wind mass loss rate, entropy per baryon and outflow expansion time-scale, given by (see Appendix \ref{sec:appendNotation}, Table~\ref{table:symbols} for a description of all symbols used in this work): 
\begin{flalign}\label{mdot}
    \Dot{M}=4\pi r^2\rho vf_{\rm open}f_{\rm cent},&&
\end{flalign}
\begin{flalign}\label{s}
    S=\frac{4\pi^2m_Nk_B}{45}\frac{T^3}{\rho}\frac{1}{(\hbar c)^3},&&
\end{flalign}
\begin{flalign}\label{tau}
    \tau_{\textrm{exp}}=\frac{r}{v}\Bigr\rvert_{T_{\textrm{rec}}},&&
\end{flalign}
where the outflow quantities $\Dot{M},~r,~\rho,~v,~S,~T,~\tau_{\textrm{exp}}$, and $T_{\textrm{rec}}$, are the mass loss rate, radial coordinate of the outflow, mass density at the radial coordinate, velocity at radial coordinate, entropy, temperature, expansion time-scale, and recombination temperature, respectively (see QW96). We modify the spherical mass loss rate with $f_{\rm open}$, the fraction of the PNS surface that is threaded by open magnetic field lines, and with $f_{\rm cent}$, the factor that enhances mass loss rate due to a magnetocentrifugal slinging effect. Here, $f_{\rm open}$ is approximated as $(1+\sin^2\chi)^{1/2}R_{\rm NS}/2R_Y$ where, in this work, obliquity angle $\chi = \pi/2$, $R_{\rm NS}$ is the NS radius, and $R_Y$ is the last radius where the magnetic field lines are closed \citep{Metzger_2011} and $f_{\rm cent}=f_{\rm cent,max}(1-\exp[-R_A/R_s])+\exp[-R_A/R_s]$ where $R_A=R_L$min$(\sigma_0^{1/3},1)$ is the Alfven radius, $R_L=c/\Omega$ is the light cylinder radius, $\Omega=2\pi/P$ is the PNS angular velocity, $P$ is the spin period, and $R_s=(GM/\Omega^2)^{1/3}$ is the sonic radius (\citealt{1999isw..book.....L,2007metzger}). $f_{\rm cent,max}$ = $\exp[(P_{\rm cent}/P)^{1.5}$] where $P_{\textrm{cent}} =(2.1\, {\rm ms})\sin\alpha$ ($R_{\textrm{NS}}/10\, {\rm km})^{3/2} (M_{\rm NS}/1.4\, M_{\odot})^{1/2}$, $\alpha$ = max($\theta_{\textrm{open}}/2,\chi$), and $\theta_{\textrm{open}}$ is the opening angle of the polar cap ($M_{\rm NS}$ is the NS baryonic mass) \citep{2008metzger}. 

The $\dot{M}$ correction terms, $f_{\rm open}$ and $f_{\rm cent}$, both evolve while the PNS cools down as $R_Y$, $R_A$, and $R_s$ are all time-dependent quantities. These corrections are valid only within the light cylinder. This step is relevant for our purposes as nucleosynthesis concludes around the light cylinder radius. In reality, the outflow may be collimated due to its interaction with the surrounding cocoon as the jet propagates through the stellar material, but this typically occurs on much larger radii as compared to nucleosynthesis.

\begin{figure}
\centering
\includegraphics[width=\linewidth]{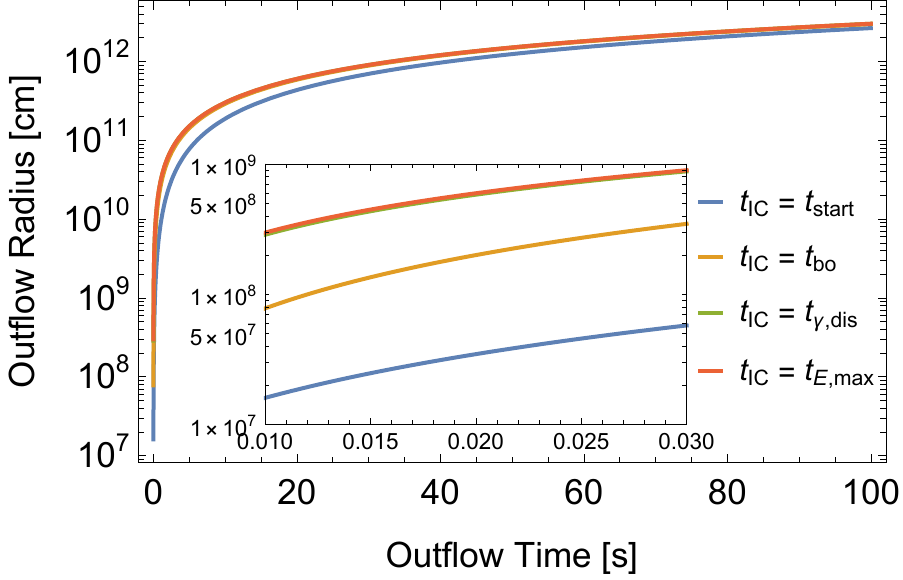}
\vspace{-0.5cm}
\caption{Outflow radius over outflow time for $B_{\textrm{dip}} = 5\times10^{14}\, {\rm G}$, $P_0 = 1.5\, {\rm ms}$ for four outflow launch times of interest (initial condition times $t_{\rm IC}$, see $\S$\ref{modelparam} for description). Inset figure shows the outflow radius at the earliest outflow times. The time when the outflow velocity approaches the speed of light depends on the initial conditions, but generally takes only a few seconds. The starting point of each outflow at $t=0\,{\rm seconds}$ is from the PNS surface at $R_{\rm NS} \sim 12-30\,{\rm km}$, depending on the IC time when each outflow begins.}
\label{fig:rad}
\end{figure}

Combining equations (\ref{mdot}) and (\ref{tau}) gives: 
\begin{flalign}\label{dens}
    \rho=\frac{\tau_{\textrm{exp}}\Dot{M}}{4\pi r^3} (f_{\rm open}f_{\rm cent})^{-1},&&
\end{flalign}
\begin{flalign}\label{temp}
    T=\left(\frac{45\rho S}{4\pi^2m_Nk_B}\right)^{1/3}\hbar c,&&
\end{flalign}
where $m_N$ is the nucleon mass. Different corrections to the mass loss rate and outflow geometry can result in different density profiles. For instance, \citet{MGH11} uses an areal function of the flux tube correction that leads to $\rho \propto r^{-3}$ inside the light cylinder in contrast to $\rho \propto r^{-2}$ for larger radial distances $r \gtrsim R_{\rm Y}$. This transition in the mass density profile at the light cylinder radius can also influence the nucleosynthesis yields (see \citealt{Vlasov_2017}, for detailed discussions). However, in this work we adopt the areal function of the dipolar flux tube from \citet{Metzger_2011} for consistency. We estimate the protomagnetar wind mass loss rate, outflow entropy, and outflow expansion time-scale (equations (\ref{mdot}), (\ref{s}), and (\ref{tau})) with equations (58a), (48a), and (61) from QW96 and updated in \citet{Metzger_2011}. We also include a $\sim$ 20\% general relativistic enhancement to the entropy:
\begin{flalign}\label{mdot2}
    \Dot{M}&=(5\times10^{-5}~\textrm{M}_\odot~\textrm{s}^{-1})f_{\rm open}f_{\rm cent}\notag\\
    &\times C_{\rm es}^{5/3}L_{\nu,52}^{5/3}\epsilon_{\nu,10}^{10/3}R_{10}^{5/3}M_{1.4}^{-2},&&
\end{flalign}
\begin{flalign}\label{s2}
    S=(88.5\ {\rm k_B\ nuc^{-1}})\ C_{\rm es}^{-1/6}L_{\nu,52}^{-1/6}\epsilon_{\nu,10}^{-1/3}R_{10}^{-2/3}M_{1.4},&&
\end{flalign}
\begin{flalign}\label{tau2}
    \tau_{\textrm{exp}}=(68.4\ {\rm ms})\ C_{\rm es}^{-1}L_{\nu,52}^{-1}\epsilon_{\nu,10}^{-2}R_{10}M_{1.4}f_{\rm open},&&
\end{flalign}

\noindent where $C_{\rm es}$ is a heating correction from neutrino-electron scattering (see QW96 equations (50) and (51a)), $L_{\nu,52}=L_{\nu}/10^{52}\, {\rm erg\, s^{-1}}$, $\epsilon_{\nu,10}=\epsilon_{\nu}/10\, {\rm MeV}$, $R_{10}=R_{\textrm{NS}}/10\, {\rm km}$, and $M_{1.4}=M_{\textrm{NS}}/1.4\, M_{\odot}$. 

We adopt the dynamical neutrino quantities for a $1.4\, M_{\odot}$ PNS given in \citet{Pons_1999} and assume that the electron and anti-electron neutrinos have similar $L_{\nu}$ and $\epsilon_{\nu}$ evolution with time. Given that \citet{Pons_1999} does not account for a rotating progenitor, stretching the neutrino quantities accounts for the increased time-scale for NS cooling due to rotation. We follow \citet{Metzger_2011} for this procedure: we take $L_{\nu}\rightarrow L_\nu|_{\Omega=0}~\eta_s^{-1}, t\rightarrow t|_{\Omega=0}~\eta_s, \epsilon_{\nu}\rightarrow \epsilon_{\nu}|_{\Omega=0}~\eta_s^{-1/4}$ where $\Omega=0$ represents the non-rotating cases and $\eta_s$ is the stretch factor fixed to 3. Rapidly rotating NSs with misaligned magnetic fields and rotation axes also experience a suppression in entropy. To account for the reduced neutrino heating as a result of centrifugal acceleration due to rapid rotation, we apply an exponential suppression factor: $S_{\textrm{rot}}=S\times \textrm{exp}[-P_{\textrm{cent}}/P]$.

\subsection{Post-launch evolution}\label{postlaunchevolution}

The distance of the jet-head from the central engine (where, specifically, the jet refers to the collimated, relativistic outflow) is determined from\footnote{Note the original expression is in the relativistic limit; here we multiply by an additional $\beta_j$ factor, which allows us to model the outflow through the sub-relativistic regime as well.}
\citep{drenkhahn}:
\begin{flalign}\label{gammaj}
\beta_j\Gamma_j=
\begin{cases} 
      \sigma_0(r/R_{\textrm{mag}})^{1/3}, & r<R_{\textrm{mag}} \\
      \sigma_0, & r>R_{\textrm{mag}},
\end{cases}&&
\end{flalign}
where:
\begin{flalign}\label{rmag}
    R_{\textrm{mag}}=(5\times10^{12}\, \rm{cm})\left(\frac{\sigma_0}{10^2}\right)^2\left(\frac{P}{ms}\right)\left(\frac{\epsilon}{0.01}\right)^{-1},&&
\end{flalign}
is the magnetic saturation radius. Here, $\sigma_0$ is the magnetization as described below, and $\epsilon$ is a parameter used to describe the reconnection speed (here, $\epsilon$ is assumed to be 0.01). Rearranging further gives $dr/dt= c(\alpha/\sqrt{1+\alpha^2})$, where $\alpha=\sigma_0(r/R_{\textrm{mag}})^{1/3}$. The outflow velocity quickly becomes relativistic, with a weak dependence on the initial outflow conditions $\sigma_0$ and $R_{\textrm{mag}}$ (see Fig.~\ref{fig:rad}). At later launch times, the magnetar outflows become ultrarelativistic more quickly than at early launch times, but in general, $dr/dt$ approaches $c$ within $\sim 1-5\, {\rm seconds}$. The initial magnetization of the outflow at launch is defined as 
\begin{flalign}
    \label{sigma0}
    \sigma_0=\frac{\phi^2\Omega^2}{\Dot{M}c^3},&&
\end{flalign}
where $\phi=(f_{\textrm{open}}/4\pi)B_{\textrm{dip}}R_{\textrm{NS}}^2$ is the magnetic flux threading the open magnetosphere and $B_{\rm dip}$ is the dipolar magnetic field strength. 

If the central engine and outflow are active for long enough, the jet eventually breaks out of the progenitor. 
The jet breakout time is then given by \citep{MB_2021} 
\begin{flalign}
    t_{\textrm{bo}}=(4.2~\rm{s})\left(\frac{B_{dip}}{3\times10^{15}~\rm{G}}\right)^{-1/3}\left(\frac{P_0}{3~\rm{ms}}\right).&&
    \label{tbo}
\end{flalign}
This is based on the breakout-time analytical fit given by \citet{Bromberg_2015} and assumes a magnetized, Poynting-flux dominated jet with opening angle $\theta_j=7^{\circ}$ that can be collimated by the interactions with the stellar material. Equation (\ref{tbo}) is obtained assuming a $15\, M_{\odot}$ and $4\, R_{\odot}$ Wolf-Rayet star with stellar density profile $\rho \propto r^{-2.5}$, and is applicable for $B_{\rm dip}$ between $[3\times10^{14},3\times10^{16}]\, {\rm G}$ and $P_0$ between $[1.0,5.0]\, {\rm ms}$, as considered in \citet{MB_2021}, which includes the parameter range considered in this work.

At large radial distances after breakout ($r \sim 10^{13}-10^{16}\, {\rm cm}$), the jet dissipates its energy either through internal shocks or magnetic reconnection. During this dissipation phase, the jet powers high energy gamma-ray emission (the GRB). It is also during this dissipation that the nuclei synthesized in the neutrino-driven outflows are accelerated to ultra-high energies \citep{MGH11}. We use the magnetic reconnection model of \citet{2002drenkhahn}, where dissipation occurs from small outflow radius up to the saturation radius, $R_{\rm mag}$ (equation (\ref{rmag})). We consider first-order Fermi acceleration with a characteristic time-scale similar to the Larmor gyration time-scale. For nuclei to be accelerated, this acceleration time-scale must be smaller than the time-scales for jet expansion and synchrotron cooling; this condition determines the maximum UHECR energy. Soon after core collapse, synchrotron cooling is more restrictive but in the regime that concerns us, the expansion time-scale limits the maximum energy (see \citealt{MB_2021}, Fig. 9, for further details).

For heavy nuclei to survive and escape the GRB, they must first endure some critical number of photodisintegration interactions. This number of interactions, or photodisintegration optical depth $\tau_{\gamma-N}$, is estimated assuming a Band prompt emission spectrum for the surrounding photon field \citep{MB_2021}, and is given as
\begin{flalign} 
    \tau_{\gamma-N} \approx \frac{\dot{E}_{\rm iso}\varepsilon_{\rm rad}C\sigma_{\rm GDR} (\Delta \epsilon_{\rm GDR}/\overline{\epsilon}_{\rm GDR})}{4\pi\epsilon_p rc\Gamma_j^2},&&
    \label{photointeractions}
\end{flalign}
where $\Dot{E}_{\rm iso}$ is the isotropic jet luminosity, $\varepsilon_{\rm rad} \sim 0.1 - 1$ is the radiative efficiency of the jet, and $C \sim 0.2$ is the fraction of gamma-ray photons released below the peak energy $\epsilon_p \sim 0.1-1\, {\rm MeV}$ of the Band spectrum \citep{MGH11}. Since giant dipole resonances dominate the energy loss processes, the relevant resonance width
$\Delta \epsilon_{\rm GDR}/\overline{\epsilon}_{\rm GDR} \sim 0.4\, (A/56)^{0.21}$ and cross section $\sigma_{\rm GDR} \sim 8\times10^{-26}\,(A/56)\, {\rm cm^2}$ are used (see \citealt{Khan_2005,Murase_2008}). We conservatively use this collisional optical depth since the inelasticity of collisions is typically $\lesssim 10\%$. Including a treatment for inelastic collisions generally broadens the time by about 10 seconds between our photodisintegration and max acceleration IC times (see \S\ref{modelparam}) thereby allowing more heavy nuclei to be synthesised and accelerated to UHECR energies (see \citealt{MB_2021}). Also, our expression assumes a pure iron composition, and thus is a conservative estimate, since one photodisintegration event does not fully destroy the heavy nuclei. In fact, since the nuclei scattering inelasticity factor is typically $\lesssim 10\%$, we can expect that a photodisintegration optical depth of a few will still maintain a heavy ultra-high energy cosmic-ray population.

\subsection{Model parameters}\label{modelparam}

We next describe the parameter choices for our systematic study of nucleosynthesis in protomagnetar outflows. The nucleosynthesis depends on the properties of the outflow, which in turn depend on the protomagnetar and its time evolution. 

Our semi-analytic protomagnetar model primarily depends on the surface dipole field strength $B_{\textrm{dip}}$ and initial spin period $P_0$ of the protomagnetar. We adopt six $B_{\textrm{dip}}$ values between $[5\times 10^{14},1\times 10^{16}]$ G and five $P_0$ values between $[1.5,3.5]$ ms. We must also specify the electron fraction $Y_e = n_p/(n_n+n_p)$ of the outflow, where $n_p(n_n)$ are proton(neutron) densities. There is currently significant uncertainty in the value of $Y_e$ and its evolution over time, and our protomagnetar model does not self-consistently compute them. Instead, we adopt four fixed electron fraction values around and including 0.5, and assume them to not depend on $B_{\rm dip}$, $P_0$ or vary over time. In our study, we do not account for the $\nu p$-process (\citealt{Fr_hlich_2006,Arcones_2011}) that would change the electron fraction and increase the neutron-to-seed ratio. 

Once these parameters are set, the protomagnetar evolves over time according to our semi-analytic model as described in the previous sections. We describe this evolution using the notation of ``Initial Condition'' (IC) time: IC times are characteristic post-core-collapse times that represent physical stages in the protomagnetar evolution. Early IC times define the head of the jet, while later IC times define later parts of the jet. In this way, IC time can also be thought of as different locations within the magnetized jet. 

We explore nucleosynthesis at four specific IC times (consequently, locations in the jet) corresponding to four physical epochs, as described below (also see \citealt{MB_2021}). In reality, the system ejects material continuously, not just at four epochs, and the overall jet composition is the superposition of this continuous outflow. The discretization of the continuous outflow into our four IC times serves to estimate the composition at four physically different epoch/locations. Although, in principle, the outflows can last for longer time-scales (up to $\sim10^3\,{\rm s}$ for GRB emission and $\sim10^4\,{\rm s}$ for X-ray plateaus), our final IC time marks the end of the epoch relevant for accelerating UHECRs. Around $t_{\rm E,max}$ (details below), the outflows become optically thin to neutrinos \citep{MB_2021} and the mass loss rate begins to fall off significantly.

To summarize, we consider the following values: 
\begin{itemize}[leftmargin=*]
    \item \underline{\textbf{$B_{\textrm{dip}}$ (G)}}: $5\times10^{14}$, $7.5\times10^{14}$, $1\times10^{15}$, $2.5\times10^{15}$, $5\times10^{15}$ and $1\times10^{16}$
    \item \underline{\textbf{$P_0$ (ms)}}: 1.5, 2, 2.5, 3, 3.5
    \item \underline{\textbf{$Y_e$}}: 0.45, 0.475, 0.5, 0.55
    \item \underline{\textbf{IC Time}}: we consider the following four times
    \begin{enumerate}[leftmargin=2em]
        \item Start ($t_{\textrm{start}}$): $\sim$ 0.5 seconds. This is when the outflow starts, assumed to be the same for all protomagnetar configurations.
        \item Jet breakout ($t_{\textrm{bo}}$): The time when the bipolar outflow breaks out of the progenitor envelope. Depends on $B_{\textrm{dip}}$ and $P_0$ as well as the progenitor and jet dynamics assumptions. We adopt equation (\ref{tbo}). Outflows before this time have not escaped from the star.
        \item Photodisintegration time ($t_{\textrm{$\gamma$,dis}}$): Even if the outflow contains nuclei they can be destroyed during the outflow evolution. We define the photodisintegration time to be when the majority of synthesized heavy elements survive photodisintegration during particle acceleration at the dissipation radius, i.e.,  the number of interactions that can destroy nuclei ($\tau_{\gamma-N}$, equation (\ref{photointeractions})) $\approx$ 1. This   depends on $B_{\textrm{dip}}$ and $P_0$. 
        \item Max acceleration time ($t_{\textrm{E,max}}$): The maximum energy reached by accelerated nuclei is determined by the balance of the acceleration time and energy loss times, and generally shows a peak before falling with time. We define the max acceleration time as the final time when heavy nuclei can be accelerated above a maximum energy of $10^{20}\, {\rm eV}$; after this time, the maximum energy reached is $<10^{20}\, {\rm eV}$. We assume a pure iron composition for this calculation, and the resulting time depends on $B_{\textrm{dip}}$ and $P_0$ (see \citealt{MB_2021}).
    \end{enumerate}
\end{itemize}

Prior to $t_{\rm bo}$, any nuclei synthesized will not make it out of the star unscathed, and hence we assume they cannot become ultra-high energy cosmic rays. However, $t_{\rm start}$ has its uses. Analysing the nucleosynthesis products at a specific time (independent of $B_{\textrm{dip}}$ and $P_0$, at $\sim 0.5\, {\rm seconds}$) provides a benchmark to extract some of the qualitative dependencies of initial magnetar properties to the final results. This specific time is chosen to reduce the effect of numerical artifacts at very early times. 

The most significant epoch at which protomagnetar outflows can contribute to the UHECR composition is between the photodisintegration and max acceleration times - where nuclei can be synthesized, avoid photodisintegration, and be accelerated to ultra high energies. We stress, however, our conditions are conservative, and that in reality, some heavy nuclei are expected to survive even between $t_{\textrm{bo}}$ and $t_{\textrm{$\gamma$,dis}}$. 

Although our adopted photodisintegration and maximum acceleration times depend on the assumption that the outflow is dominated by a pure iron composition, this is not a strong dependence (see \citealt{MB_2021}) and therefore they can be used as a proxy for the true time. Improving the IC times and resolution in parameter space is left for future study.

There is an inherent uncertainty about the time evolution of $Y_e$ in magnetars' neutrino-driven outflows. The true electron fraction in protomagnetar outflows may vary as the equilibrium $Y_e$ (see QW96, equation (77)), but will still depend implicitly on the neutrino luminosities, the neutrino mean energies, and the dynamical evolution of the PNS. Furthermore, it is impacted by the effects of neutrino oscillations, which undergo potentially time-varying collective oscillations. In light of these uncertainties, we test nucleosynthesis for fixed $Y_e$ as an independent model parameter. Since we consider four IC times independently of our choices of $Y_e$, we can use different combinations to make qualitative statements for any assumed time evolution of $Y_e$ (within the range considered here).

\subsection{Nuclear reaction network}\label{network}

\begin{figure}
\centering
\includegraphics[width=\linewidth]{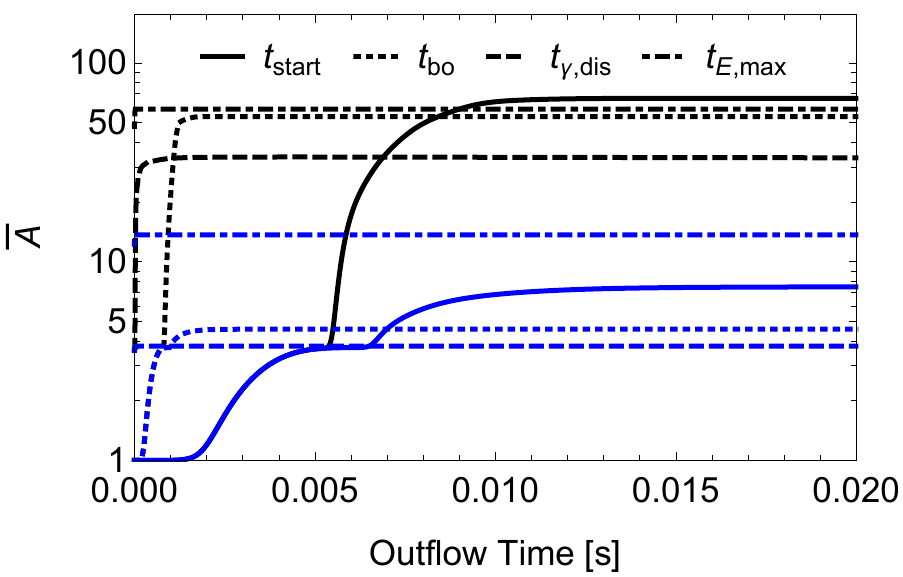}
\vspace{-0.5cm}
\caption{Mean mass number of the outflow as a function of outflow time for $B_{\textrm{dip}} = 5\times10^{14}\, {\rm G}$, $P_0=1.5\, {\rm ms}$ model. Evolution is shown for two extremal electron fractions (black lines for $Y_e=0.45$ and blue lines for $Y_e= 0.55$) and four IC times (as labeled). $\overline{A}$ first saturates to helium-4 before heavier elements are synthesized. For all launch times, nucleosynthesis concludes by $\sim 10\, {\rm ms}$ post launch.} 
\label{fig:abartime}
\end{figure}

To calculate nucleosynthesis in protomagnetar outflows we use the nuclear reaction network {\tt SkyNet} \citep{Lippuner_2017}. {\tt SkyNet} uses a modified Helmholtz equation of state \citep{timmesswesty} that supports initial densities up to $10^{11}\, {\rm g\, cm^{-3}}$. All models tested in this work begin with densities lower than this limit. 
Most configurations of $B_{\textrm{dip}}$ and $P_0$ begin their thermodynamic trajectories with temperatures above the $7\, {\rm GK}$ threshold for nuclear statistical equilibrium (NSE); after outflow temperatures decrease beyond this threshold, full network evolution kicks in.

We use 93,271 forward reaction rates from JINA REACLIB \citep{reaclib} and inverse reactions are calculated with detailed balance to ensure consistency with NSE. Masses and partition functions for nuclear species are used from WebNucleo XML included with REACLIB. Finally, 7836 nuclear species are evolved up to $Z = 112$, $A = 337$.

\begin{figure*}
\includegraphics[width=0.49\textwidth]{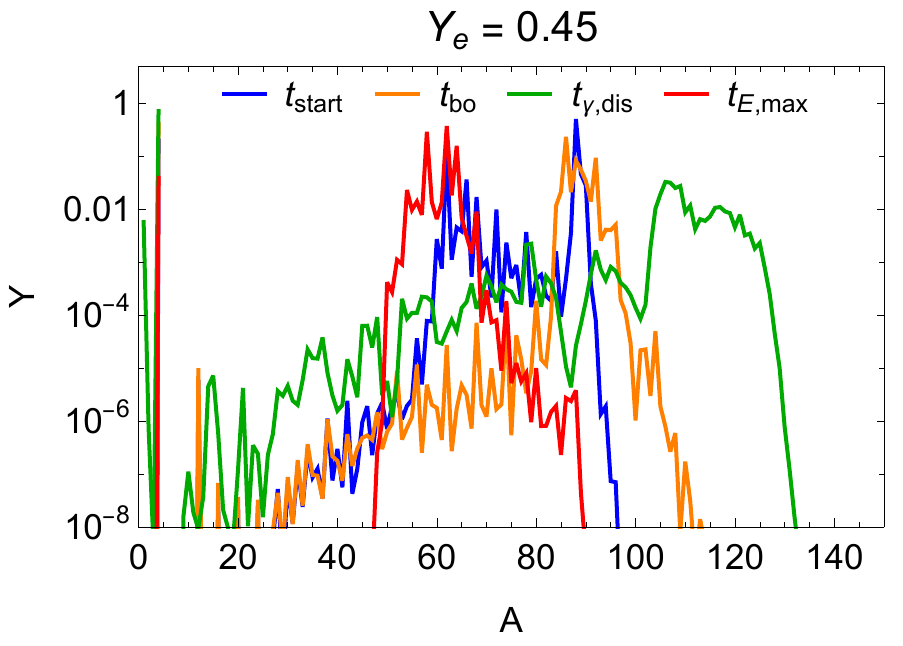} 
\includegraphics[width=0.49\textwidth]{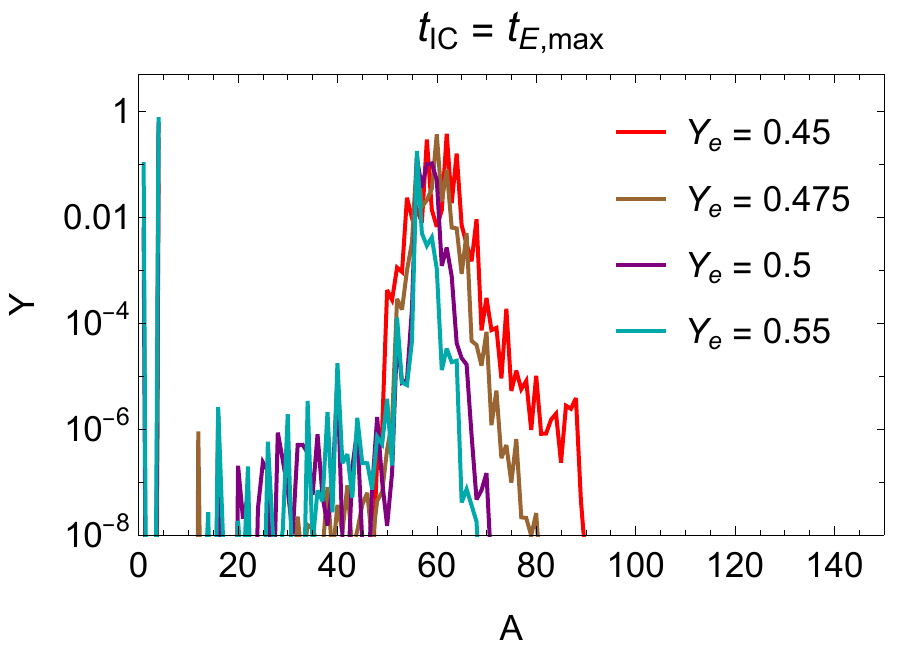}
\vspace{-0.2cm}
\caption{\textit{Left panel:} Abundance pattern for an electron fraction of $Y_e = 0.45$ is shown for the $B_{\textrm{dip}} = 5\times10^{14}\, {\rm G}$, $P_0 = 1.5\,{\rm ms}$ model at four IC times. \textit{Right panel:} Abundance pattern at the final IC time when max nuclei acceleration occurs is shown for four $Y_e$ values. In both panels, if isotopes have the same mass number, their abundances are added.} 
\label{fig:ybya}
\end{figure*}

To produce the thermodynamic trajectories (density, temperature and electron fraction with time) necessary for {\tt SkyNet}, we make a few simplifying assumptions. The magnetization, entropy, mass loss rate, and expansion time-scale ($\sigma_0,~S,~\Dot{M},~\tau_{\rm exp}$) all evolve with time. However, since we want to probe the outflow composition at four epochs, we evaluate these quantities at the four IC times and keep them constant in Outflow time. Hence, we call them `ICs'; they determine the outflow properties at a given IC time, but remain constant in Outflow time. This is primarily justified because nucleosynthesis concludes after $\sim 10\, {\rm ms}$ in general (see Fig.~\ref{fig:abartime}) and these quantities are approximately constant in that time. We account for the evolution of the thermodynamic trajectories due to the expansion of the outflow into the progenitor by numerically solving the $dr/dt$ equation for outflow radius over time. In this way, the initial densities and temperatures are set by the ICs and then evolve according to the radial coordinate time evolution. For all trajectories, we evolve the network until $100\, {\rm seconds}$ in Outflow time. Appendix \ref{sec:appendIC} discusses how these ICs depend on the protomagnetar parameters. 

\section{Results}\label{results}

\subsection{Abundance patterns}\label{abundancepatterns}

\begin{figure}
\centering
\includegraphics[width=\linewidth]{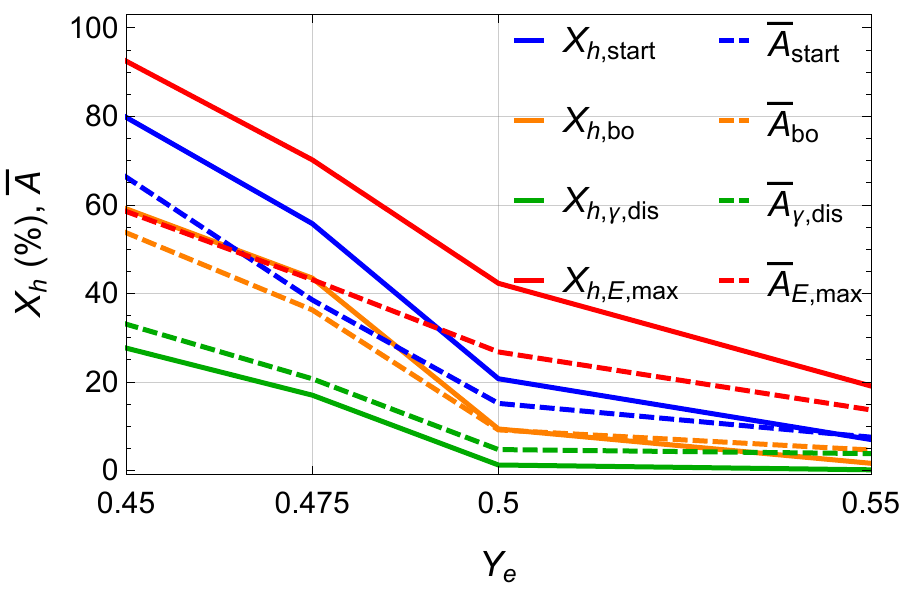}
\vspace{-0.5cm}
\caption{The mass fraction of nuclei heavier than iron ($X_h$, in $\%$, solid lines) and the mean mass number ($\overline{A}$, dashed lines) as functions of electron fraction (see equations (\ref{abar}) and (\ref{xh})). The results are shown for four IC times for the $B_{\textrm{dip}} = 5\times10^{14}\, {\rm G}$, $P_0 = 1.5\,{\rm ms}$ model.}
\label{fig:xhaB514P15}
\end{figure}

In Fig.~\ref{fig:ybya}, we show the distributions of abundances $Y_i$ for species $i$, where $\sum_iY_i$ is normalized to 1. We show for a fixed $Y_e=0.45$ at four different IC times (left panel) and at max acceleration time for four $Y_e$ (right panel) for a magnetar with $B_{\textrm{dip}} = 5\times10^{14}\, {\rm G}$ and $P_0=1.5\, {\rm ms}$. In all scenarios, significant amounts of free nucleons and helium are formed, but in proton-rich conditions and at late times they become more significant compared to heavy elements. At early IC times after core collapse, the abundance pattern peaks at nickel-62 ($t_{\rm start}$) and above the first r-process peak at strontium-88 ($t_{\rm start}$ and $t_{\rm bo}$). At late IC times ($t_{\rm \gamma,dis}$ and $t_{\rm E,max}$), heavier elements are formed, but in smaller quantities, including nuclei near the second r-process peak ($A \sim 130$). This trend continues as time progresses. 

As a function of $Y_e$, more neutron-rich conditions favor heavy nucleosynthesis with larger relative abundances, while with proton-rich conditions the opposite is true. Generally, across all other configurations of $B_{\textrm{dip}}$ and $P_0$, these trends hold true. There is an important exception in this particular configuration: at the max acceleration time, $t_{\rm E,max}$, the abundance pattern resembles more closely the pattern at $t_{\rm start}$ (see $\S$\ref{meancomptrends} for more details). Although the trends in IC time are generally monotonic among other configurations, the ICs at this time result in an overall higher mean mass number (see Appendix \ref{sec:appendIC} for the ICs of all configurations).

The majority of configurations do not synthesize third r-process peak elements ($A \sim 196$) because the outflow entropies are much too low, their expansion time-scales too long, and their neutron-to-seed ratios are quite small (see \citealt{Hoffman_1997,nagataki2001,Thompson_2001}). In other words, the figure of merit parameter for these outflows is always below the critical value required for the synthesis of third r-process peak elements \citep{Hoffman_1997}. However, for configurations with $Y_e = 0.5$ at $t_{\rm \gamma,dis}$ and $t_{\rm E,max}$, third r-process peak nuclei are synthesized, albeit in very small abundances by mass. For some regions of parameter space in entropy and expansion time-scale, a particular process occurs. Below NSE, there is a persistent disequilibrium between free nucleons and helium nuclei. Even a small abundance of free nuclei can then become seeds for the available free nucleons. These are captured readily by the seeds and what is left is dominated by mass in alpha particles, but with almost negligible abundances of very heavy nuclei. These include r-process elements, but do not match with solar abundances. This process is sensitive not just to the entropy and expansion time-scale, but also to electron fraction. It can also occur for outflows that are only slightly neutron- or proton-rich, but becomes less effective when $Y_e$ moves farther away from 0.5 (see \citealt{PhysRevLett.89.231101} for details on the process and \citealt{2016fujibayashi} for another example).

Our results are broadly consistent with studies of nucleosynthesis under similar conditions. For example, \citet{Vlasov_2017} explored nucleosynthesis with {\tt SkyNet} under a comparable protomagnetar parameter space. They adopted a force-free derived field structure in contrast to our simple analytic model, but found abundance patterns similar to those shown in Fig.~\ref{fig:ybya}. A notable feature common to both is the negligible abundance of $Z = 41$ (present in some, not all configurations of the model parameters explored here). For the model with $B_{\textrm{dip}}=1\times10^{16}\, {\rm G}$, $P_0=3\, {\rm ms}$, $Y_e=0.45$, at $t_{\rm start}$, we carried the calculation out to $10^9\, {\rm seconds}$ in Outflow time. Between $100\, {\rm seconds}$ and $10^9\, {\rm seconds}$, isotopes of niobium with mass numbers $A = 94,\, 96,\, 98,\, 99,\, 100,\, 101,\, 102,\, 103,\, {\rm and}\, 105$ underwent complete decay. In addition, \citet{2014surman} and \citet{2017bliss} study the weak r-process under similar astrophysical conditions and comparable entropies and electron fraction. These studies confirm an abundance pattern that peaks around these nickel- and strontium-like nuclei. \citet{2017bliss} also confirms a negligible abundance of nuclei near $Z = 41$. Finally, other magnetorotational studies (e.g. \citealt{2014PhDT.......206W,Halevi_2018,M_sta_2018}) also show successful synthesis of heavy elements, although the magnetic field strengths and electron fractions are generally lower than the corresponding values considered in this work.

\subsection{Mean composition trends}\label{meancomptrends}

To quantify how much nucleosynthesis occurs, we express the results in terms of abundance-weighted mean mass number and mass fraction of nuclei heavier than iron ($A \geq 56$), that is:

\begin{flalign}\label{abar}
    \overline{A}=\sum_iY_iA_i/\sum_iY_i,&&
\end{flalign}
\begin{flalign}\label{xh}
    X_h=\sum_{A\geq56}X_i/\sum_iX_i,&&
\end{flalign}
where $X_i=Y_i\times A_i/\sum_iY_i\times A_i$ is the mass fraction for species $i$ (see Appendix \ref{sec:appendXh} for a discussion on the analytic estimation of $X_h$ with the parameter space considered here).

In Fig.~\ref{fig:xhaB514P15}, we show the mass fraction above iron and mean mass number for the same model as in Fig.~\ref{fig:ybya}. $X_h$ (\%) and $\overline{A}$ show the same trends: significant nucleosynthesis in neutron-rich conditions and then a sharp decrease as the outflow becomes more and more proton-rich. As IC time increases, synthesis reaches higher mass numbers (see Fig.~\ref{fig:ybya}), but in small quantities, resulting in smaller $X_h$ and $\bar{A}$ than at earlier times. The exception to this is at $t_{\rm E,max}$, when $\overline{A}$ is higher than expected, as we saw also in Fig.~\ref{fig:ybya}. We have found that this does occur for other configurations with $P_0=1.5\, {\rm ms}$, but is especially pronounced for the case of $B_{\textrm{dip}} = 5\times10^{14}\, {\rm G}$. In proton-rich conditions, the nucleosynthesis products are increasingly dominated by free nucleons and helium as IC time increases.

\begin{figure*}
\begin{minipage}{0.245\textwidth}
\begin{tikzpicture}
\node (img) {\includegraphics[width=0.95\linewidth]{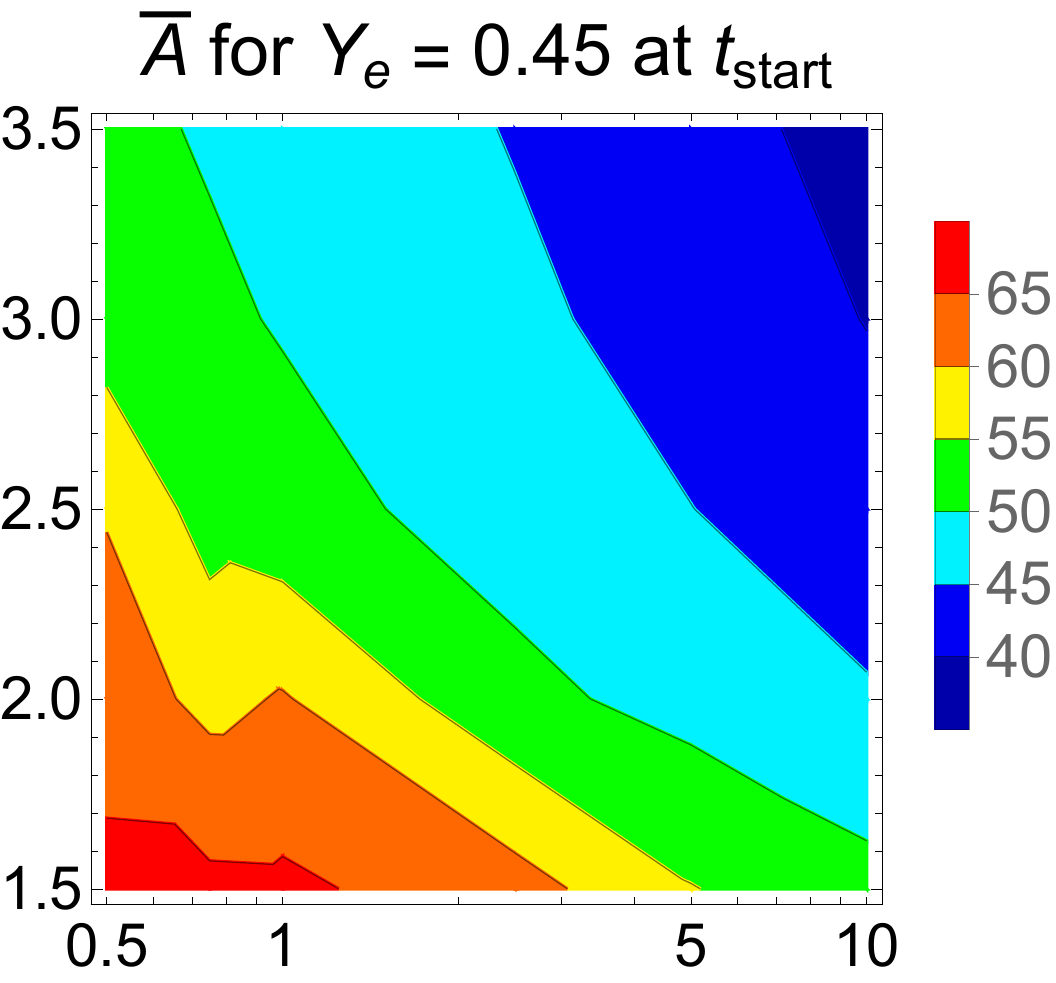}};
\node[left=67pt, node distance=0cm, rotate=90, anchor=center,font=\color{white}] {$|$};
\end{tikzpicture}
\end{minipage}
\begin{minipage}{0.245\textwidth}
\begin{tikzpicture}
\node (img) {\includegraphics[width=0.95\linewidth]{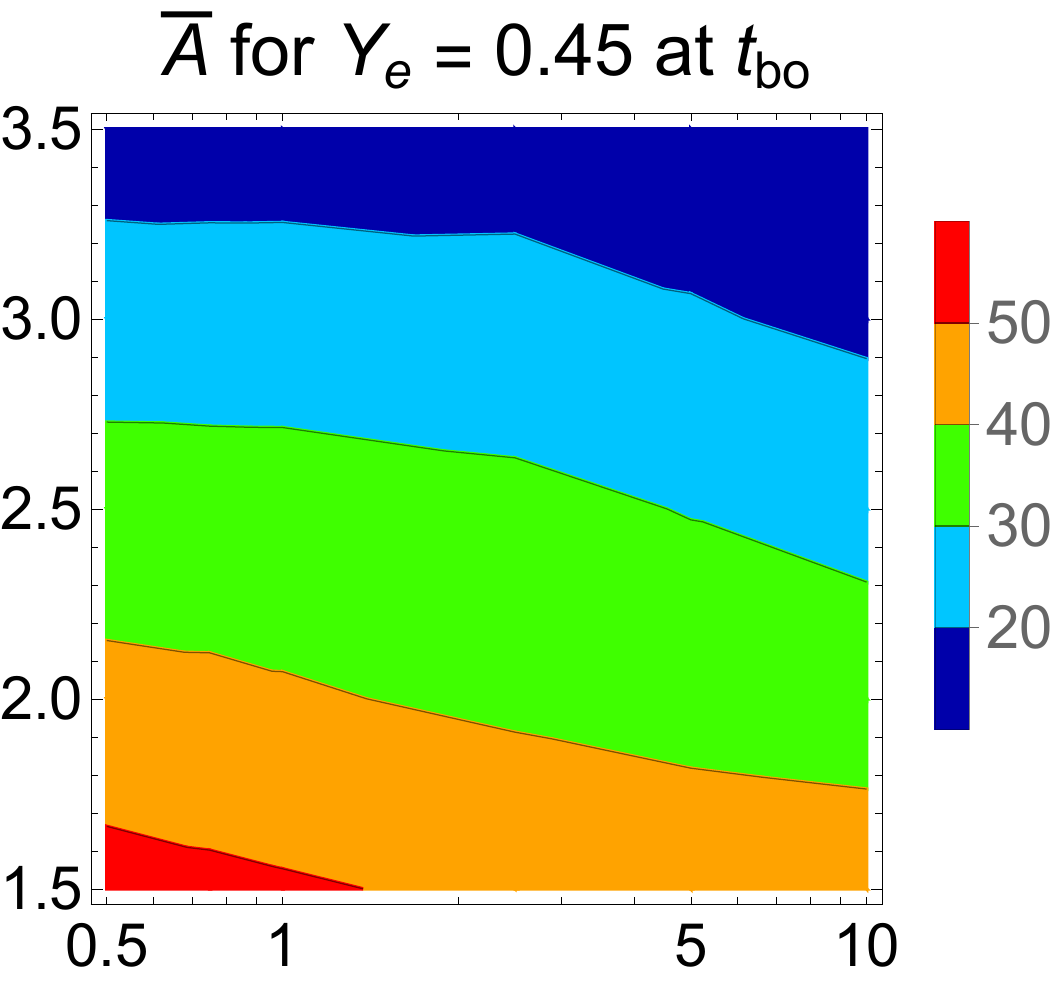}};
\node[left=67pt, node distance=0cm, rotate=90, anchor=center,font=\color{white}] {~$|$};
\end{tikzpicture}
\end{minipage}
\begin{minipage}{0.245\textwidth}
\begin{tikzpicture}
\node (img) {\includegraphics[width=0.95\linewidth]{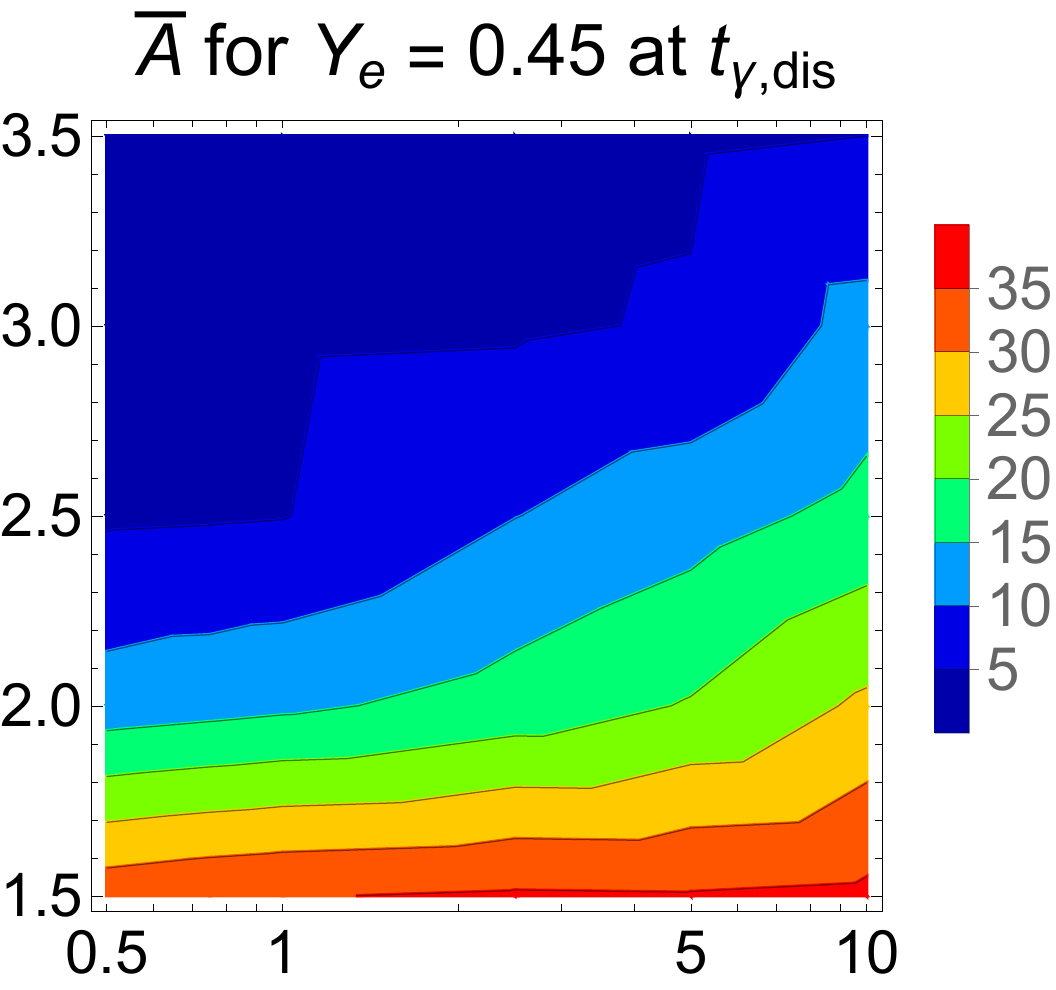}};
\node[left=67pt, node distance=0cm, rotate=90, anchor=center,font=\color{white}] {~$|$};
\end{tikzpicture}
\end{minipage}
\begin{minipage}{0.245\textwidth}
\begin{tikzpicture}
\node (img) {\includegraphics[width=0.95\linewidth]{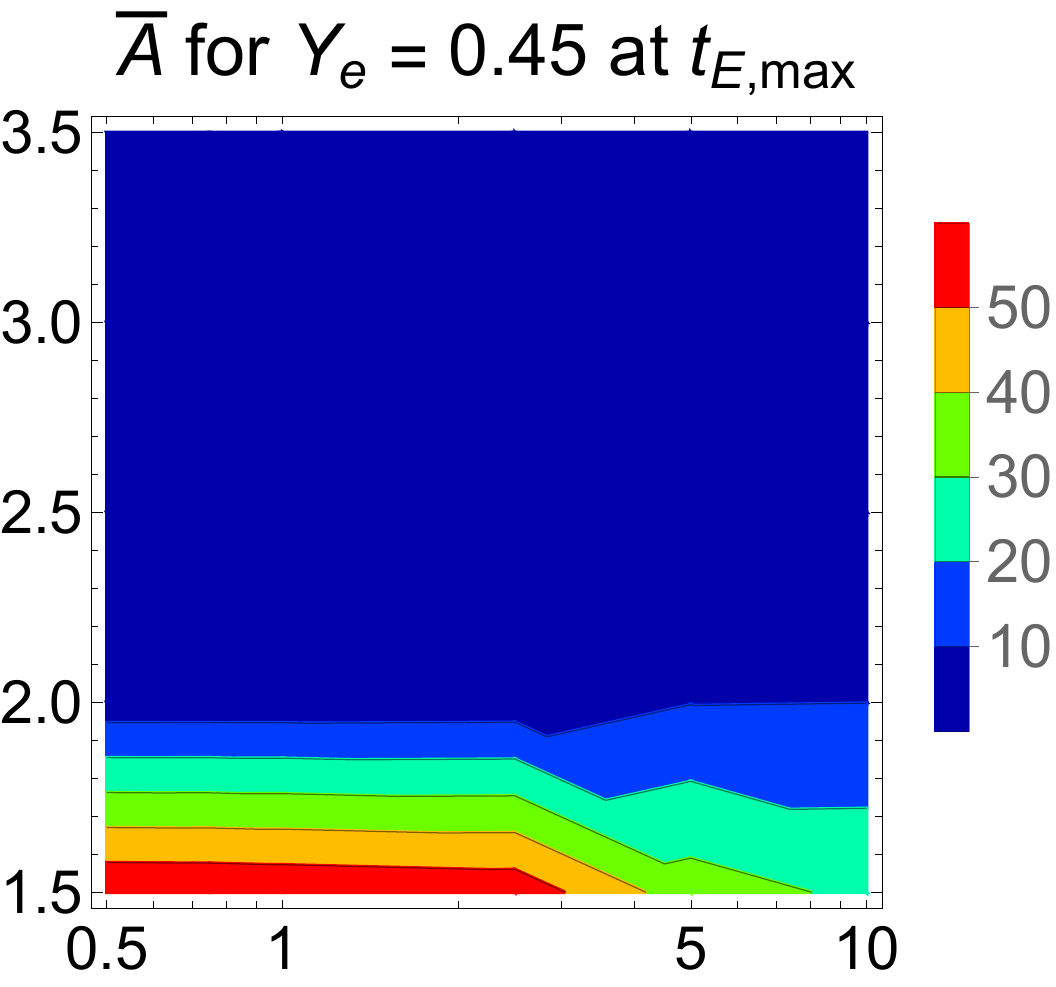}};
\node[left=67pt, node distance=0cm, rotate=90, anchor=center,font=\color{white}] {~~~~$|$};
\end{tikzpicture}
\end{minipage}
\newline
\begin{minipage}{0.245\textwidth}
\begin{tikzpicture}
\node (img) {\includegraphics[width=0.95\linewidth]{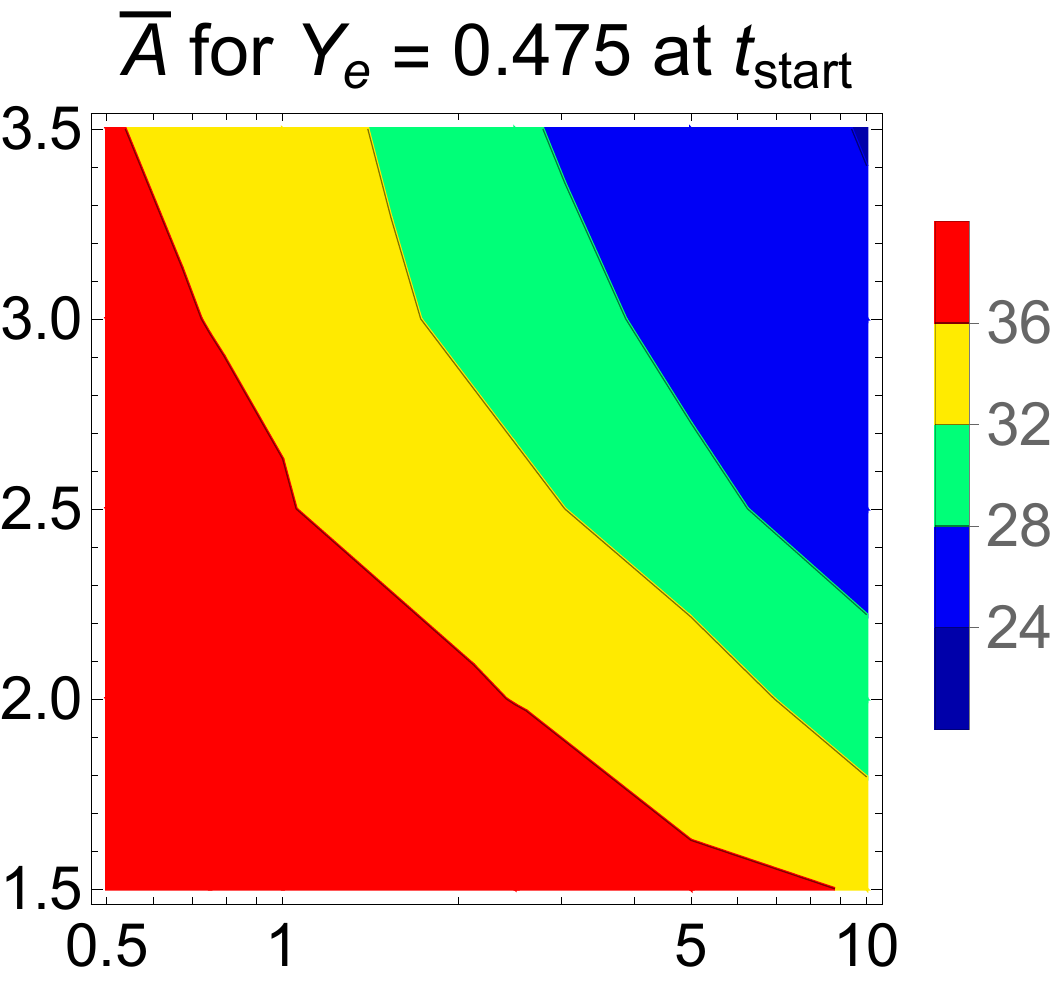}};
\node[left=67pt, node distance=0cm, rotate=90,xshift=-40pt]{[ms]};
\end{tikzpicture}
\end{minipage}
\begin{minipage}{0.245\textwidth}
\begin{tikzpicture}
\node (img) {\includegraphics[width=0.95\linewidth]{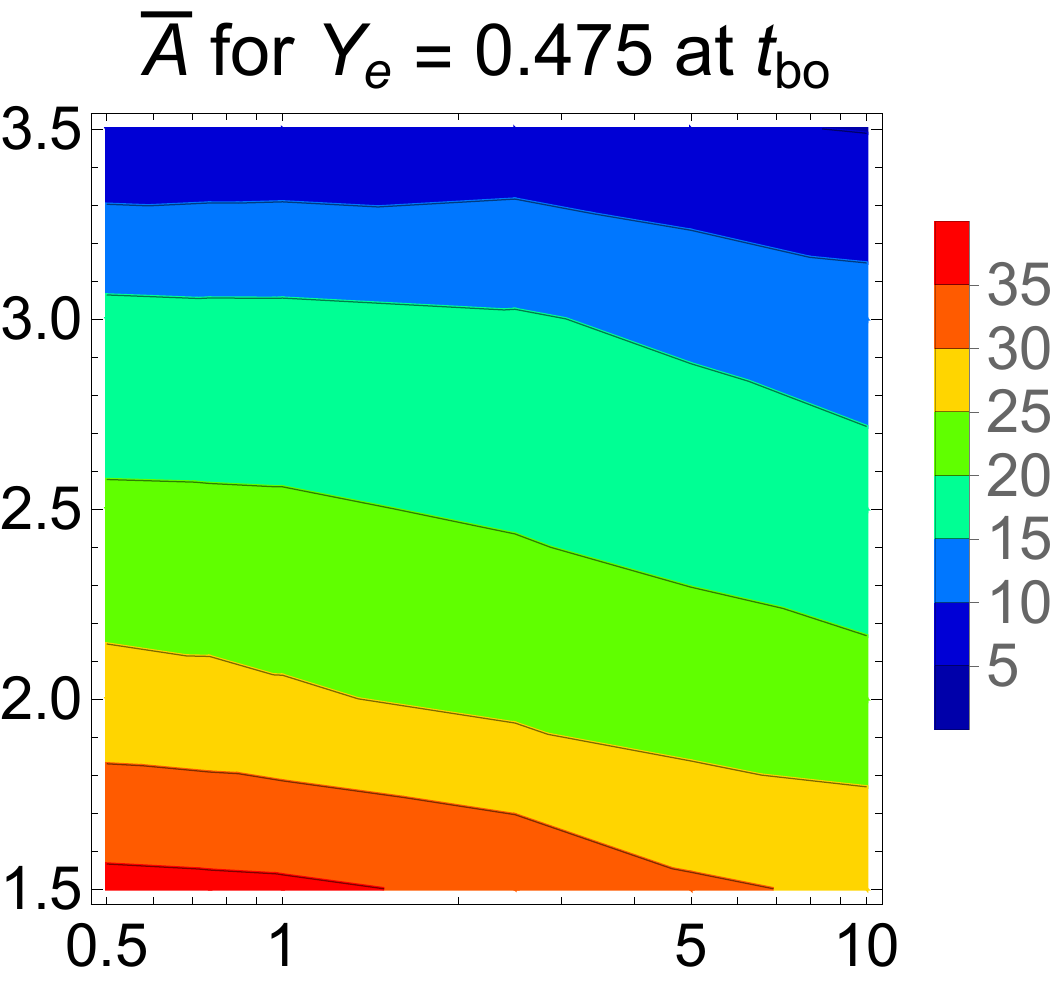}};
\node[left=67pt, node distance=0cm, rotate=90, anchor=center,font=\color{white}] {~~$|$};
\end{tikzpicture}
\end{minipage}
\begin{minipage}{0.245\textwidth}
\begin{tikzpicture}
\node (img) {\includegraphics[width=0.95\linewidth]{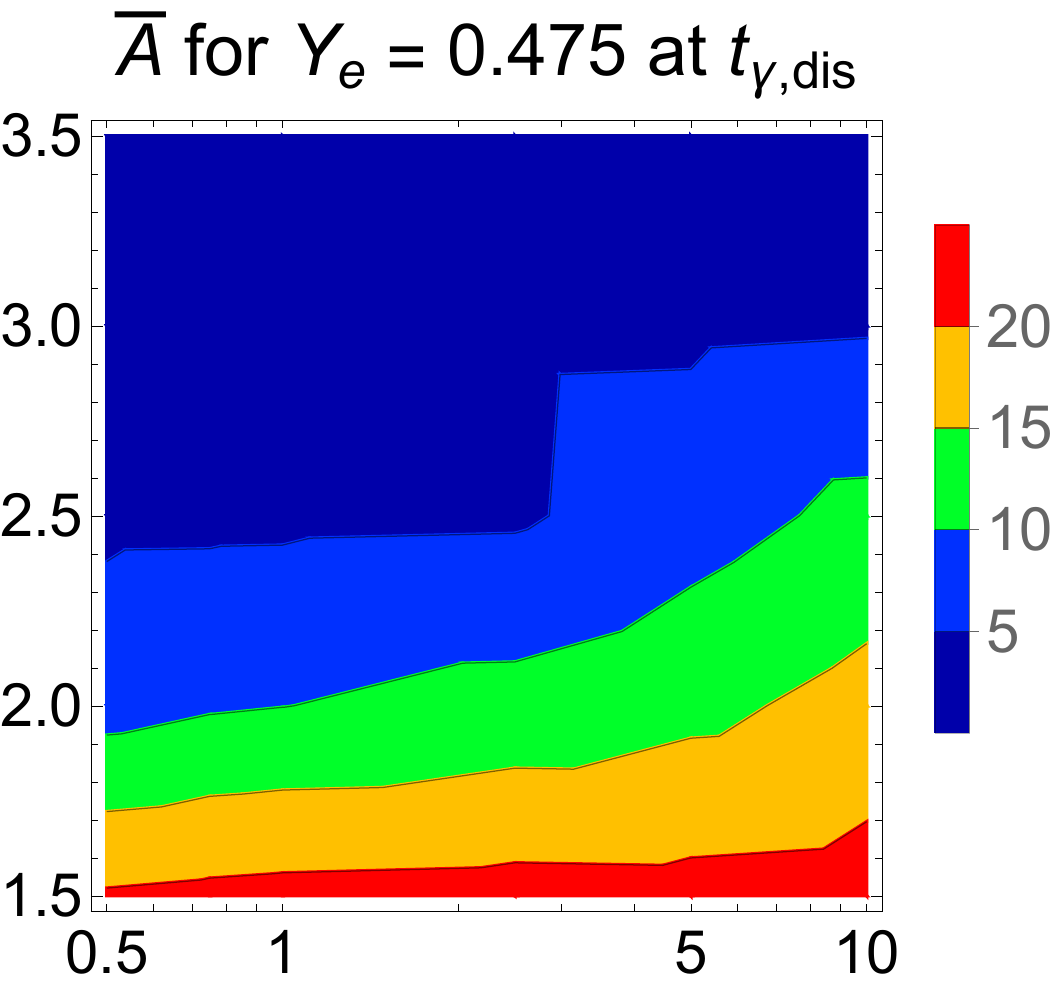}};
\node[left=67pt, node distance=0cm, rotate=90, anchor=center,font=\color{white}] {~~~~$|$};
\end{tikzpicture}
\end{minipage}
\begin{minipage}{0.245\textwidth}
\begin{tikzpicture}
\node (img) {\includegraphics[width=0.95\linewidth]{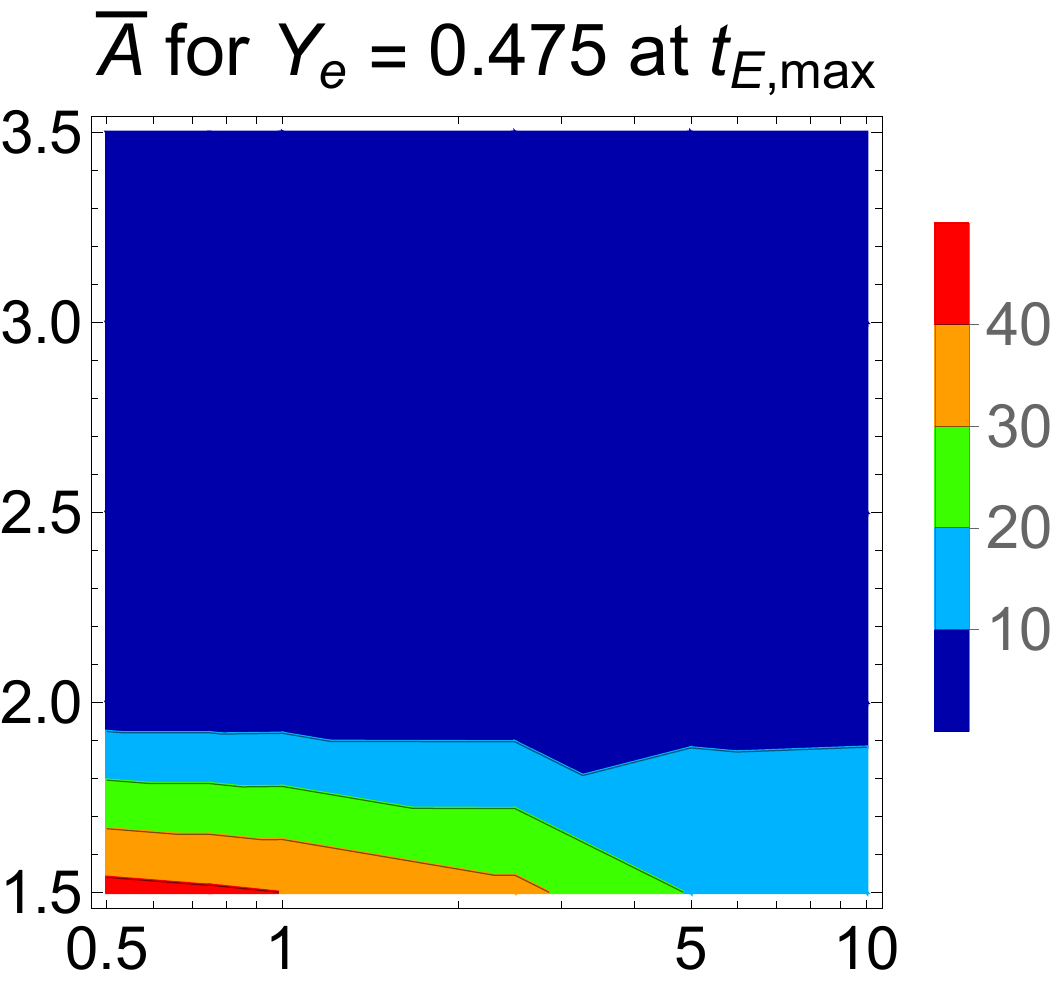}};
\node[left=67pt, node distance=0cm, rotate=90, anchor=center,font=\color{white}] {$|$};
\end{tikzpicture}
\end{minipage}
\newline
\begin{minipage}{0.245\textwidth}
\begin{tikzpicture}
\node (img) {\includegraphics[width=0.95\linewidth]{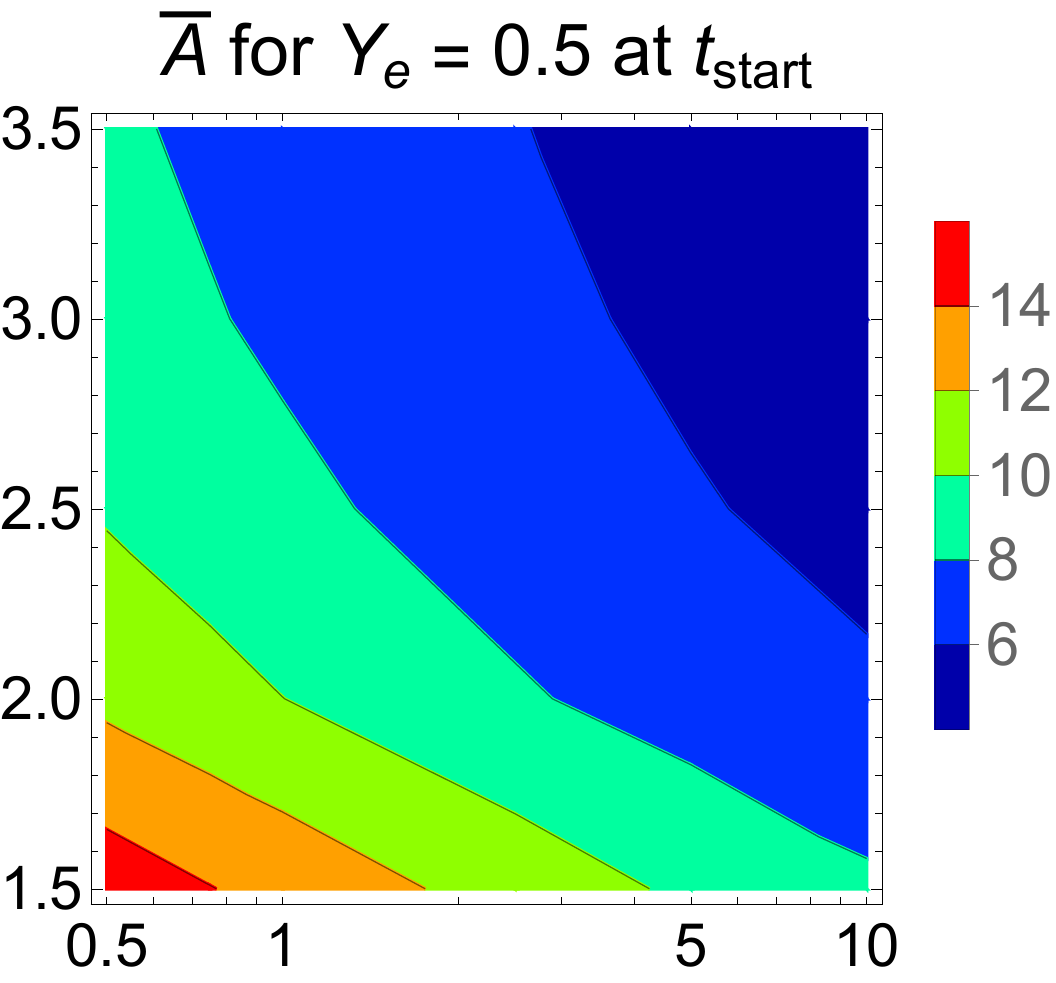}};
\node[left=67pt, node distance=0cm, rotate=90,xshift=60pt]{$P_0$};
\end{tikzpicture}
\end{minipage}
\begin{minipage}{0.245\textwidth}
\begin{tikzpicture}
\node (img) {\includegraphics[width=0.95\linewidth]{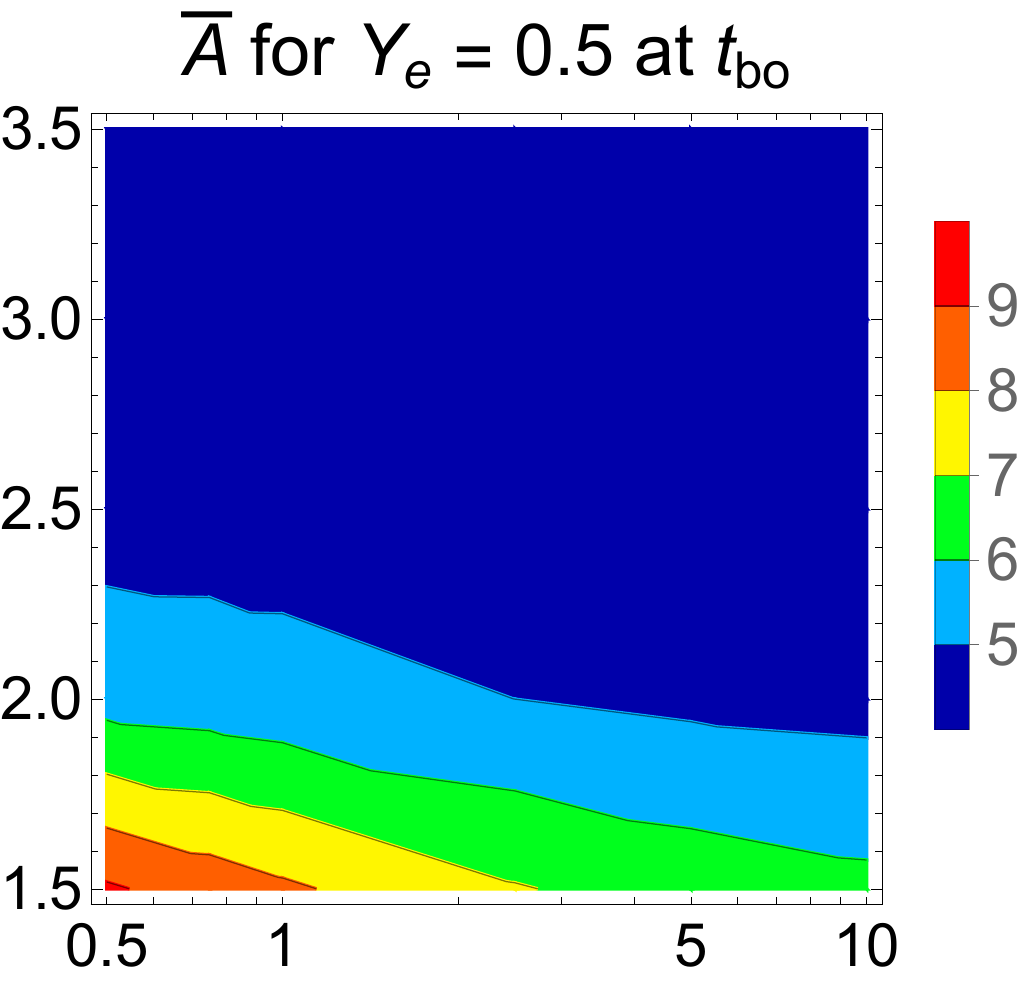}};
\node[left=67pt, node distance=0cm, rotate=90, anchor=center,font=\color{white}] {~~~$|$};
\end{tikzpicture}
\end{minipage}
\begin{minipage}{0.245\textwidth}
\begin{tikzpicture}
\node (img) {\includegraphics[width=0.95\linewidth]{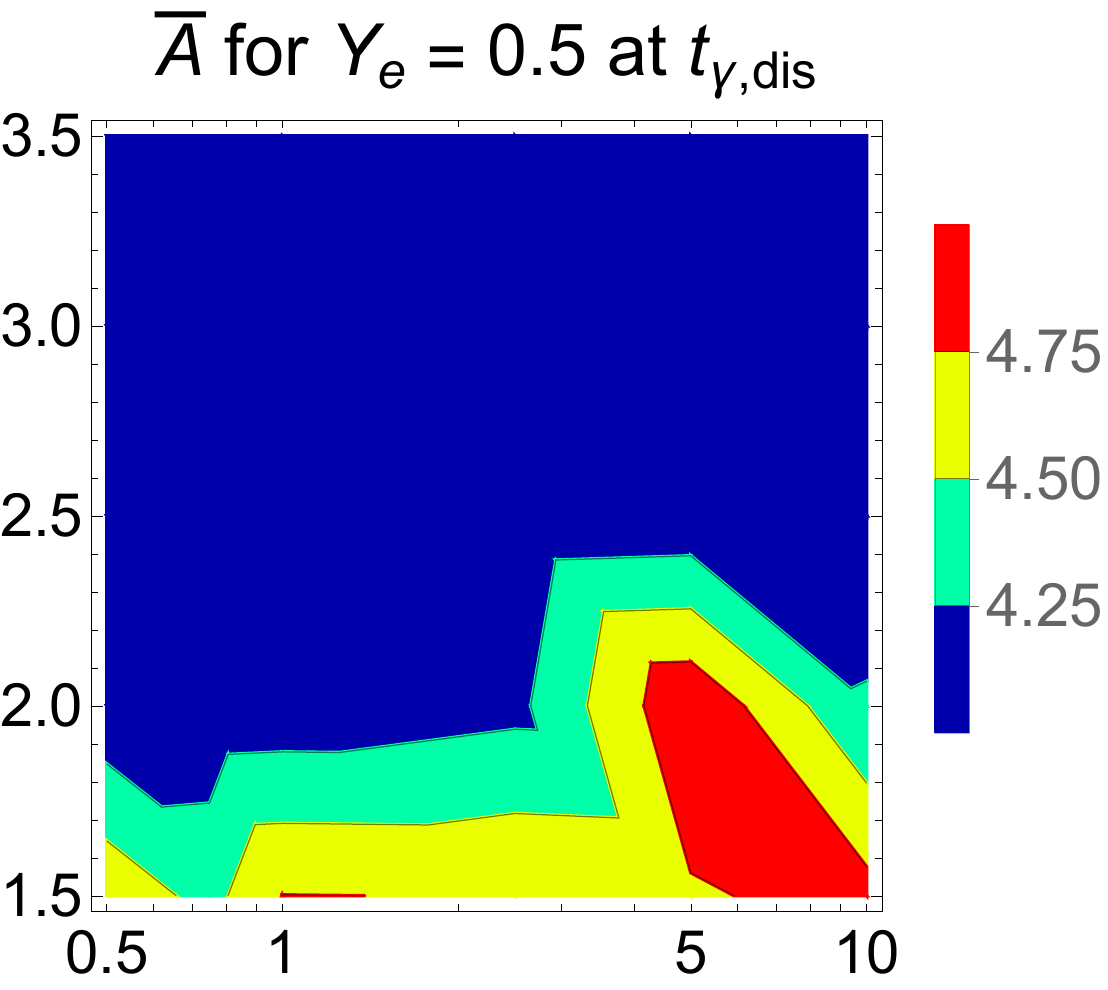}};
\node[left=67pt, node distance=0cm, rotate=90, anchor=center,font=\color{white}] {~~~~$|$};
\end{tikzpicture}
\end{minipage}
\begin{minipage}{0.245\textwidth}
\begin{tikzpicture}
\node (img) {\includegraphics[width=0.95\linewidth]{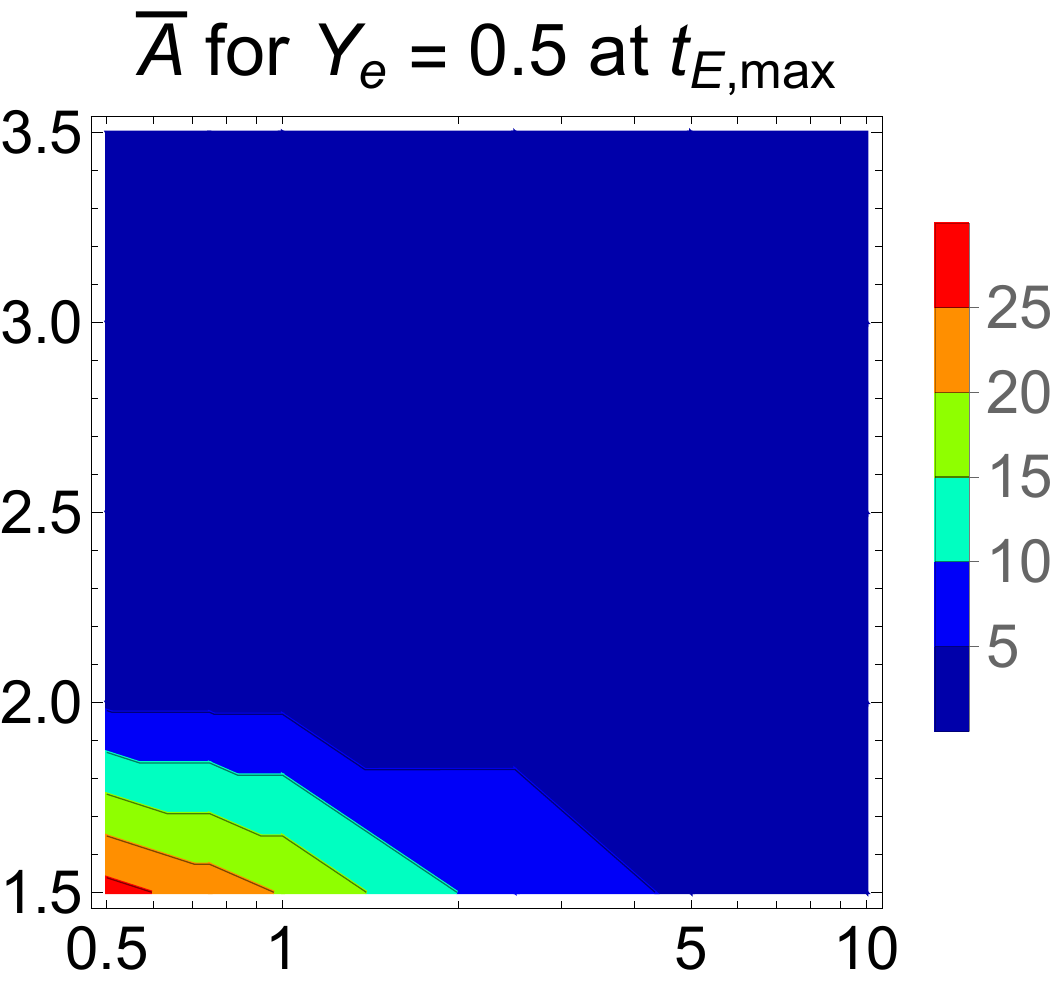}};
\node[left=67pt, node distance=0cm, rotate=90, anchor=center,font=\color{white}] {~~~~$|$};
\end{tikzpicture}
\end{minipage}
\newline
\begin{minipage}{0.245\textwidth}
\begin{tikzpicture}
\node (img) {\includegraphics[width=0.95\linewidth]{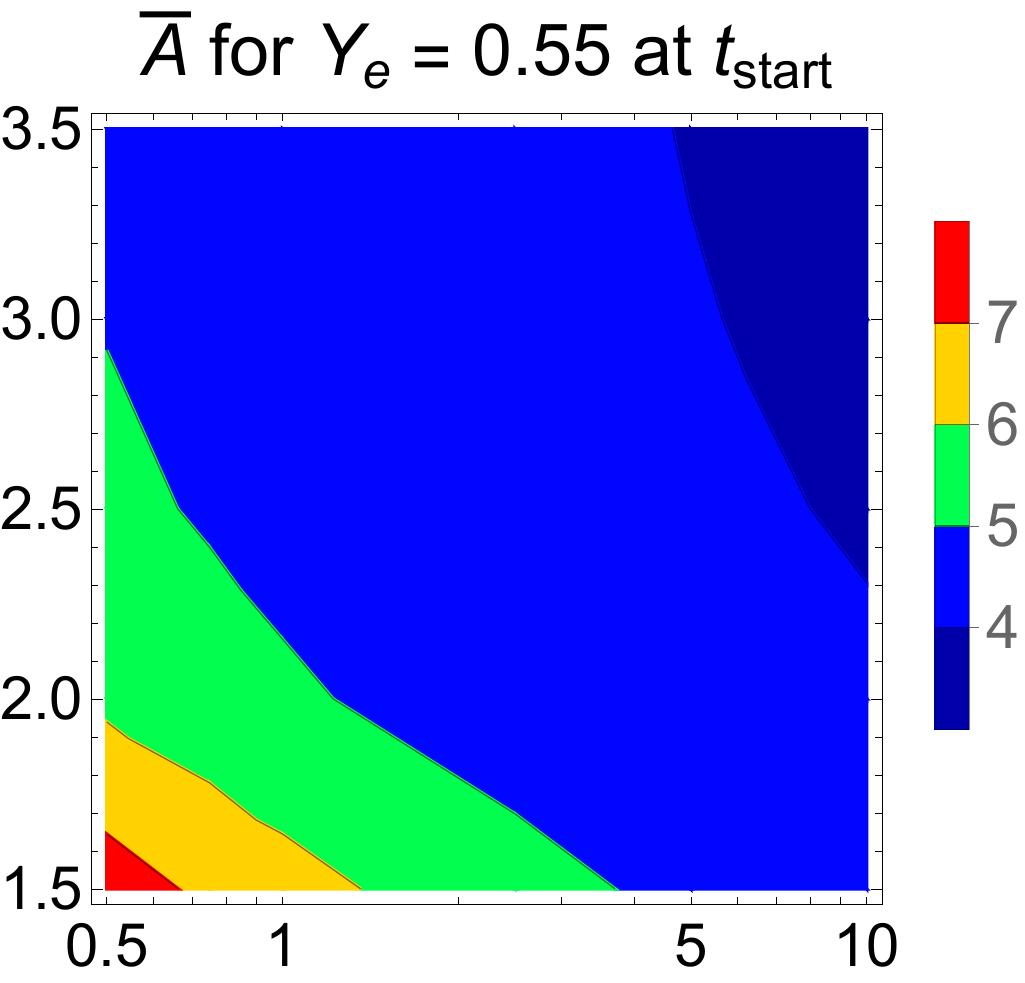}};
\node[below=55pt, node distance=0cm, font=\color{white}] {$|$};
\node[left=67pt, node distance=0cm, rotate=90, anchor=center,font=\color{white}] {$|$};
\end{tikzpicture}
\end{minipage}
\begin{minipage}{0.245\textwidth}
\begin{tikzpicture}
\node (img) {\includegraphics[width=0.95\linewidth]{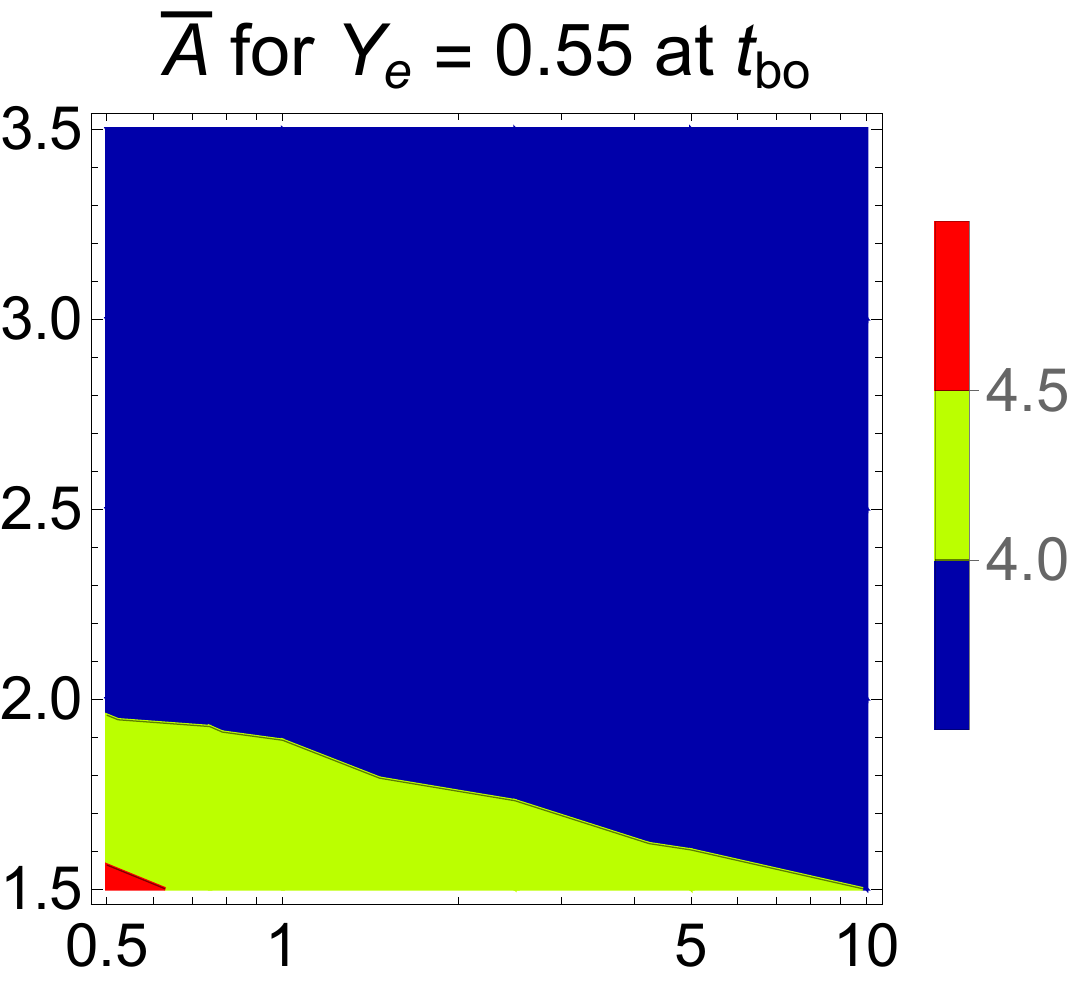}};
\node[below=54pt, node distance=0cm, xshift=2.1cm] {$B_{\textrm{dip}}$ [$10^{15}$ G]};
\node[left=67pt, node distance=0cm, rotate=90, anchor=center,font=\color{white}] {$|$};
\end{tikzpicture}
\end{minipage}
\begin{minipage}{0.245\textwidth}
\begin{tikzpicture}
\node (img) {\includegraphics[width=0.95\linewidth]{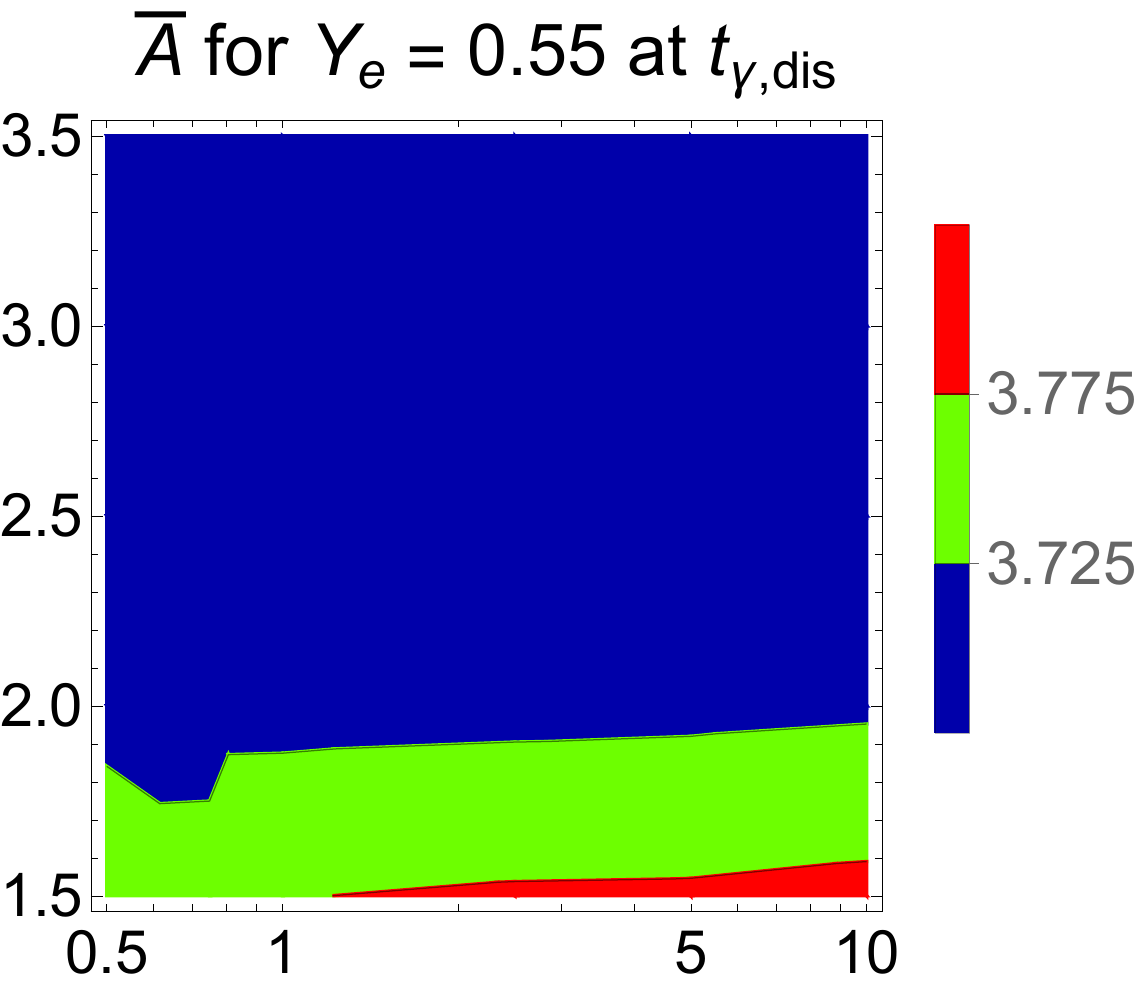}};
\node[below=55pt, node distance=0cm,font=\color{white}] {$|$};
\node[left=67pt, node distance=0cm, rotate=90, anchor=center,font=\color{white}] {$|$};
\end{tikzpicture}
\end{minipage}
\begin{minipage}{0.245\textwidth}
\begin{tikzpicture}
\node (img) {\includegraphics[width=0.95\linewidth]{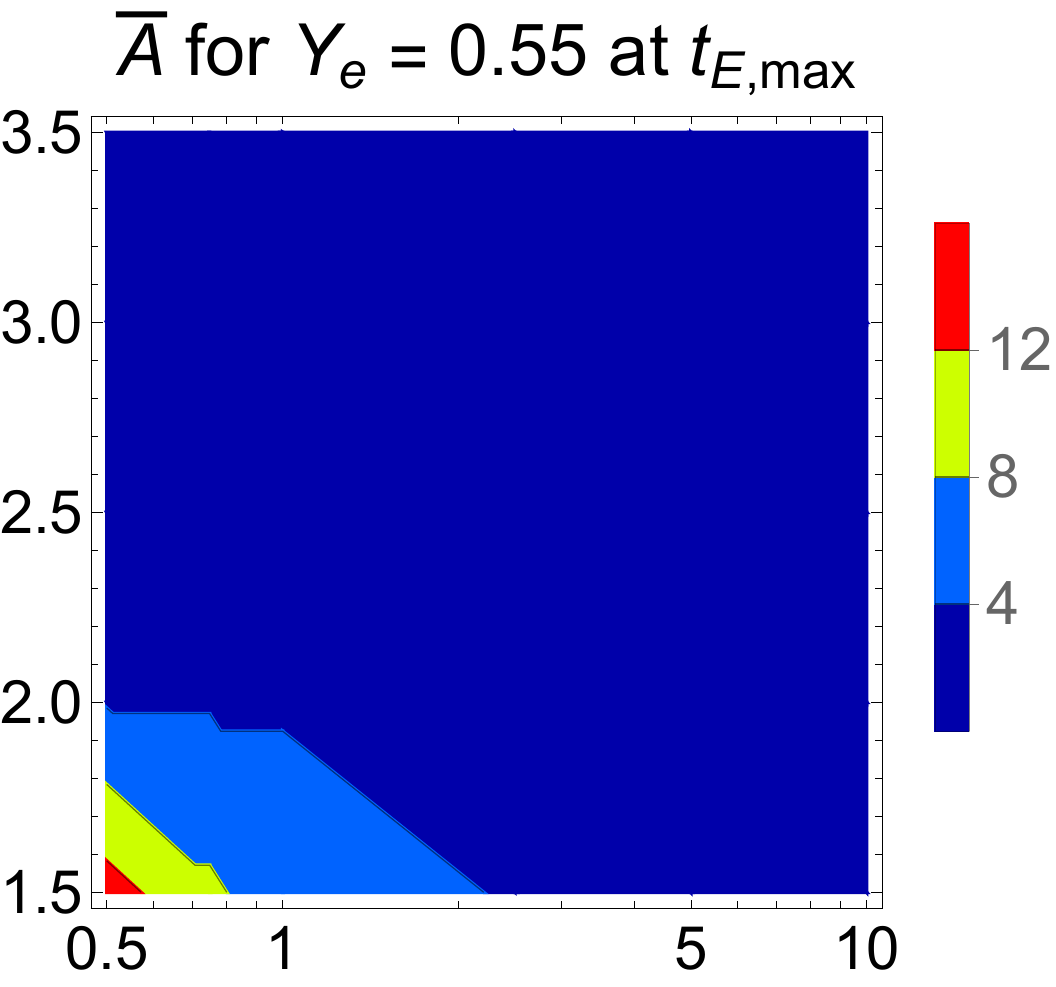}};
\node[below=55pt, node distance=0cm,font=\color{white}] {$|$};
\node[left=67pt, node distance=0cm, rotate=90, anchor=center,font=\color{white}] {$|$};
\end{tikzpicture}
\end{minipage}
\newline
\vspace{-0.2cm}
\caption{In each panel, we show the mean mass number $\bar{A}$ in the $B_{\textrm{dip}}-P_0$ plane. \textit{Top-to-bottom:} for electron fraction $Y_e = 0.45,\, 0.475,\, 0.5,\, 0.55$. \textit{Left-to-right:} for our four chosen IC times: start time, breakout, photodisintegration, and max acceleration time. Note that the color range in each panel shows a different scale.} 
\label{fig:mainresults}
\end{figure*}
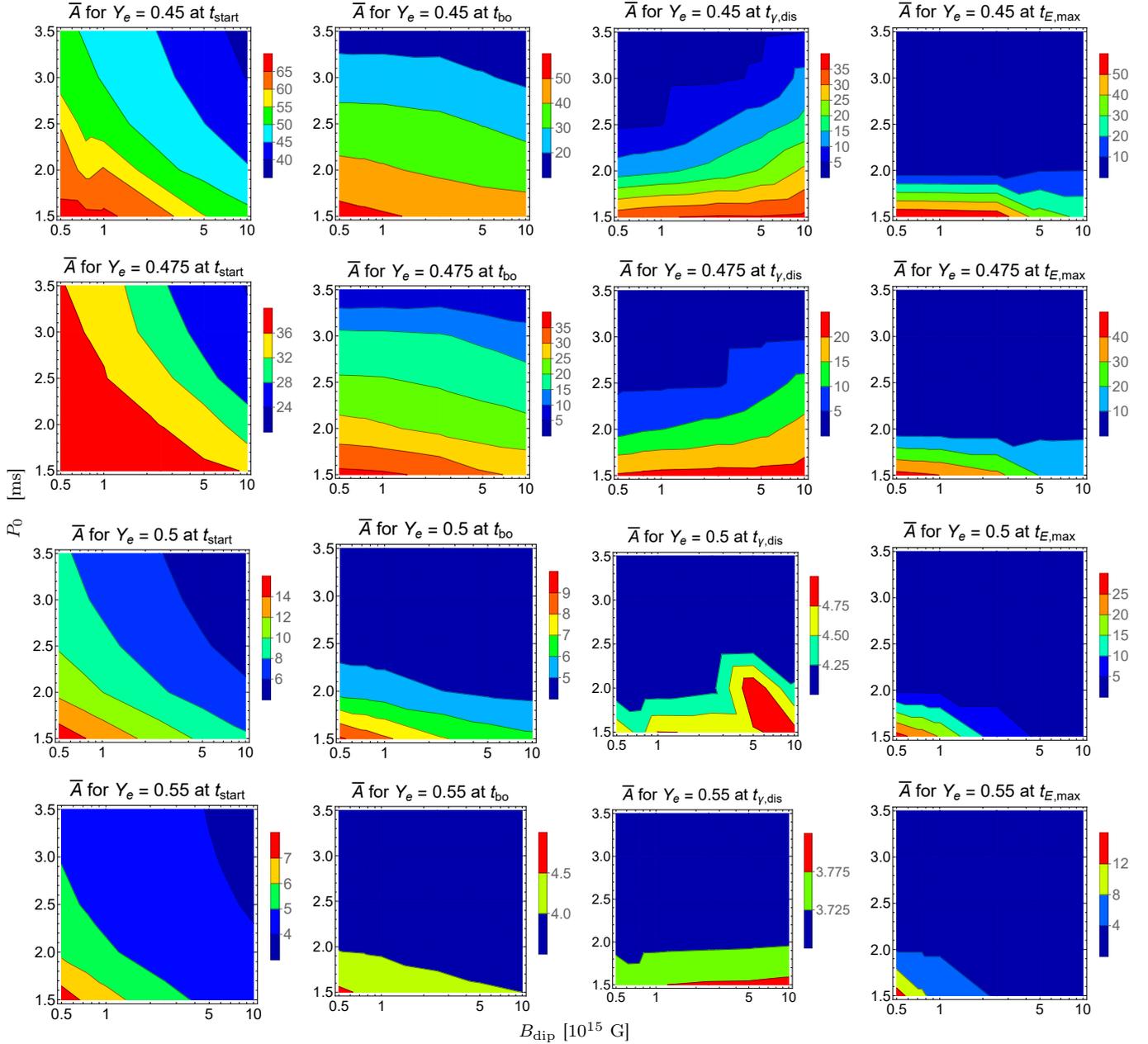

Finally, in Fig.~\ref{fig:mainresults}, we show the results of nucleosynthesis in protomagnetar outflows in terms of mean mass number as a function of the four model parameters: $B_{\rm dip}$, $P_0$, $Y_e$, and IC time. Note that the contours, i.e., the range and scale of $\overline{A}$, are different for each panel. These panels continue to show the same trends as in the previous figures, but also display the effects of the dipolar field and initial spin period on the final abundance of heavy elements.

At early IC times ($t_{\textrm{start}}$ and $t_{\textrm{bo}}$), a smaller magnetic field is more conducive to nucleosynthesis and this trend generally switches at late IC times ($t_{\textrm{$\gamma$,dis}}$ and $t_{\textrm{E,max}}$). {\tt SkyNet} determines these trends from the density and temperature trajectories: generally, at $t_{\textrm{start}}$ and $t_{\textrm{bo}}$, the densities are higher for the low-$B_{\rm dip}$ configurations and the opposite is true at $t_{\textrm{$\gamma$,dis}}$ and $t_{\textrm{E,max}}$ (see Appendix \ref{sec:appendIC}). An increased initial density leads to increased neutron-capture reaction rates and, with longer expansion time-scales, there is more time for the free neutrons to capture onto seed nuclei (see \citealt{HIX1999321,2015lippuner,2016fujibayashi}). On the other hand, entropy does not vary much in $B_{\rm dip}$ (see Appendix \ref{sec:appendIC}, Fig.~\ref{fig:initialconditions}). Since $T$ $\sim$ $S^{1/3}$, outflow entropy does not play a significant role in determining the trend of $\overline{A}$ relative to $B_{\rm dip}$. The trends we observe then must come from $\tau_{\rm exp}$ and $\dot{M}$; in turn, the $B_{\rm dip}$ dependence for these quantities come from $f_{\rm open}$ and $f_{\rm cent}$. Traditionally, shorter expansion time-scales are required for strong r-process to occur, however in our outflow system, a longer expansion time-scale gives rise to larger outflow densities (see equation (\ref{dens})). These larger densities (at moderate initial temperatures) give rise to the aforementioned enhanced neutron-capture rate and higher mean mass numbers. Since $\tau_{\rm exp}$ and $\dot{M}$ are greater for low-$B_{\rm dip}$ configurations at early times, this increased density gives higher values of $\overline{A}$. After $t_{\rm bo}$, centrifugal slinging increases for higher magnetic field strengths (as $f_{\rm cent}$ increases), driving a high-$B_{\rm dip}$ preference for $\dot{M}$. This gives the transition in the magnetic field preference for $\overline{A}$ (and $\rho$).

Entropy plays a larger role as the rotation rate is changed: since the entropy suppression is directly related to the spin period, we have higher initial temperatures for slower rotators. These higher initial temperatures lead to lower $\overline{A}$ values from increased photodisintegration effects. The final $\overline{A}$ also depends on the outflow density, whose $P_0$ trends are set by the factor $\sim$ 5 difference in $f_{\rm open}$ as the rotation rate changes structure of the open magnetic field lines. With decreased photodisintegration and enhanced nucleon-capture reaction rates (from the temperature and density trends, respectively), shorter rotation periods are more conducive to heavy element nucleosynthesis (in agreement with \citealt{Vlasov_2017}). Additionally, Fig.~\ref{fig:mainresults} highlights that at $t_{\rm E,max}$, rotators with $P_0=1.5\, {\rm ms}$ synthesize heavy elements in large mass fractions, compared to similar configurations.

\begin{figure}
\centering
\includegraphics[width=\linewidth]{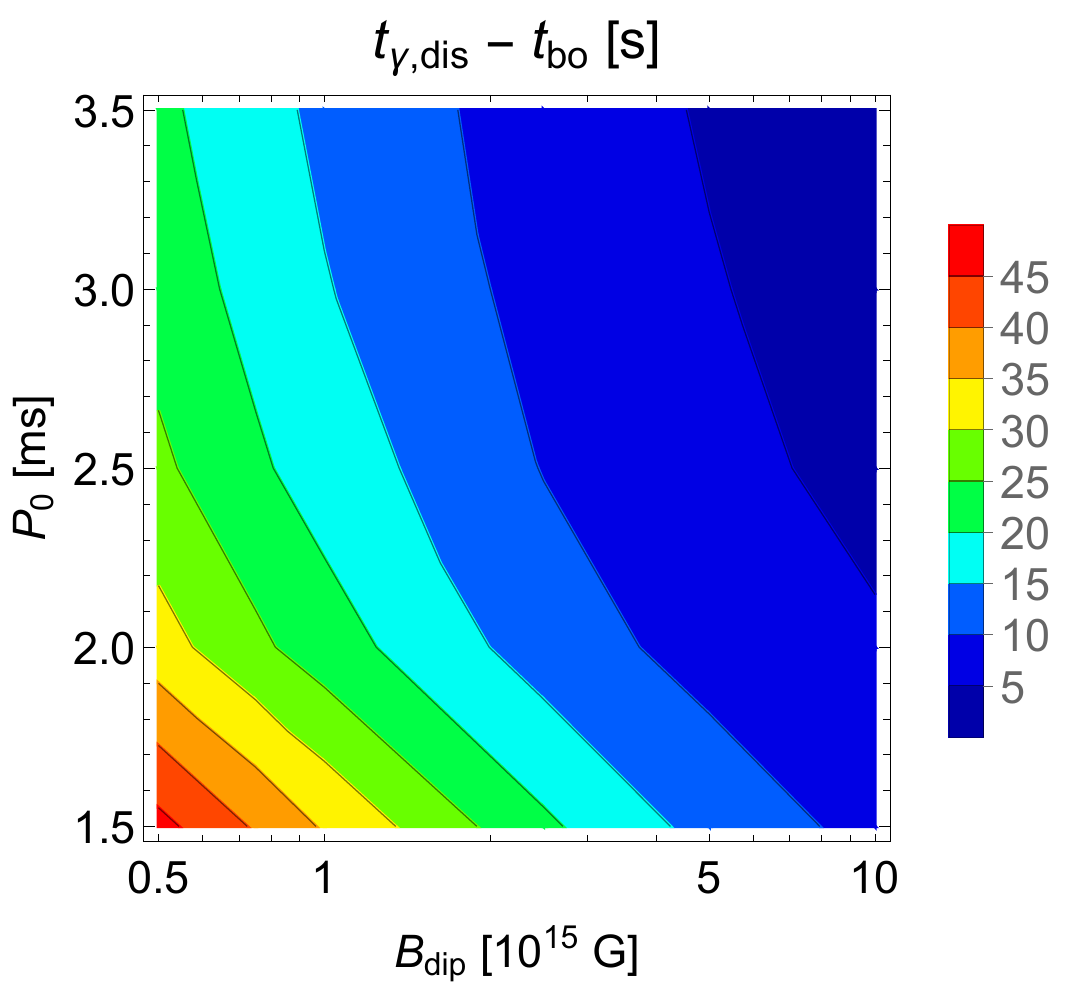}
\vspace{-0.5cm}
\caption{The time elapsed between jet breakout and when photodisintegration is no longer important (see Appendix B, Fig.~\ref{fig:times} for the individual time values). In this time window, any nuclei synthesized can be accelerated but the majority will be disintegrated, though some amount may survive as heavy nuclei. This epoch is particularly long for the low-$B_{\textrm{dip}}$, low-$P_0$ models and short for the high-$B_{\textrm{dip}}$ models; this trend in behavior leads to the field strength turnover in heavy element nucleosynthesis, as discussed in $\S$\ref{results}.}
\label{fig:delphobo}
\end{figure}

Since the IC times tested here are $B_{\textrm{dip}}$ and $P_0$ dependent, they vary significantly depending on our protomagnetar parameter choices. This difference in IC time between $t_{\rm bo}$ and $t_{\textrm{$\gamma$,dis}}$ further drives a transition in the trends of $B_{\textrm{dip}}$; at IC times before $t_{\rm \gamma,dis}$, a smaller field strength results in a higher $\overline{A}$, while after $t_{\rm bo}$ this trend is reversed. This is supported in Fig.~\ref{fig:delphobo} where we see that the low-field strength models have the largest difference between these IC times. This time window contributes to the decrease in $\rho$ and $T$, i.e.,  $\rho$ and $T$ decrease more between $t_{\rm bo}$ and $t_{\rm \gamma,dis}$ for low-field strength models and less for high-field models. Since the initial densities and temperatures are somewhat predictive of the trends in $\overline{A}$, we have this transition.

Changing $Y_e$ gives expected results: neutron-rich outflows lead to greater heavy element nucleosynthesis than proton-rich outflows. This variation in $Y_e$ naturally sets the scale of $\overline{A}$ and leads to monotonic trends in electron fraction.

\begin{figure*}
\begin{minipage}{0.245\textwidth}
\begin{tikzpicture}
\node (img) {\includegraphics[width=0.95\linewidth]{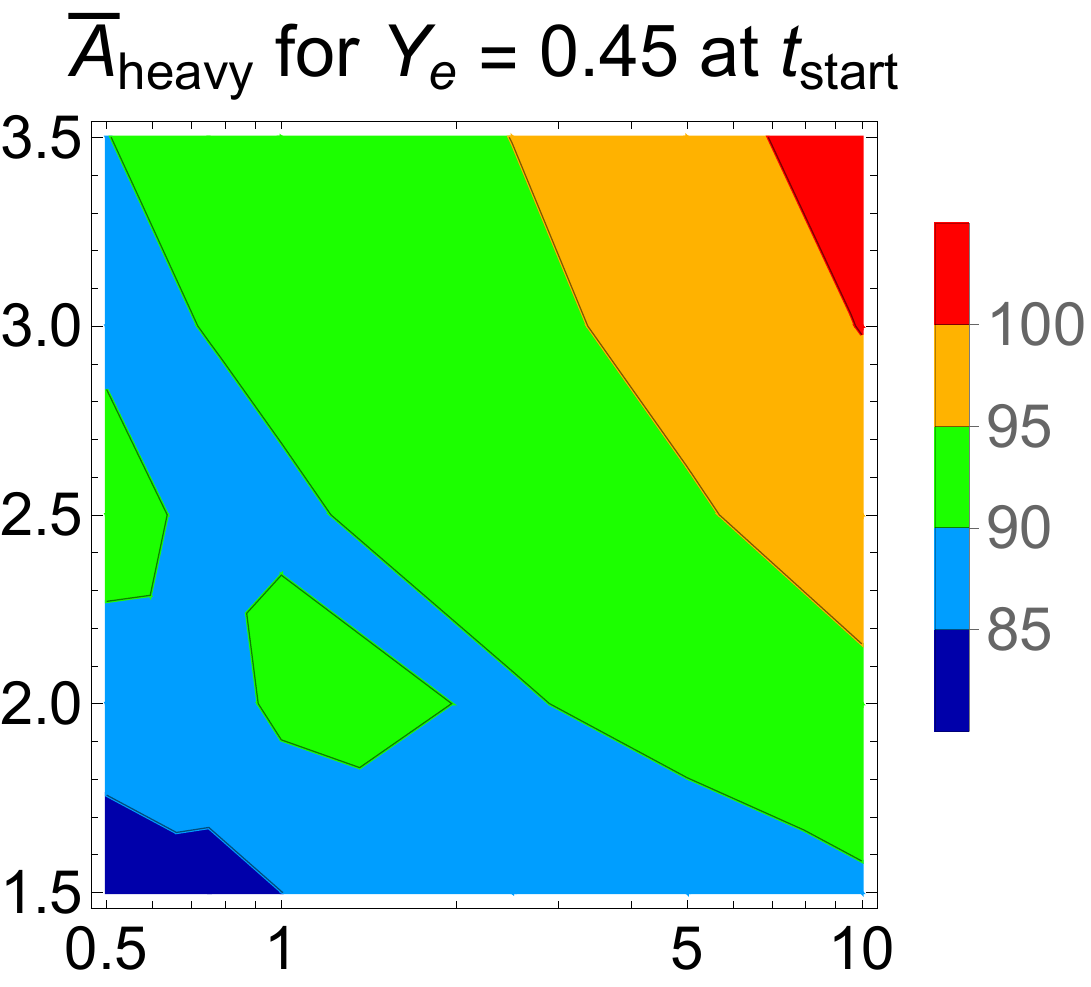}};
\node[below=55pt, node distance=0cm, xshift=1.9cm,font=\color{white}] {$A_{blank}$};
\node[left=67pt, node distance=0cm, rotate=90, anchor=center]{$P_0~$[ms]};
\end{tikzpicture}
\end{minipage}
\begin{minipage}{0.245\textwidth}
\begin{tikzpicture}
\node (img) {\includegraphics[width=0.95\linewidth]{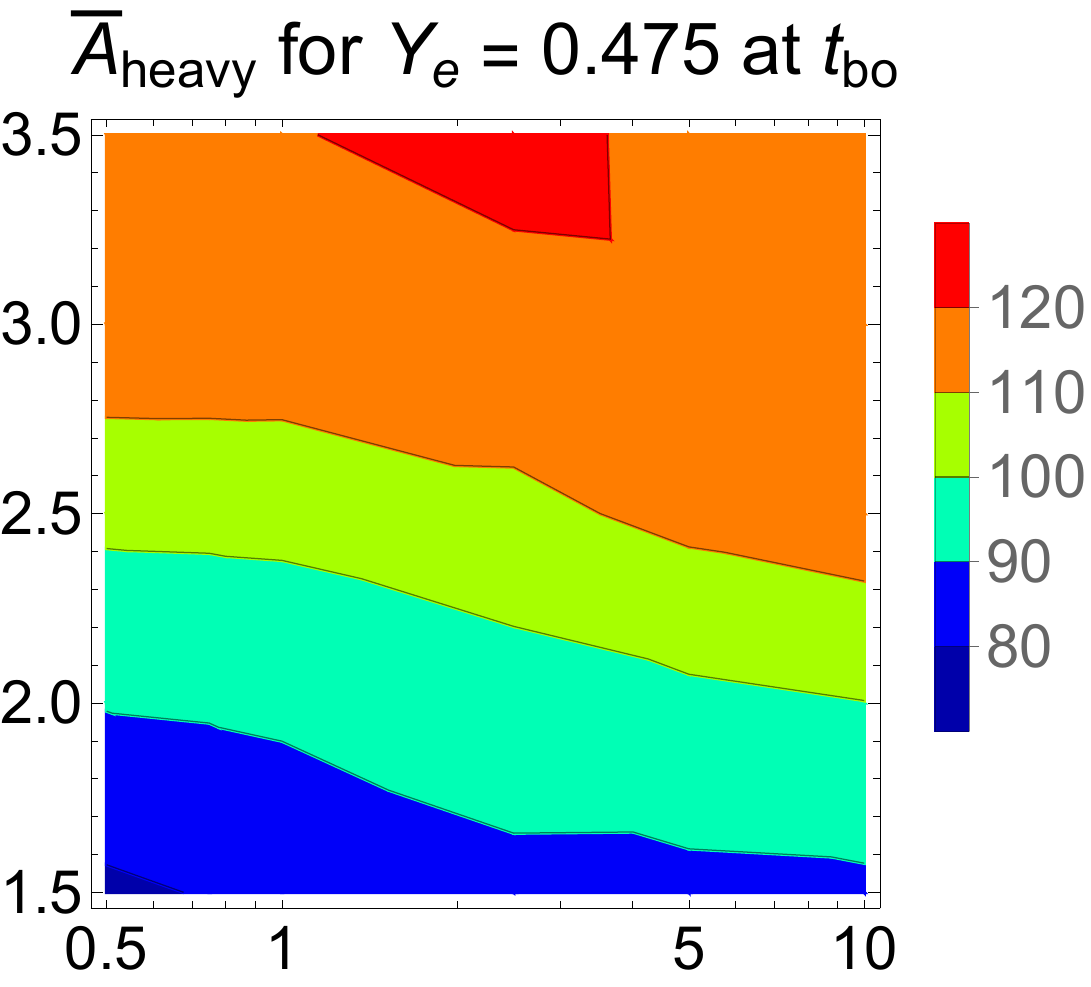}};
\node[below=55pt, node distance=0cm, xshift=1.9cm] {$B_{\textrm{dip}}$ [$10^{15}$ G]};
\node[left=67pt, node distance=0cm, rotate=90, anchor=center,font=\color{white}]{$~~|$};
\end{tikzpicture}
\end{minipage}
\begin{minipage}{0.245\textwidth}
\begin{tikzpicture}
\node (img) {\includegraphics[width=0.95\linewidth]{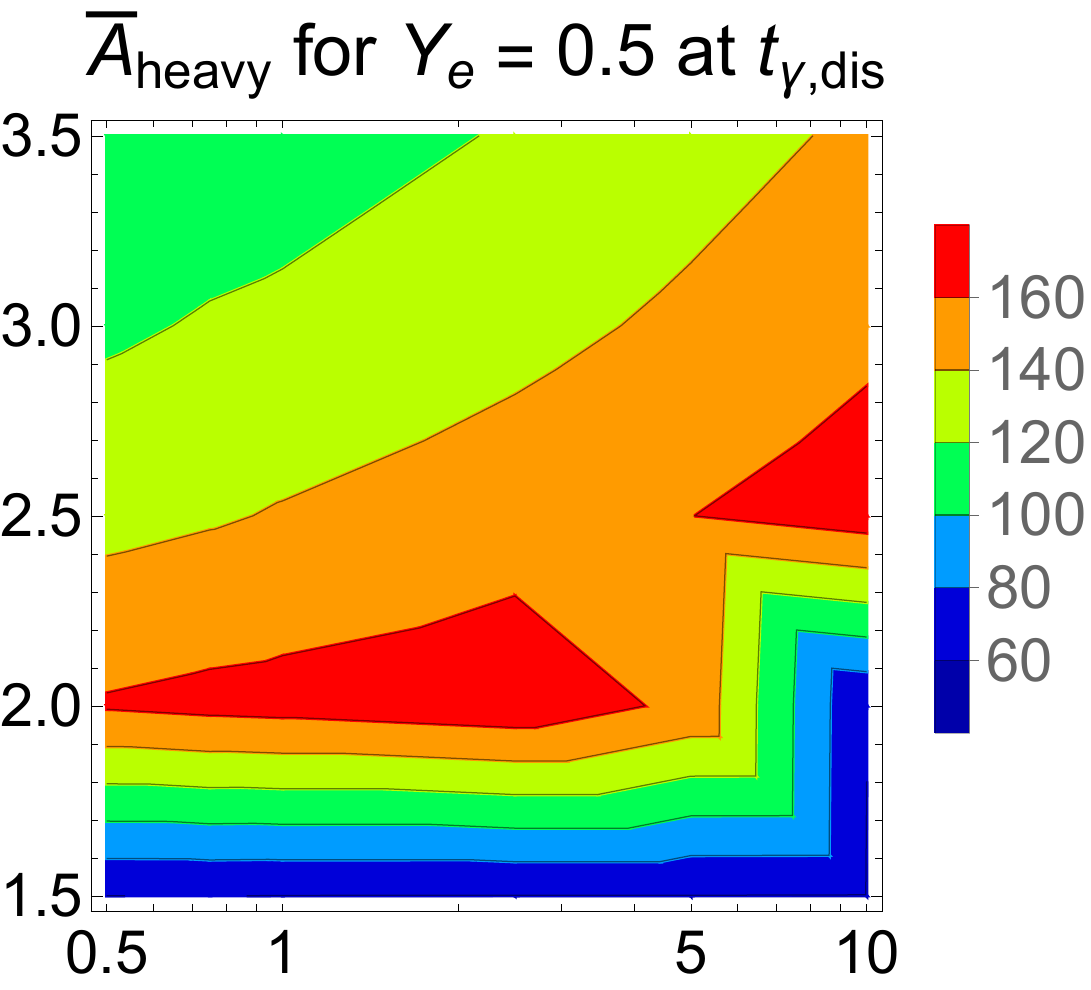}};
\node[below=55pt, node distance=0cm, xshift=1.9cm,font=\color{white}] {$A_{blank}$};
\node[left=67pt, node distance=0cm, rotate=90, anchor=center,font=\color{white}]{$|$};
\end{tikzpicture}
\end{minipage}
\begin{minipage}{0.245\textwidth}
\begin{tikzpicture}
\node (img) {\includegraphics[width=0.95\linewidth]{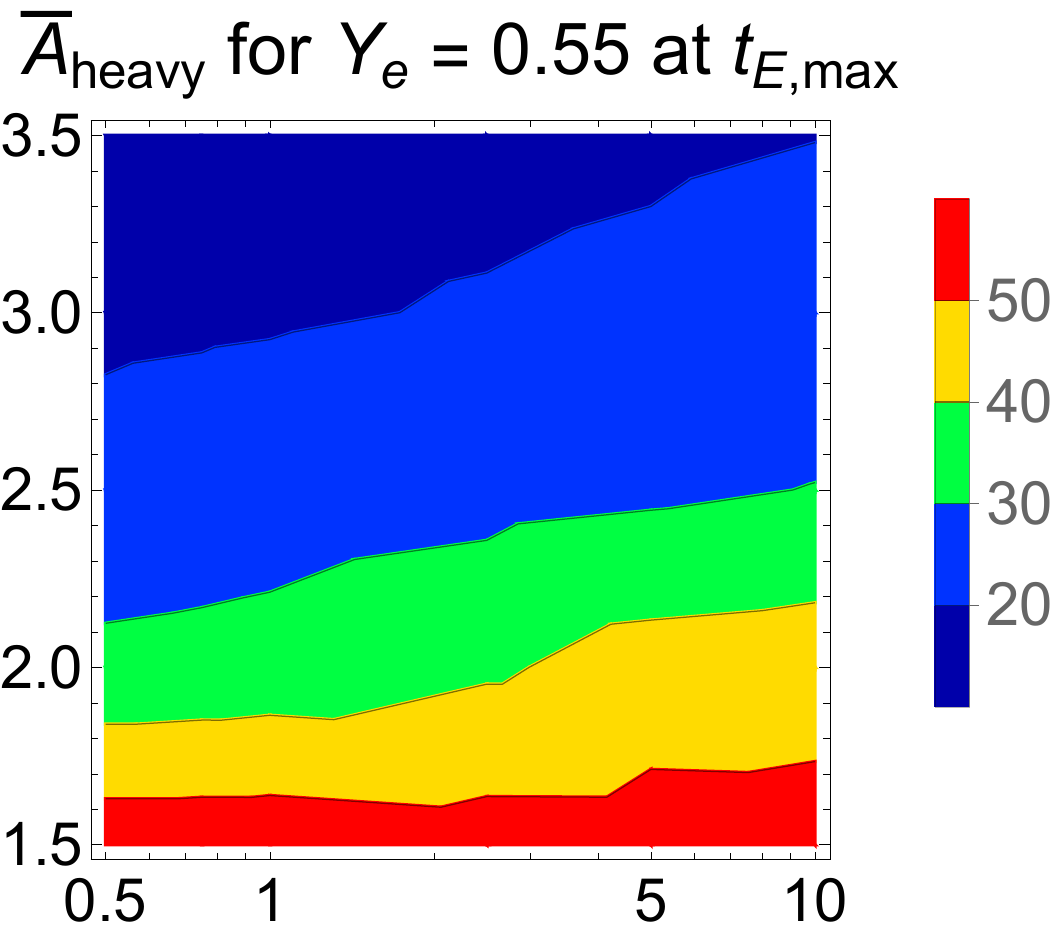}};
\node[below=55pt, node distance=0cm, xshift=1.9cm,font=\color{white}] {$A_{blank}$};
\node[left=67pt, node distance=0cm, rotate=90, anchor=center,font=\color{white}]{$~|$};
\end{tikzpicture}
\end{minipage}
\caption{Contours of mean mass number calculated after excluding free nuclei and helium isotopes, $\overline{A}_{\rm heavy}$. This provides a useful measure of the distribution in abundance of heavy elements since much of the mass contributing to $\overline{A}$ comes from free nucleons and helium. The figures are shown for a diagonal section of Figure \ref{fig:mainresults}: four IC times and at four electron fractions to quantify most of the trends, although they are not so monotonic and each have exceptions. Note that the color range in each figure is different.}
\label{fig:aheavy}
\end{figure*}

Interestingly, the qualitative trends change somewhat if we discuss $\overline{A}$ as a function of the model parameter space, but exclude the contribution from free neutrons, protons, and helium nuclei. Since these light nuclei make up the majority of the mass synthesized in proton-rich conditions and later IC times, excluding them allows us to focus on the trends seen in higher $A$ nuclei. For example, $\overline{A}$ is an especially useful measure for cosmic ray physics, but does not say much about the distribution of the abundance pattern created. In some configurations, elements up to and exceeding the third r-process peak are created, but in negligible amounts by mass fraction. These nuances are more readily captured by excluding protons, neutrons, and helium nuclei, which we denote by $\overline{A}_{\rm heavy}$. Figure \ref{fig:aheavy} shows the subfigures along the diagonal of Fig.~\ref{fig:mainresults} but expressed instead with $\overline{A}_{\rm heavy}$. This diagonal shows some of the general features of $\overline{A}_{\rm heavy}$ as a function of the model parameters, but is not as purely monotonic as the $\overline{A}$ plot and has several exceptions. Similar to the trends in $\overline{A}$, initial spin period generally has a greater effect than magnetic field strength for $\overline{A}_{\rm heavy}$. At early IC times (with the exception of $Y_e=0.55$), $\overline{A}_{\rm heavy}$ has an inverse magnetic field strength-initial spin relation. Most of the mass for the high period configurations goes into free nucleons and helium, driving a lower $P_0$ preference in $\overline{A}$, but in terms of $\overline{A}_{\rm heavy}$, a higher $P_0$ leads to more massive nuclei. At late IC times (with the exception of $Y_e=0.55$), $\overline{A}_{\rm heavy}$ peaks in a band between $P_0 \sim 2-3\, {\rm ms}$, as shown in the third panel of Fig.~\ref{fig:aheavy}. The trends in $Y_e$ and IC time are strictly not monotonic. In neutron-rich conditions ($Y_e=0.45,~0.475$), a lower electron fraction leads to heavier nucleosynthesis - as with $\overline{A}$. The lowest values of $\overline{A}_{\rm heavy}$ are reached in proton-rich conditions ($Y_e=0.55$) and this trend is also present in $\overline{A}$. 
However, when the electron fraction is maintained at 0.5, much higher mass numbers can be synthesized (due to the process described earlier from \citealt{PhysRevLett.89.231101}). In IC time, generally the most nucleosynthesis occurs at photodisintegration time ($t_{\rm \gamma,dis}$), and $\overline{A}_{\rm heavy}$ decreases both before and after. Finally, at $Y_e=0.55$, the trends in time, $B_{\rm dip}$, and $P_0$ generally agrees with the trends of $\overline{A}$, but at higher mass numbers.

Note that the rigid changes in the contour curves are not necessarily a feature of the numerical analysis. They are a result of a simple order-one interpolation on the 30 $B_{\textrm{dip}}-P_0$ configurations presented here; adding additional data points would improve the resolution and, thus, the smoothness of the figures. Nevertheless, the qualitative conclusions discussed here should remain valid.

\section{Discussion}\label{discussion}

We first discuss the trends in model parameter space that explain the conditions conducive to heavy element synthesis, explore the validity of the parameter range, and describe the validity of other assumptions made in this work.

\textit{Magnetic field trends and considerations:} At early IC times (start and breakout times), a lower magnetic field is more conducive to heavy element nucleosynthesis. The low-magnetic field yields initial outflow densities that are higher than those for high-magnetic field configurations; slower expansions and high mass loss rates initially enhance the neutron-capture rate and suggest the synthesis of heavier elements. The transition between a preference of the low-$B_{\rm dip}$ and high-$B_{\rm dip}$ models is driven by the transition in preference of the mass loss rate. This is, in turn, caused by increased magnetocentrifugal slinging for high-$B_{\rm dip}$ models after $t_{\rm bo}$. This is true regardless of the choice of $Y_e$ and $P_0$ (except in the case for the fastest rotators, with $P_0=1.5\, {\rm ms}$).

Given that the later IC times (photodisintegration and max acceleration times) are generally more relevant for UHECR implications, it seems that the higher magnetic field configurations are more important for heavy element nucleosynthesis (with the notable exception of $B_{\rm dip}=5\times10^{14}\,{\rm G}$ and $P_0=1.5\,{\rm ms}$). However, we do not time-integrate $\overline{A}$ nor do we consider additional photodisintegration interactions as the nuclei make their way out of the star and are accelerated. It may turn out, then, that the protomagnetars with low/moderate $B_{\rm dip}$ are important as well, since some amount of the heavy nuclei synthesized between $t_{\rm bo}$ and $t_{\rm \gamma,dis}$ are only partially photodisintegrated. Finally, the variation in magnetic field strength has a weaker effect on the final $\overline{A}$ compared to the variation in initial spin period at all IC times.

The range of magnetic field strengths considered here should be a representative sample of the possible stable field strengths. For example, magnetic field strengths of order $\sim10^{17}\, {\rm G}$ can be reached if the magnetic energy in the dipole field is approximately equal to the rotational energy. However, stable field configurations require a total magnetic field strength at least 10 times larger than the dipole component, which results in a dipole field strength of $\sim10^{16}\, {\rm G}$ \citep{Metzger_2011}. We can also look at recent observations to inform us on the magnetic field and spin periods of millisecond magnetars. Multiple recent works have shown that millisecond magnetar central engines can give rise to long-duration GRBs (see e.g., \citealt{2014lu,2014arXiv1401.1601Y,2015rea}). The values chosen here fit reasonably well within the range of beam-corrected dipolar magnetic field strengths inferred from these studies. Below $\sim10^{14}\, {\rm G}$, the protoneutron star is no longer generally considered a protomagnetar. Also, mildly magnetized pulsars may not have the collimation power to produce jets that break through the stellar material; for example, in the mild magnetic field scenarios studied in \citealt{2012fang}, particle acceleration occurs in the vicinity of the compact object and therefore hadronuclear interactions as nuclei escape from the supernova remnant significantly reduce the final nuclear composition.

\textit{Initial spin period trends and considerations:} At all times and $Y_e$, an increasing period yields decreased heavy element nucleosynthesis. Initial spin period, $P_0$ is more dominant with regard to $\overline{A}$ synthesized than the field strength is. $P_0$ has such an important impact because of its effect on the magnetic field structure and the fraction of the PNS surface threaded by the open magnetic field lines (i.e.,  in the correction factor $f_{\rm open}$). There is $\sim 2$ times more variation in $f_{\rm open}$ over the range of $P_0$ values compared to the range of $B_{\rm dip}$ values chosen here. This gives a decreased density which correlates with a decreased $\overline{A}$. Further, and more importantly, as $P_0$ increases, entropy increases by a factor $\gtrsim$ 2. The larger entropy increases the temperature, resulting in more photodisintegration during the nucleosynthesis epoch and lower $\overline{A}$ values.

These initial spin period trends may break down for $P_0 \leq 1\, {\rm ms}$. Although spin period is treated as an independent parameter in this study, it is coupled to the electron fraction, the mass loss rate, and the overall stability of the neutron star (e.g. \citealt{1999A&A...350..497S,2008metzger,Metzger_2011}). At initial spin periods greater than $3.5\, {\rm ms}$, these trends are likely to continue monotonically (see also \citealt{Vlasov_2017} for trends in protomagnetar spin period and \citealt{MB_2021} for analysis up to $5\, {\rm ms}$). Furthermore, we can look at the X-ray plateau data for long-duration GRBs \citep{2014lu,2014arXiv1401.1601Y,2015rea}, to see that the $P_0$ range chosen here fits well within the inferred, beam-corrected spin periods assuming a magnetar central engine.

\textit{IC time trends and considerations:} Earlier times post core-collapse are more favorable for the synthesis of heavy elements; the initial time and breakout times yield similar results. After breakout time, there is a more significant drop in the ability for the magnetar outflow to synthesize heavy elements. At later times relevant for sourcing UHECR (between photodisintegration time and max acceleration times), rapid rotators could create large quantities of intermediate/heavy mass nuclei, but only under neutron-rich conditions. 

However, some fraction of heavy elements should survive photodisintegration between $t_{\textrm{bo}}$ and $t_{\textrm{$\gamma$,dis}}$ to be important for UHECR considerations. Since this time difference (Fig.~\ref{fig:delphobo}) is parameter dependent, it may turn out that the low-$B_{\textrm{dip}}$ configurations are still more important for heavy element survival if partial survival is taken into account. This analysis is left for future work.

As previously mentioned, between the early and late IC times a transition occurs where high-$B_{\textrm{dip}}$ protomagnetars are better at synthesizing heavier nuclei. Finally, at IC times later than $t_{\rm E,max}$, the outflows would be dominated by free neutrons, protons, and helium nuclei, although the mass ejected is much lower at such late times.

\textit{$Y_e$ trends and considerations:} Overall heavy nucleosynthesis depends monotonically on the choice of $Y_e$; in terms of the mean mass number (and fraction above iron), $\overline{A}_{Y_e=0.45} > \overline{A}_{Y_e=0.475} > \overline{A}_{Y_e=0.5} > \overline{A}_{Y_e=0.55}$. In neutron-rich conditions ($Y_e=0.45,~0.475$), large fractions of heavy elements are created. As soon as there are equal parts of neutrons and protons, there is a noticeable drop in the amount of heavy elements synthesized (see the change in slope past $Y_e=0.5$ in Fig.~\ref{fig:xhaB514P15}).

Although it is not explored in this work, it is expected that under even more neutron rich conditions ($Y_e \lesssim 0.4$), we would see a higher $\overline{A}$ and $X_h$. In order to reach significant amounts of third r-process peak nuclei, studies suggest $Y_e \lesssim 0.25$ is required (\citealt{2015kasen,2015lippuner}). In this study we adopt a range of $Y_e$ that may represent more typical values for magnetorotational CCSNe (see numerical simulations of \citealt{Vlasov_2017}, \citealt{Grimmett_2020}, and \citealt{Reichert_2021}), but in rarer cases, e.g., where the rotation period is less than $1\, {\rm ms}$, $Y_e$ at the base of the outflow may be significantly lower and reach $\sim 0.1-0.3$ (\citealt{2008metzger,2012ApJ...750L..22W}). In general, under more proton-rich conditions ($Y_e \gtrsim 0.6$), it is expected that we would see less heavy element synthesis (although the $\nu$p-process could induce interesting departures). 

Finally, we hold $Y_e$ fixed throughout the thermodynamic trajectories given that there is a degree of uncertainty in its dynamical evolution for the magnetized outflows considered here. With that in mind, choosing to test a range in electron fraction, $0.45 \leq Y_e \leq 0.55$ allows us to understand the important trends in $Y_e$ and with a given $Y_e$ trajectory, we are also able to make qualitative statements on the composition.

\textit{Other considerations:} For this study, we fix the PNS mass to $1.4\, M_{\odot}$ and magnetic obliquity angle to $\pi/2$. A higher PNS mass leads to an elevated energy loss rate, but a smaller $\sigma_0$ and $\tau_{\rm exp}$ which generally leads to less nucleosynthesis. At early times post-CC this can result in up to a $\sim20\%$ decrease in $X_h$, but becomes marginal at late times \citep{MB_2021}. The magnetic field of the PNS can also play a role in baryon loading of the outflow, the amount of mass ejected (see e.g. \citealt{2011shibata}), and therefore its heavy element yield, but in this work we assume the neutrino-driven wind mass loss of \citet{Metzger_2011} as the dominant mechanism. Changes in the magnetic obliquity angle can also have significant implications for the nucleosynthesis products of these outflows (see \citealt{Halevi_2018}, up to a $\sim40\%$ difference in $X_h$ at early times post-CC in \citealt{MB_2021}). Decreasing this angle tends to reduce the amount of entropy suppression in the outflow (in the case of \citealt{MB_2021}, Fig. 7, this results in decreased heavy element nucleosynthesis). Increased heating near the seed formation radius, however, may increase the neutron-to-seed ratio and improve nucleosynthesis (see \citealt{Vlasov_2017} for discussion of aligned vs. misaligned rotators).

We also evolve $\Dot{M},~S,~\tau_{\textrm{exp}},$ and other outflow properties in IC time, but consider them as fixed parameters in Outflow time for our thermodynamic trajectories. We do not expect these to strongly affect the products of nucleosynthesis, since these quantities do not change appreciably in the nucleosynthesis time-scale ($\sim 10\, {\rm ms}$, see Fig.~\ref{fig:abartime}). With these assumptions, we can still extract some important results in terms of the magnetar model parameters. To estimate the outflow radius as a function of time and jet magnetization, we adopt the magnetic reconnection model of \citet{2002drenkhahn} (see Fig.~\ref{fig:rad} for the radial trajectories). However, the dissipation process need not be continuous within the outflow and is model dependent. Other descriptions, such as the turbulent reconnection models of \citet{2010zhang} and \citet{2019lazarian}, could yield different results for the time-dependent thermodynamic quantities derived here, but the extent of this difference is unclear. These alternative models also have important implications for the acceleration and photodisintegration of the nuclei via, e.g., the GRB photon spectrum. Thus, an investigation of the effects of the dissipation model on the composition and subsequent evolution is warranted.

{\tt SkyNet} requires an initial abundance pattern, along with thermodynamic trajectories and other input parameters, to evolve the network and calculate the abundance patterns over time. In order to determine this initial abundance pattern, we make use of the NSE evolution mode, which requires only the initial density, temperature and electron fraction. The majority of all outflows studied here begin with temperatures comfortably above the $\sim7\,{\rm GK}$ NSE threshold for full network evolution, meaning that for most configurations, this initial abundance pattern created by {\tt SkyNet} is effectively made up of free nucleons determined by the electron fraction. However, at $t_{\rm E,max}$, some low $B_{\rm dip}$ and $P_0=1.5\,{\rm ms}$ configurations have initial temperatures below $7\,{\rm GK}$.

For these $P_0=1.5\,{\rm ms}$ cases at $t_{\rm E,max}$, the lower temperatures result in initial abundance patterns that are not quite in NSE; the free nucleons are in equilibrium with some amount of heavier elements. This process is similar to the NSE and `quasi-nuclear equilibrium' (QSE) results, e.g., \citet{2018wanajo}, where, as temperatures cool, NSE/QSE result in equilibrium of free nucleons and iron group elements (see also \citealt{1968ApJS...16..299B} and \citealt{1998ApJ...498..808M} relating to silicon burning reactions). This results in initial fractions of heavy elements that are far greater than those of other configurations where the temperature is higher. These are similar methods to those of \citet{Roberts_2016} and \citet{Lippuner_2017} where the NSE evolution mode also calculates the initial abundance, despite an initial temperature of $\sim6\,{\rm GK}$, below the network threshold. 

Finally, while $\overline{A}$ is a good measure for cosmic rays and descriptive of what most of the mass is synthesized into, $\overline{A}_{\rm heavy}$ is more descriptive of the abundance distributions in each model configuration. In magnetic field strength and initial spin period, the trends are inverted, i.e.,  at early IC times, the highest mass numbers are synthesized for high $B_{\rm dip}$-high $P_0$ configurations for $Y_e=0.45,~0.475,~\textrm{and}~0.5$. At late IC times, however, the highest mass numbers are synthesized in a band around $\sim2-3$ ms spin periods. $\overline{A}_{\rm heavy}$ in terms of $Y_e$ is ordered as $\overline{A}_{\rm heavy,Y_e=0.55}<\overline{A}_{\rm heavy,Y_e=0.45}<\overline{A}_{\rm heavy,Y_e=0.475}<\overline{A}_{\rm heavy,Y_e=0.5}$. In the case of $Y_e=0.5$ at later IC times, very massive nuclei are synthesized. In the case of $Y_e=0.55$, the trends tend not to change, but higher mass numbers are reached compared to $\overline{A}$. There are several exceptions to these rules, however.

\section{Summary and Conclusion}\label{summary}
We performed a numerical study of the nucleosynthesis of protomagnetar outflows to systematically probe the conditions suitable for heavy element nucleosynthesis. We adopted a parametrized model for the protomagnetar in terms of physical and dynamical properties of the magnetar and its outflows. Our results can be summarized as:

\begin{itemize}[leftmargin=*]
    \begin{item}
    \textit{Heavy elements} ($\overline{A}\sim20-65$) are generally synthesized only in neutron-rich conditions ($Y_e<0.5$) for the more rapid rotators ($P_0\lesssim 2.5\, {\rm ms}$). In these conditions, earlier IC times ($t_{\rm start}$ and $t_{\rm bo}$) are more conducive for heavy elements and there is a generally weak dependence on $B_{\rm dip}$. This represents a smaller subspace of the total parameter space and is subjected to composition-altering photodisintegration at the dissipation radius. This population could be described as an intermediate composition after the epoch of particle acceleration, depending on photodisintegration details.
    \end{item}
    \begin{item}
    \textit{Lighter elements}  ($\overline{A}\sim10-50$) are synthesized under a broader set of conditions than heavy elements. These elements are synthesized still under neutron-rich conditions (and in some cases, under proton-rich conditions), but also at the later IC times ($t_{\rm\gamma,dis}$ and $t_{\rm E,max}$) where photodisintegration will no longer substantially alter the composition and nuclei can still be accelerated to ultra-high energies. This also generally occurs only for the more rapid rotators with a weak dependence on the magnetic field strength.
    \end{item}
    \begin{item}
    \textit{Limited synthesis} ($\overline{A}\sim4-15$) occurs for $Y_e\geq0.5$. If the outflow is launched at $t_{\rm \gamma,dis}$ or $t_{\rm E,max}$ and $P_0\gtrsim 3\, {\rm ms}$, even neutron-rich outflows produce light elements.
    \end{item}
\end{itemize}

These trends suggest that the overall outflow composition from a protomagnetar could be dominated by light, intermediate, or heavy nuclei, but depends sensitively on the spin period and time evolution of the electron fraction. Nuclei around the third r-process peak are not typically reached, even in the most favorable conditions considered here (low $B_{\rm dip}$, low $P_0$, $Y_e=0.45$, at $t_{\rm start}$). Some amount of first and second r-process peak elements are produced in these favorable conditions, so protomagnetars may be a contributor to the weak r-process abundance.

To fully understand the contribution of magnetar central engines to the entire UHECR flux requires an improved understanding of the population and distribution of such sources, both theoretical and observational. Additionally, improved understanding of the acceleration, propagation, and survival of nuclei will be important. This paper supports a growing notion of intermediate mass UHECR composition in certain magnetar model configurations.

It remains to be seen how the composition (and abundance distribution) changes when integrated over time and weighted by the distribution of magnetar properties in the whole population. After nucleosynthesis, processes such as disintegration, acceleration, and propagation must be considered if sites like these are to contribute significantly to the UHECR spectrum. Uncertainties in features like the extragalactic magnetic field will be coupled to the composition of heavy nuclei that are synthesized near the source. Each stage requires further study and is left for future work. 

\section*{Acknowledgements}
We thank Kohta Murase, David Radice and Yudai Suwa for useful discussions. We thank Ke Fang, Kunihito Ioka, Brian Metzger, Bing Theodore Zhang and the anonymous reviewer for carefully reading the manuscript and providing insightful suggestions that helped improve the paper. 
NE\ and MB\ are supported by NSF Grant No.\ PHY-1914409. 
MB\ acknowledges support from Eberly Research Fellowship at the Pennsylvania State University. 
The work of SH\ is supported by the U.S.\ Department of Energy Office of Science under award number DE-SC0020262, NSF Grant Nos.\ AST-1908960 and No.\ PHY-1914409, and JSPS KAKENHI Grant Number JP22K03630. This work was supported by World Premier International Research Center Initiative (WPI Initiative), MEXT, Japan.

\section*{Data availability}
The data underlying this article will be shared on reasonable request to the corresponding author.

\bibliography{main}

\appendix 

\section{Notation Table}\label{sec:appendNotation}

We include a table of the symbols that we use, their physical description and the first equation/section where they are used in (`Text' if the symbol comes from within the text). In some cases, the first equation is not the best location for the symbol; for instance, $\Dot{M}$ is first mentioned in equation (\ref{mdot}) but equation (\ref{mdot2}) is used for computing the thermodynamic trajectories with {\tt SkyNet}.

\begin{table*}
\centering
\begin{tabular}{c | l | c} 
 \hline
 \textbf{Symbol} & \textbf{Description} & \textbf{$1^{\rm st}$ Equation / Section} \\
 \hline\hline
 $\Dot{M}$ & Mass loss rate from neutrino-driven outflows. & (\ref{mdot}) / $\S$\ref{properties}\\ 
 \hline
 r & Radial coordinate of the outflow. & (\ref{mdot}) / $\S$\ref{properties}\\
 \hline
 $\rho$ & Outflow density. & (\ref{mdot}) / $\S$\ref{properties}\\
 \hline
 $v$ & Outflow velocity at radial coordinate. & (\ref{mdot}) / $\S$\ref{properties}\\
 \hline
 $f_{\rm open}$ & Fraction of magnetosphere open to outflows. & (\ref{mdot}) / $\S$\ref{properties}\\
 \hline
 $f_{\rm cent}$ & Fractional mass loss rate from magnetocentrifugal slinging. & (\ref{mdot}) / $\S$\ref{properties}\\
 \hline
 $S$ & Entropy per baryon. & (\ref{s}) / $\S$\ref{properties}\\
 \hline
 $T$ & Temperature of the outflow. & (\ref{s}) / $\S$\ref{properties}\\
 \hline
 $\tau_{\rm exp}$ & Expansion time-scale, used as IC. & (\ref{tau}) / $\S$\ref{properties}\\
 \hline
 $T_{\rm rec}$ & Recombination temperature. & (\ref{tau}) / $\S$\ref{properties}\\
 \hline
 $\chi$ & Obliquity angle: angle between magnetic field and rotation axes. & Text / $\S$\ref{properties}\\
 \hline
 $R_{\rm NS}$ & Neutron star radius. & Text / $\S$\ref{properties}\\
 \hline
 $R_Y$ & `Y'-point radius: final radius where magnetic field lines are still closed. & Text / $\S$\ref{properties}\\
 \hline
 $f_{\rm cent,max}$ & Maximum centrifugal slinging correction. & Text / $\S$\ref{properties}\\
 \hline
 $R_A$ & Alfven radius. & Text / $\S$\ref{properties}\\
 \hline
 $R_S$ & Sonic radius. & Text / $\S$\ref{properties}\\
 \hline
 $R_L$ & Light cylinder radius. & Text / $\S$\ref{properties}\\
 \hline
 $\Omega$ & Neutron star rotation rate. & Text / $\S$\ref{properties}\\
 \hline
 $P_{\rm cent}$ & Period factor accounting for centrifugal effects of rapid rotation. & Text / $\S$\ref{properties}\\
 \hline
 $\alpha$ & Argument to determine inclination angle & Text / $\S$\ref{properties}\\
 \hline
 $M_{\rm NS}$ & Neutron star baryonic mass. & Text / $\S$\ref{properties}\\
 \hline
 $\theta_{\rm open}$ & Opening angle of the polar cap. & Text / $\S$\ref{properties}\\
 \hline
 $C_{\rm es}$ & Heating correction factor for neutrino-electron scattering. & (\ref{mdot2}) / $\S$\ref{properties}\\
 \hline
 $L_{\nu}$ & Neutrino luminosity for $\overline{\nu}_e$ (anti-electron flavour). & (\ref{mdot2}) / $\S$\ref{properties}\\
 \hline
 $\epsilon_{\nu}$ & Mean neutrino energy for $\overline{\nu}_e$ (anti-electron flavour).  & (\ref{mdot2}) / $\S$\ref{properties}\\
 \hline
 $R_{10}$ & NS radius divided by 10 km. & (\ref{mdot2}) / $\S$\ref{properties}\\
 \hline
 $M_{1.4}$ & $M_{\rm NS}$ divided by 1.4 solar masses. & (\ref{mdot2}) / $\S$\ref{properties}\\
 \hline
 $\eta_s$ & A `stretch' factor to correct neutrino quantities for rotation. & Text / $\S$\ref{properties}\\
 \hline
 $S_{\rm rot}$ & Entropy suppressed by rapid rotation. & Text / $\S$\ref{properties}\\
 \hline\hline
 $\beta_j$ & Unit-less jet velocity. & (\ref{gammaj}) / $\S$\ref{postlaunchevolution}\\
 \hline
 $\Gamma_j$ & Jet Lorentz factor. & (\ref{gammaj}) / $\S$\ref{postlaunchevolution}\\
 \hline
 $\sigma_0$ & Outflow magnetization, corresponds to maximum Lorentz factor. & (\ref{gammaj}) / $\S$\ref{postlaunchevolution}\\
 \hline
 $R_{\rm mag}$ & Magnetic saturation radius. & (\ref{gammaj}) / $\S$\ref{postlaunchevolution}\\
 \hline
 $\phi$ & Magnetic flux threading open magnetosphere. & (\ref{sigma0}) / $\S$\ref{postlaunchevolution}\\
 \hline
 $t_{\rm bo}$ & Jet breakout time; when outflow breaks out of progenitor envelope. & (\ref{tbo}) / $\S$\ref{postlaunchevolution}\\
 \hline
 $B_{\rm dip}$ & Surface dipolar magnetic field strength. & (\ref{tbo}) / $\S$\ref{postlaunchevolution}\\
 \hline
 $P_0$ & Initial spin period. & (\ref{tbo}) / $\S$\ref{postlaunchevolution}\\
 \hline
 $\tau_{\gamma-N}$ & The number of nuclei-destroying interactions. & (\ref{photointeractions}) / $\S$\ref{postlaunchevolution}\\
 \hline
 $t_{\rm E,max}$ & Last time when nuclei can be accelerated above $10^{20}$ eV. & Text / $\S$\ref{postlaunchevolution}\\
 \hline
 $\Dot{E}_{\rm iso}$ & Isotropic jet luminosity. & (\ref{photointeractions}) / $\S$\ref{postlaunchevolution}\\
 \hline
 $\varepsilon_{\rm rad}$ & Radiative efficiency of the jet. & (\ref{photointeractions}) / $\S$\ref{postlaunchevolution}\\
 \hline
 $C$ & Fraction of gamma-ray photons released below Band peak energy. & (\ref{photointeractions}) / $\S$\ref{postlaunchevolution}\\
 \hline
 $\sigma_{\rm GDR}$ & Resonance cross section & (\ref{photointeractions}) / $\S$\ref{postlaunchevolution}\\
 \hline
 $\overline{\epsilon}_{\rm GDR}$ & Resonance energy. & (\ref{photointeractions}) / $\S$\ref{postlaunchevolution}\\
 \hline
 $\Delta\epsilon_{\rm GDR}$ & Resonance width. & (\ref{photointeractions}) / $\S$\ref{postlaunchevolution}\\
 \hline
 $\epsilon_p$ & Band peak energy. & (\ref{photointeractions}) / $\S$\ref{postlaunchevolution}\\
 \hline
 $A$ & Mass number. & Text / $\S$\ref{postlaunchevolution}\\
 \hline\hline
 $Y_e$ & Electron fraction. & Text / $\S$\ref{modelparam}\\
 \hline
 $n_p$ & Proton number density. & Text / $\S$\ref{modelparam}\\
 \hline
 $n_n$ & Neutron number density. & Text / $\S$\ref{modelparam}\\
 \hline
 $t_{\rm start}$ & Time after core-collapse when magnetar outflow begins. & Text / $\S$\ref{modelparam}\\
 \hline
 $t_{\rm \gamma,dis}$ & When synthesized nuclei will no longer be photodisintegrated. & Text / $\S$\ref{modelparam}\\
 \hline\hline
 $Y$ & Abundance of nuclei. & Text / $\S$\ref{abundancepatterns}\\
 \hline
 $Z$ & Charge number. & Text / $\S$\ref{abundancepatterns}\\
 \hline\hline
 $\overline{A}$ & Mean mass number, weighted by abundance. & (\ref{abar}) / $\S$\ref{meancomptrends}\\
 \hline
 $X_h$ & Total mass fraction of elements more massive than iron. & (\ref{xh}) / $\S$\ref{meancomptrends}\\
 \hline
 $X$ & Mass fraction. & (\ref{xh}) / $\S$\ref{meancomptrends}\\
 \hline
 $\overline{A}_{\rm heavy}$ & Mean mass number excluding free nuclei and helium. & Text / $\S$\ref{meancomptrends}\\
 \hline
\end{tabular}
\caption{Symbol, description, and first location (equation and/or section) of the symbol for the variables used in this paper. For the symbols from equations in text, `Text' is used.}
\label{table:symbols}
\end{table*}

\section{Initial Conditions}\label{sec:appendIC}

\begin{figure*}
\begin{minipage}{0.245\textwidth}
\begin{tikzpicture}
\node (img) {\includegraphics[width=0.95\linewidth]{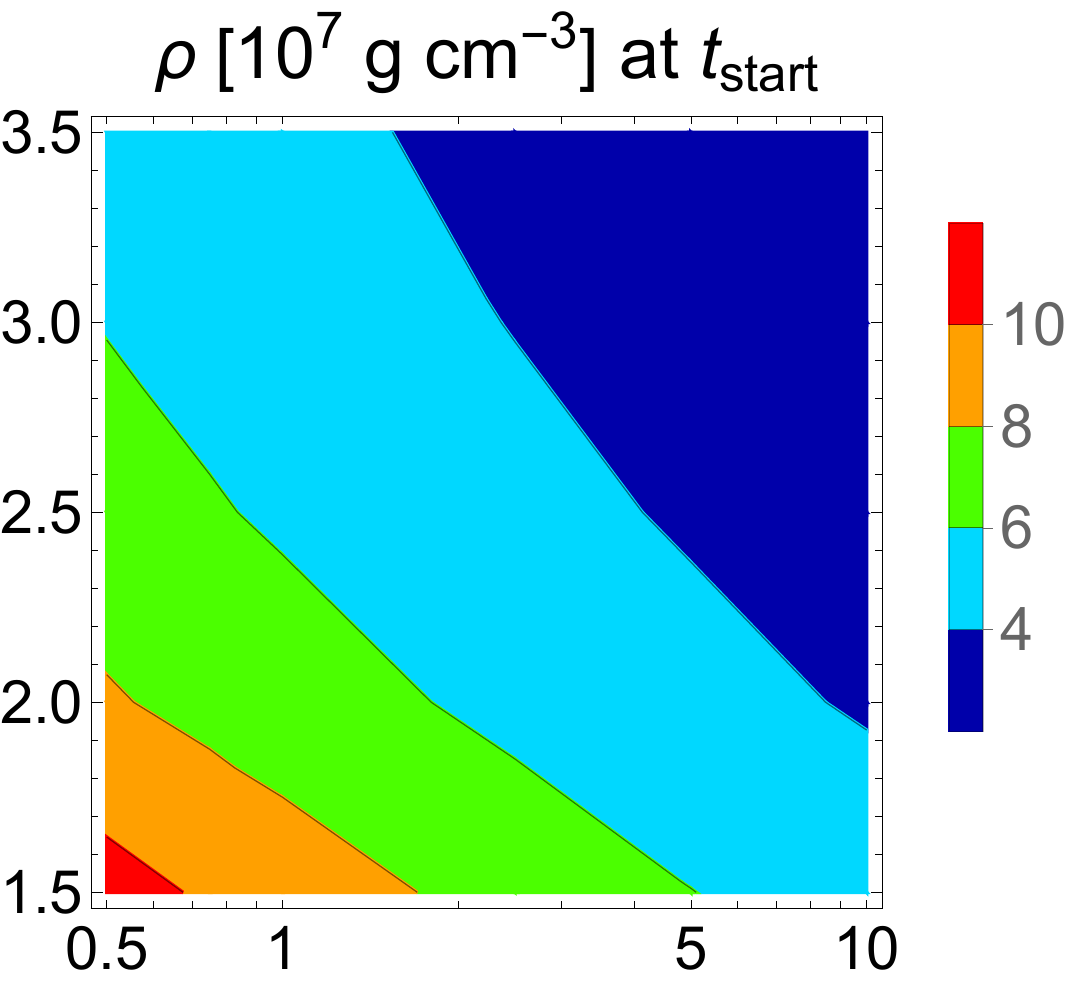}};
\node[left=67pt, node distance=0cm, rotate=90, anchor=center,font=\color{white}] {$|$};
\end{tikzpicture}
\end{minipage}
\begin{minipage}{0.245\textwidth}
\begin{tikzpicture}
\node (img) {\includegraphics[width=0.95\linewidth]{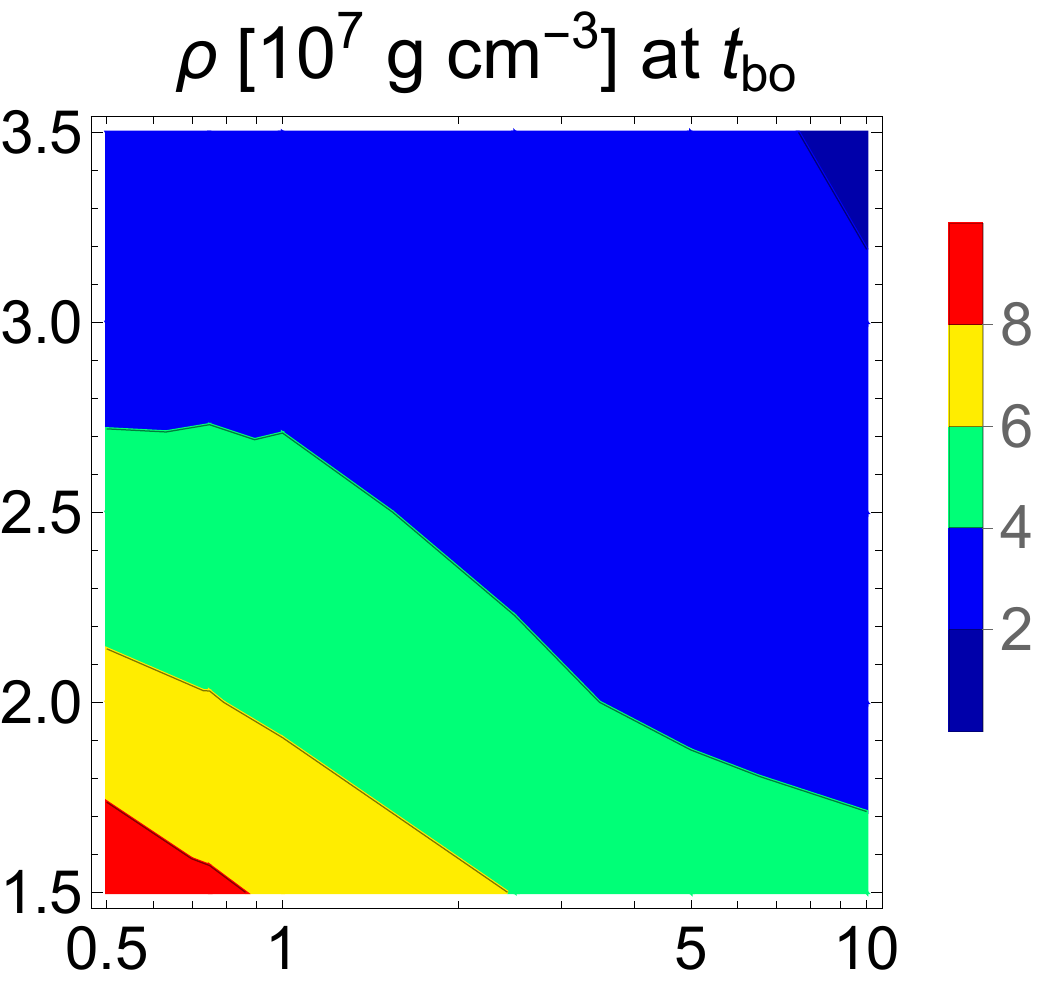}};
\node[left=67pt, node distance=0cm, rotate=90, anchor=center,font=\color{white}] {~$|$};
\end{tikzpicture}
\end{minipage}
\begin{minipage}{0.245\textwidth}
\begin{tikzpicture}
\node (img) {\includegraphics[width=0.95\linewidth]{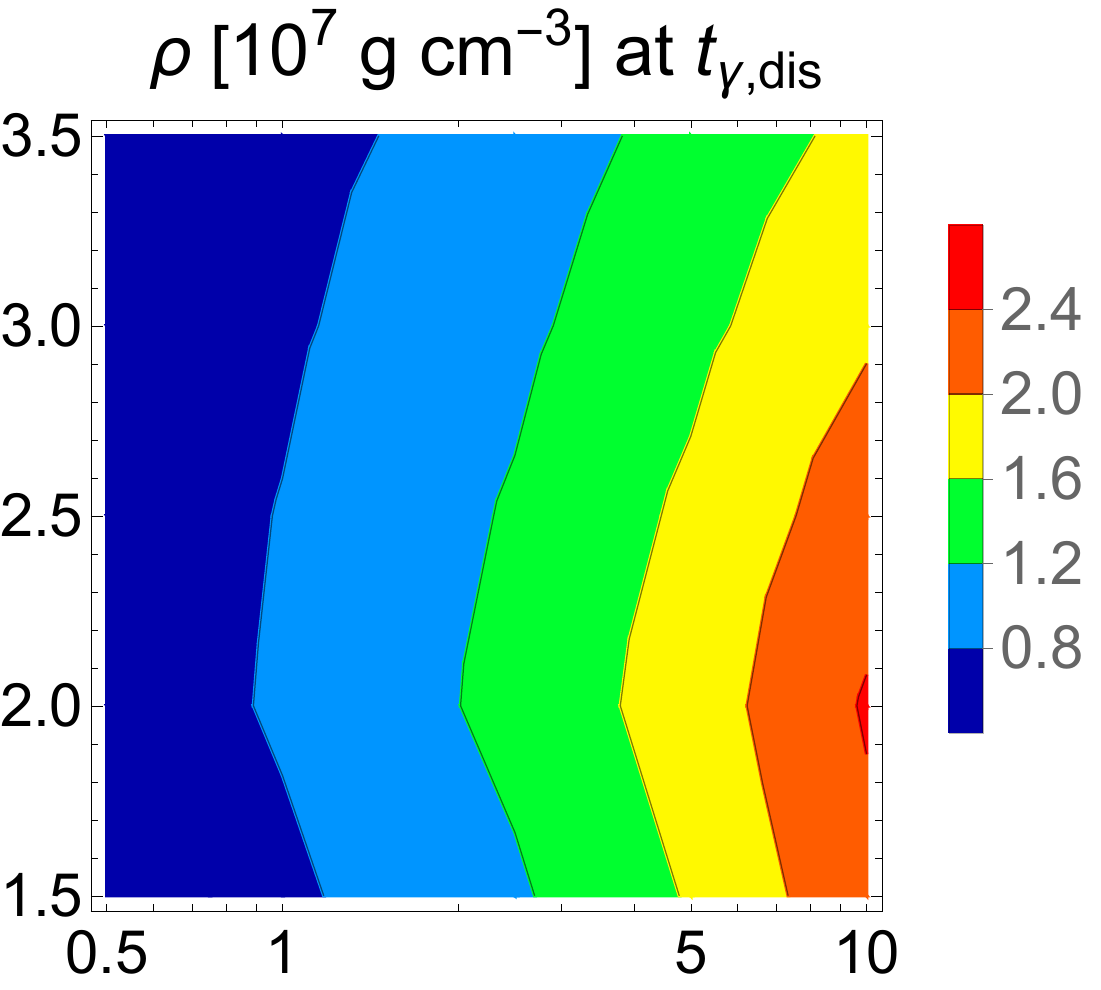}};
\node[left=67pt, node distance=0cm, rotate=90, anchor=center,font=\color{white}] {~$|$};
\end{tikzpicture}
\end{minipage}
\begin{minipage}{0.245\textwidth}
\begin{tikzpicture}
\node (img) {\includegraphics[width=0.95\linewidth]{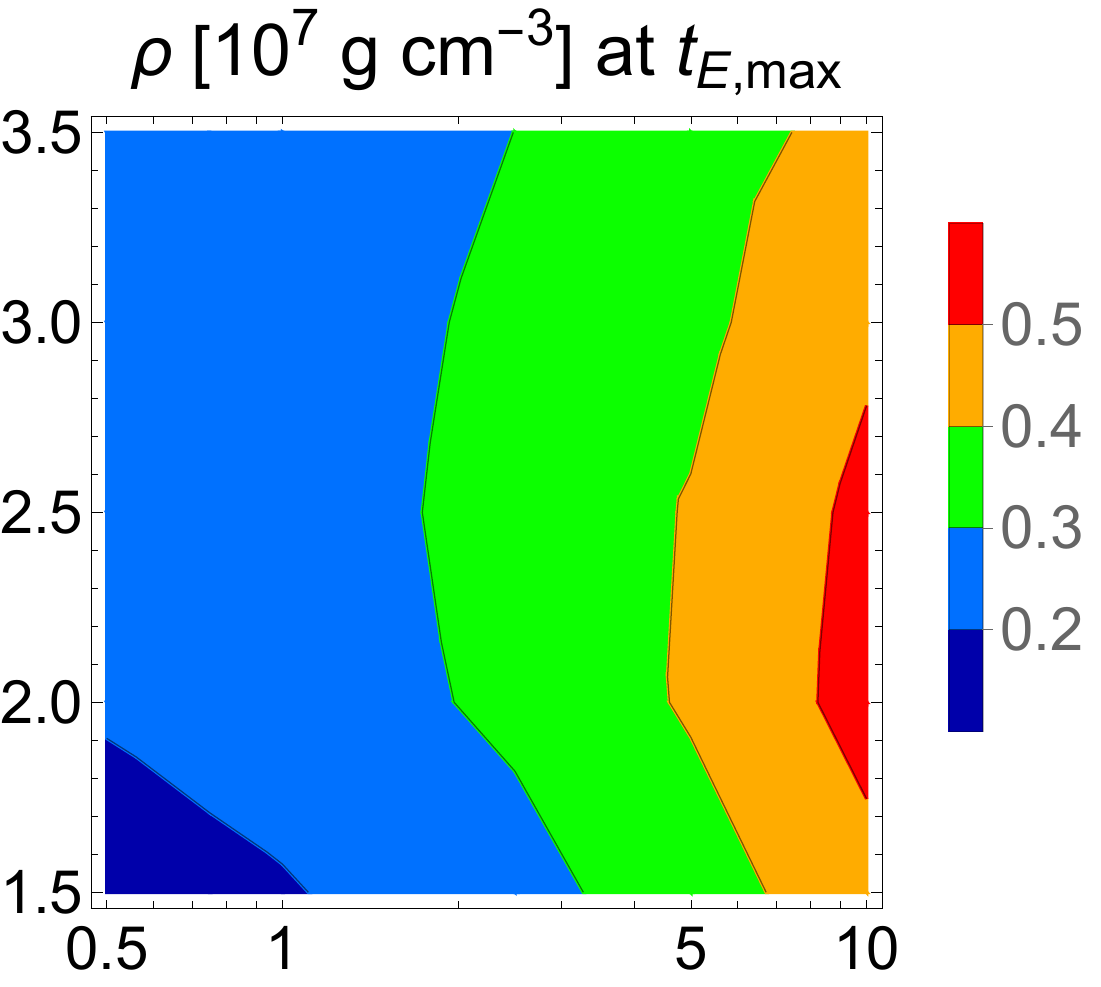}};
\node[left=67pt, node distance=0cm, rotate=90, anchor=center,font=\color{white}] {~~~~$|$};
\end{tikzpicture}
\end{minipage}
\newline
\begin{minipage}{0.245\textwidth}
\begin{tikzpicture}
\node (img) {\includegraphics[width=0.95\linewidth]{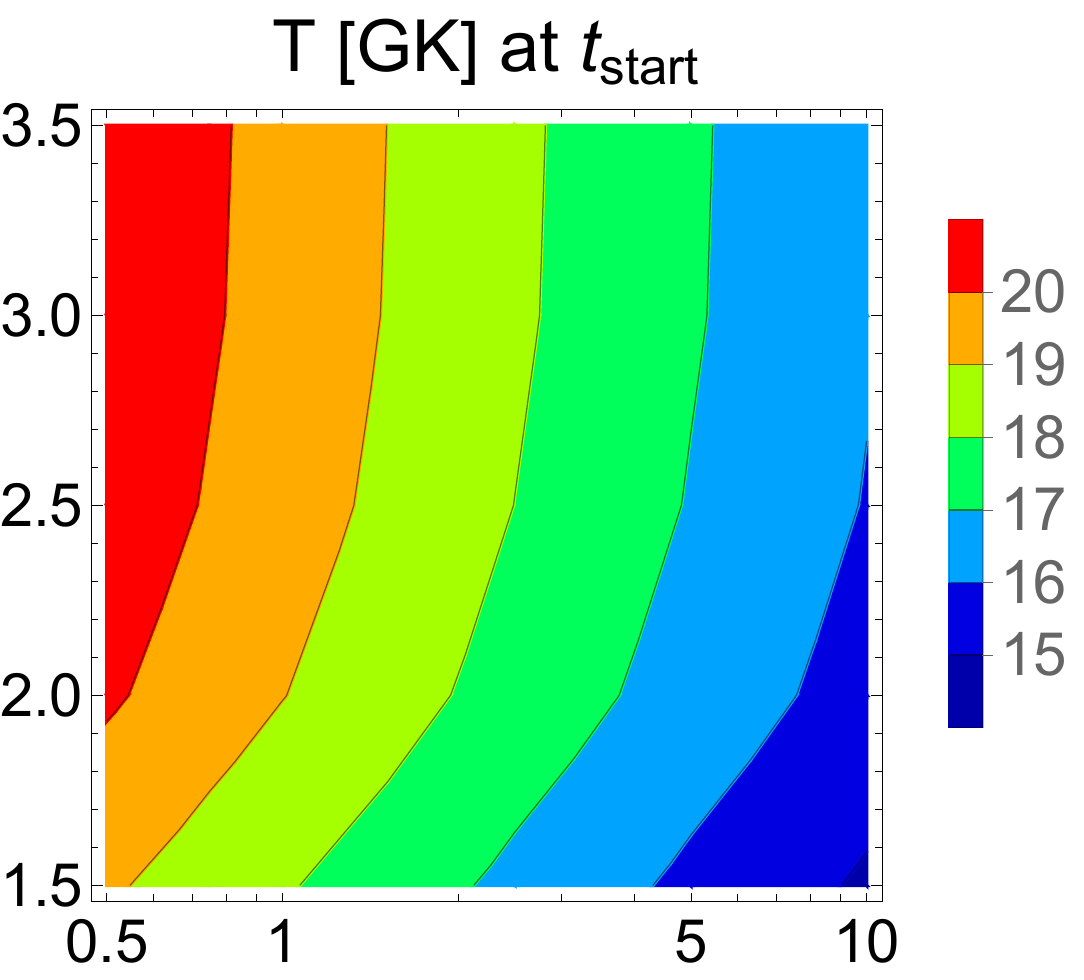}};
\node[left=67pt, node distance=0cm, rotate=90, anchor=center,font=\color{white}] {$|$};
\end{tikzpicture}
\end{minipage}
\begin{minipage}{0.245\textwidth}
\begin{tikzpicture}
\node (img) {\includegraphics[width=0.95\linewidth]{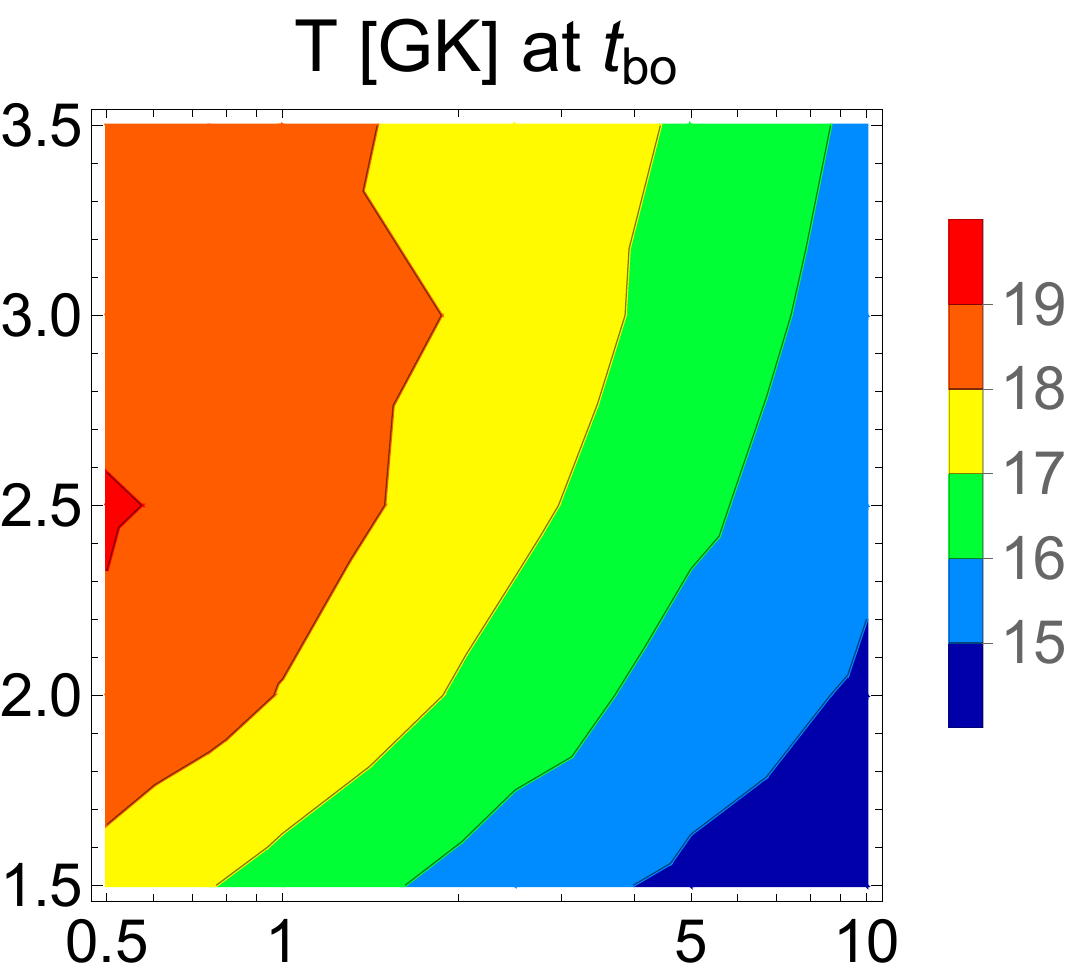}};
\node[left=67pt, node distance=0cm, rotate=90, anchor=center,font=\color{white}] {$~~|$};
\end{tikzpicture}
\end{minipage}
\begin{minipage}{0.245\textwidth}
\begin{tikzpicture}
\node (img) {\includegraphics[width=0.95\linewidth]{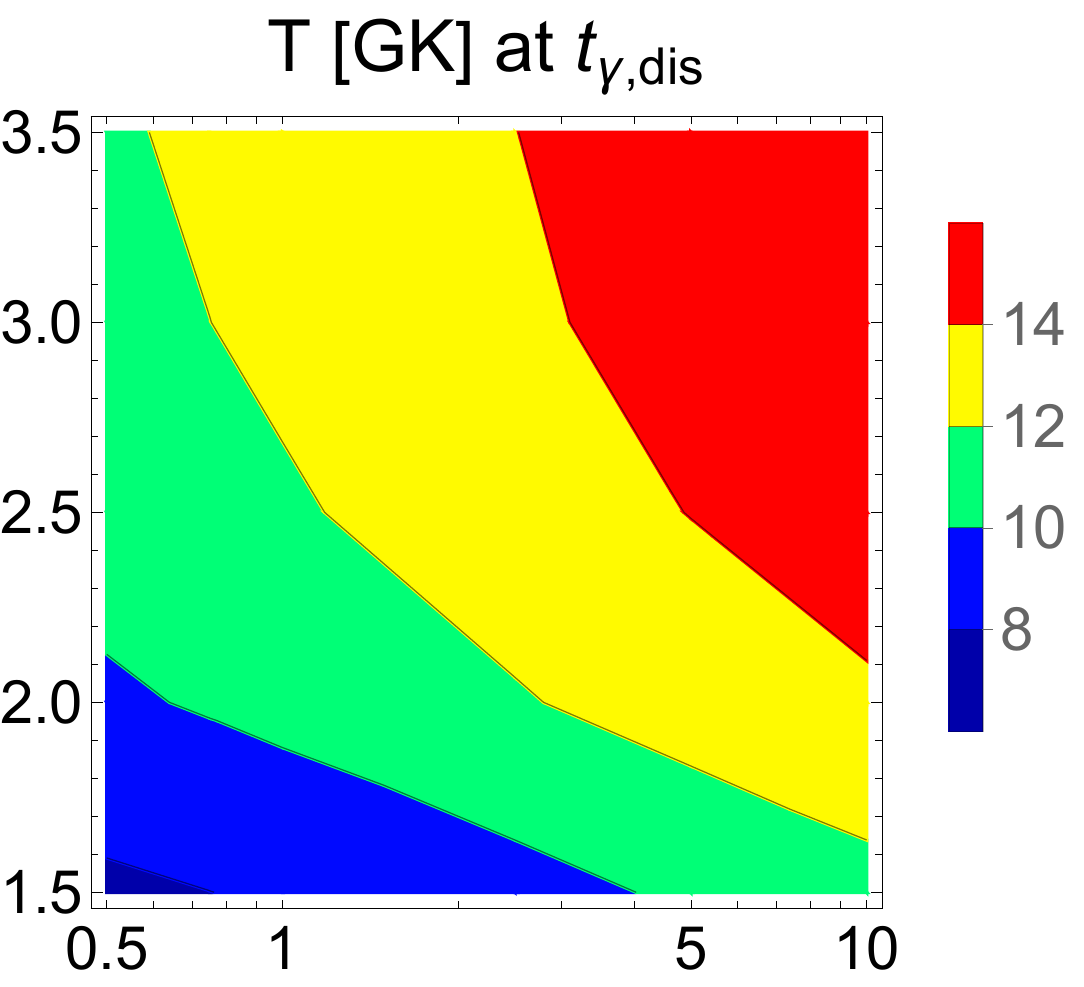}};
\node[left=67pt, node distance=0cm, rotate=90, anchor=center,font=\color{white}] {$|$};
\end{tikzpicture}
\end{minipage}
\begin{minipage}{0.245\textwidth}
\begin{tikzpicture}
\node (img) {\includegraphics[width=0.95\linewidth]{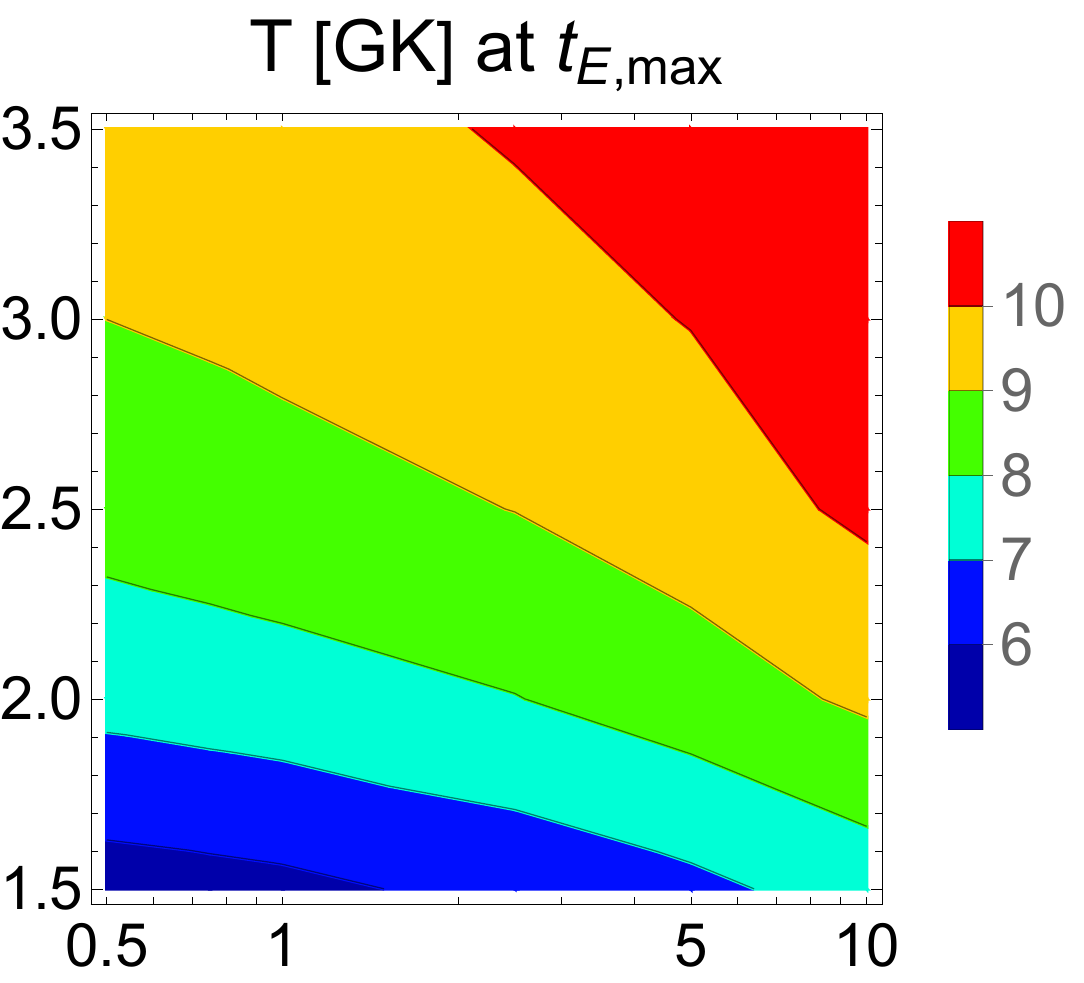}};
\node[left=67pt, node distance=0cm, rotate=90, anchor=center,font=\color{white}] {$~~~|$};
\end{tikzpicture}
\end{minipage}
\newline
\begin{minipage}{0.245\textwidth}
\begin{tikzpicture}
\node (img) {\includegraphics[width=0.95\linewidth]{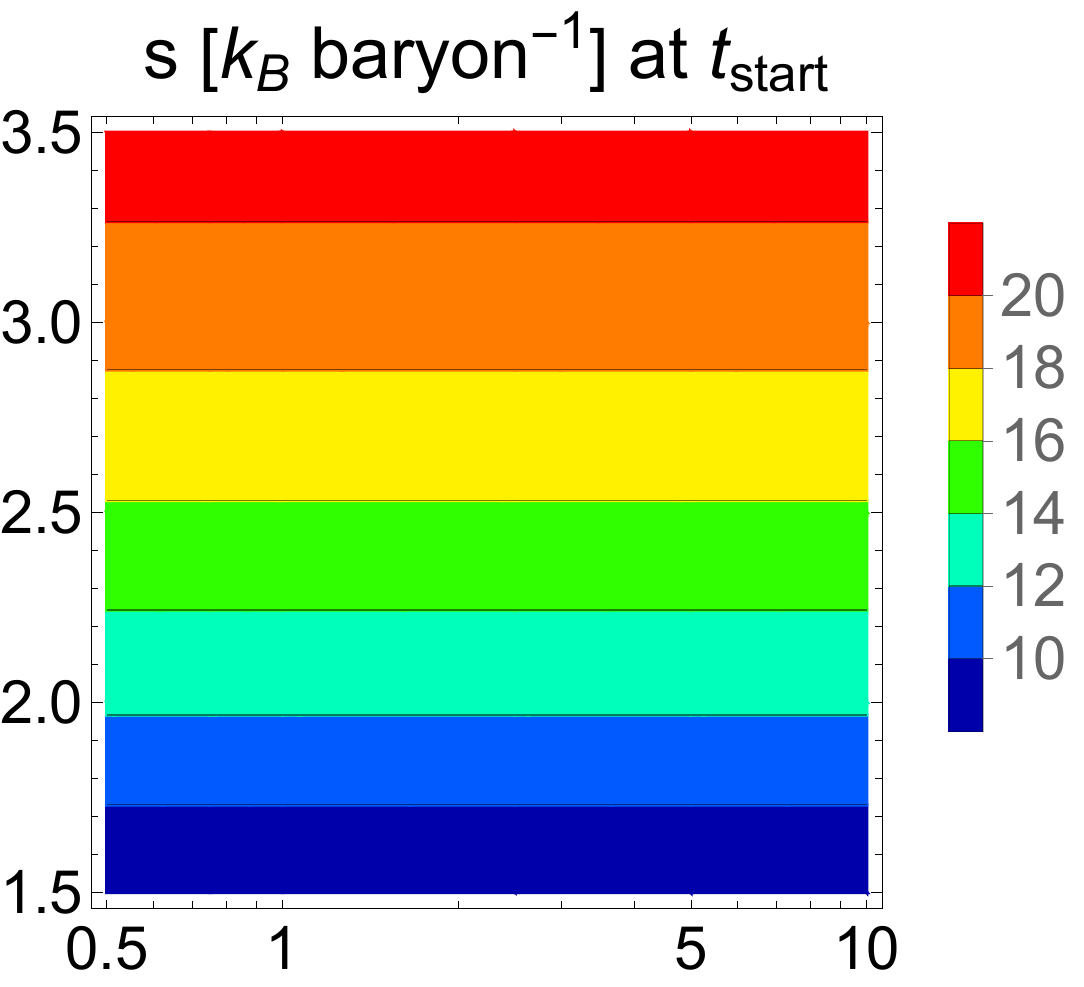}};
\node[left=67pt, node distance=0cm, rotate=90,xshift=18pt]{$P_0$ [ms]};
\end{tikzpicture}
\end{minipage}
\begin{minipage}{0.245\textwidth}
\begin{tikzpicture}
\node (img) {\includegraphics[width=0.95\linewidth]{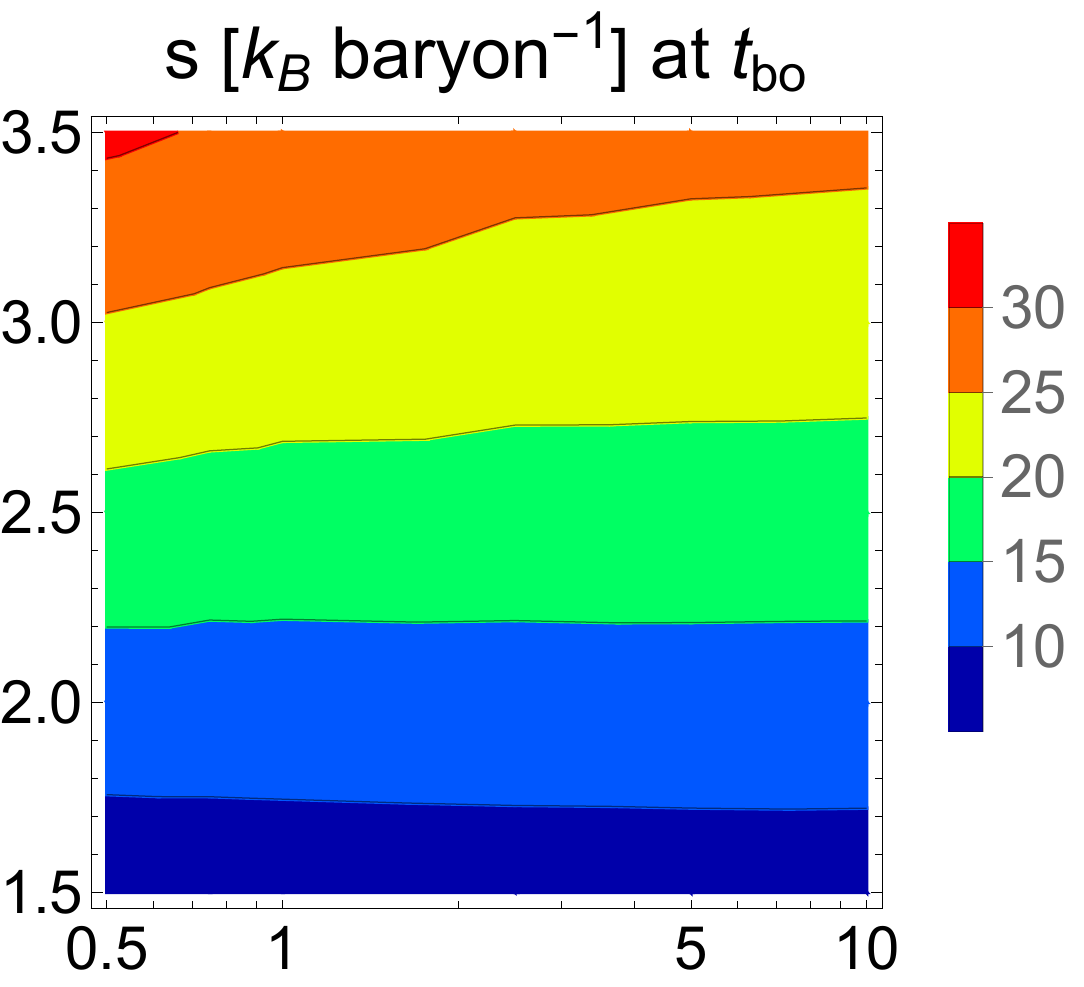}};
\node[left=67pt, node distance=0cm, rotate=90, anchor=center,font=\color{white}] {$|$};
\end{tikzpicture}
\end{minipage}
\begin{minipage}{0.245\textwidth}
\begin{tikzpicture}
\node (img) {\includegraphics[width=0.95\linewidth]{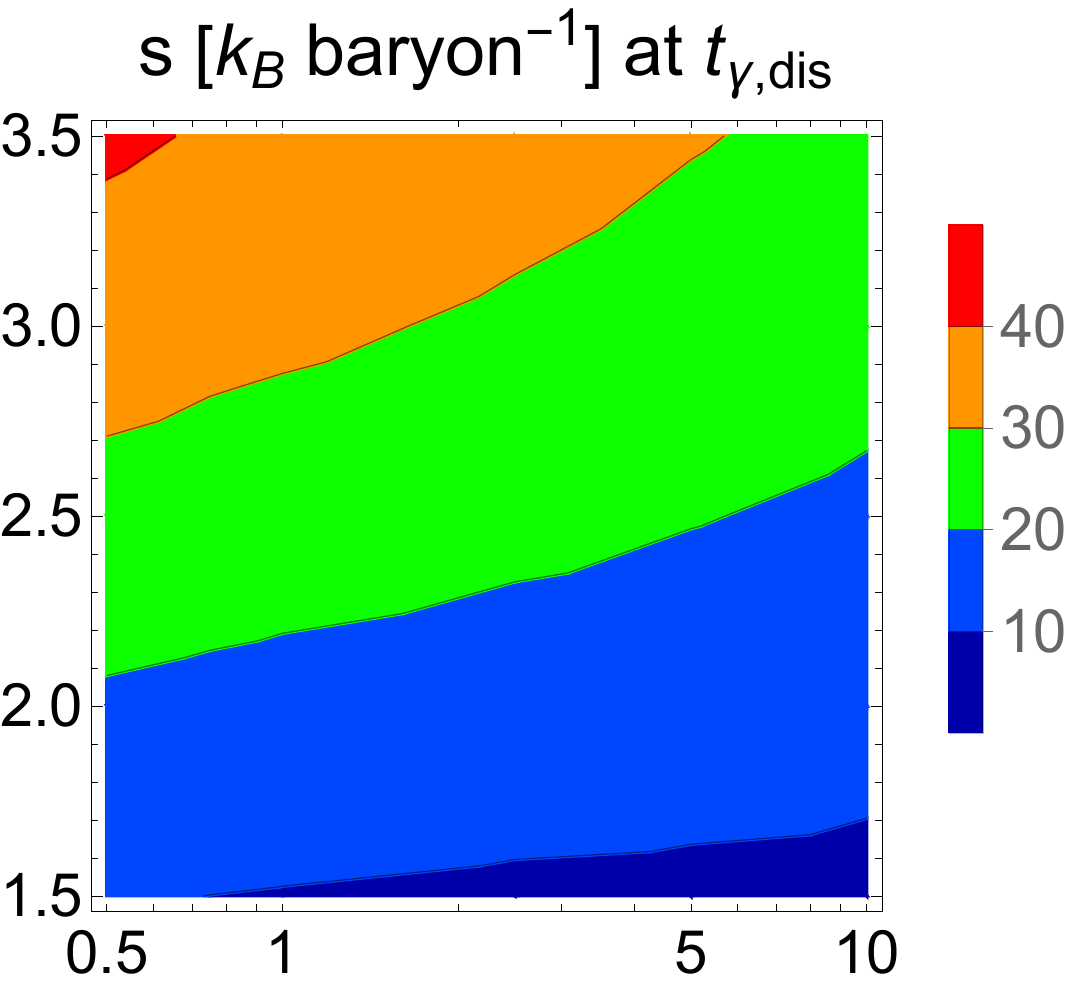}};
\node[left=67pt, node distance=0cm, rotate=90, anchor=center,font=\color{white}] {~~~~$|$};
\end{tikzpicture}
\end{minipage}
\begin{minipage}{0.245\textwidth}
\begin{tikzpicture}
\node (img) {\includegraphics[width=0.95\linewidth]{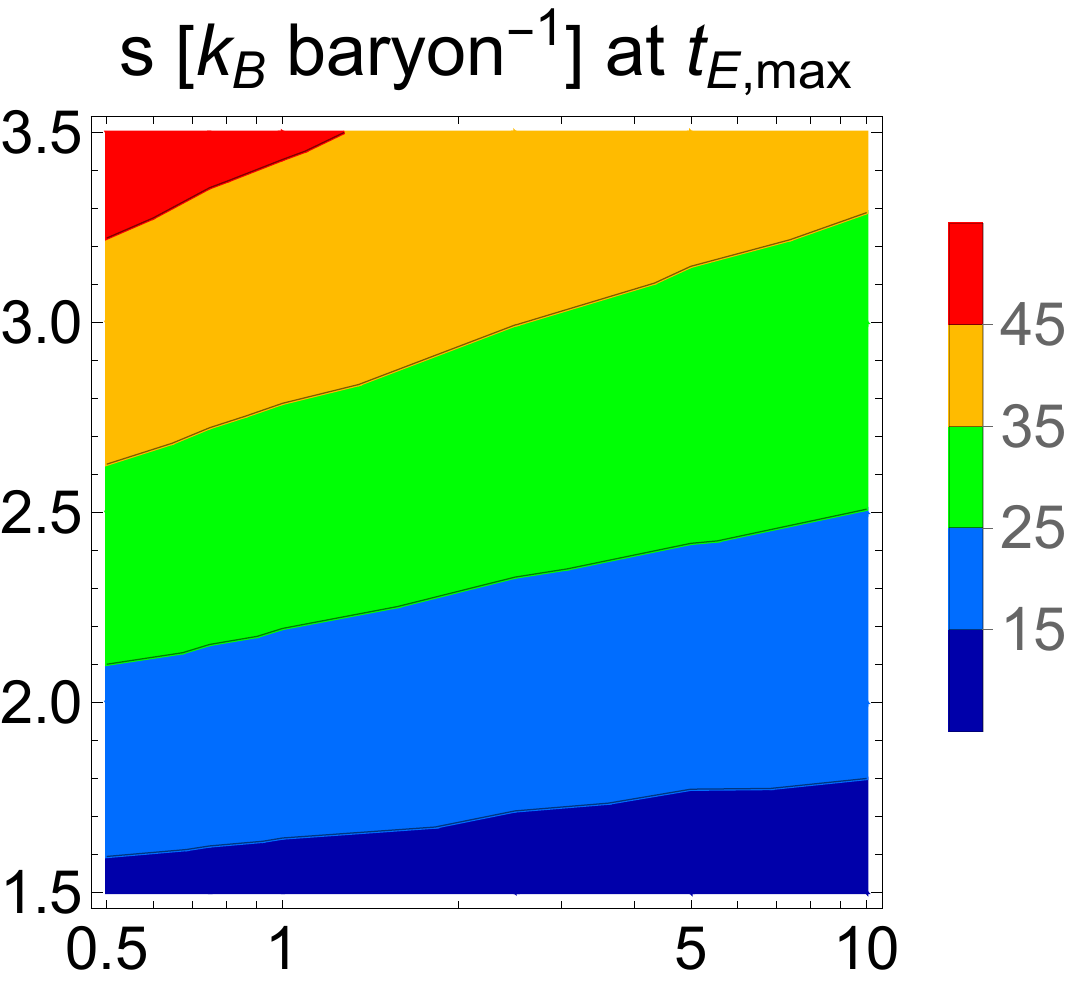}};
\node[left=67pt, node distance=0cm, rotate=90, anchor=center,font=\color{white}] {~~~$|$};
\end{tikzpicture}
\end{minipage}
\newline
\begin{minipage}{0.245\textwidth}
\begin{tikzpicture}
\node (img) {\includegraphics[width=0.95\linewidth]{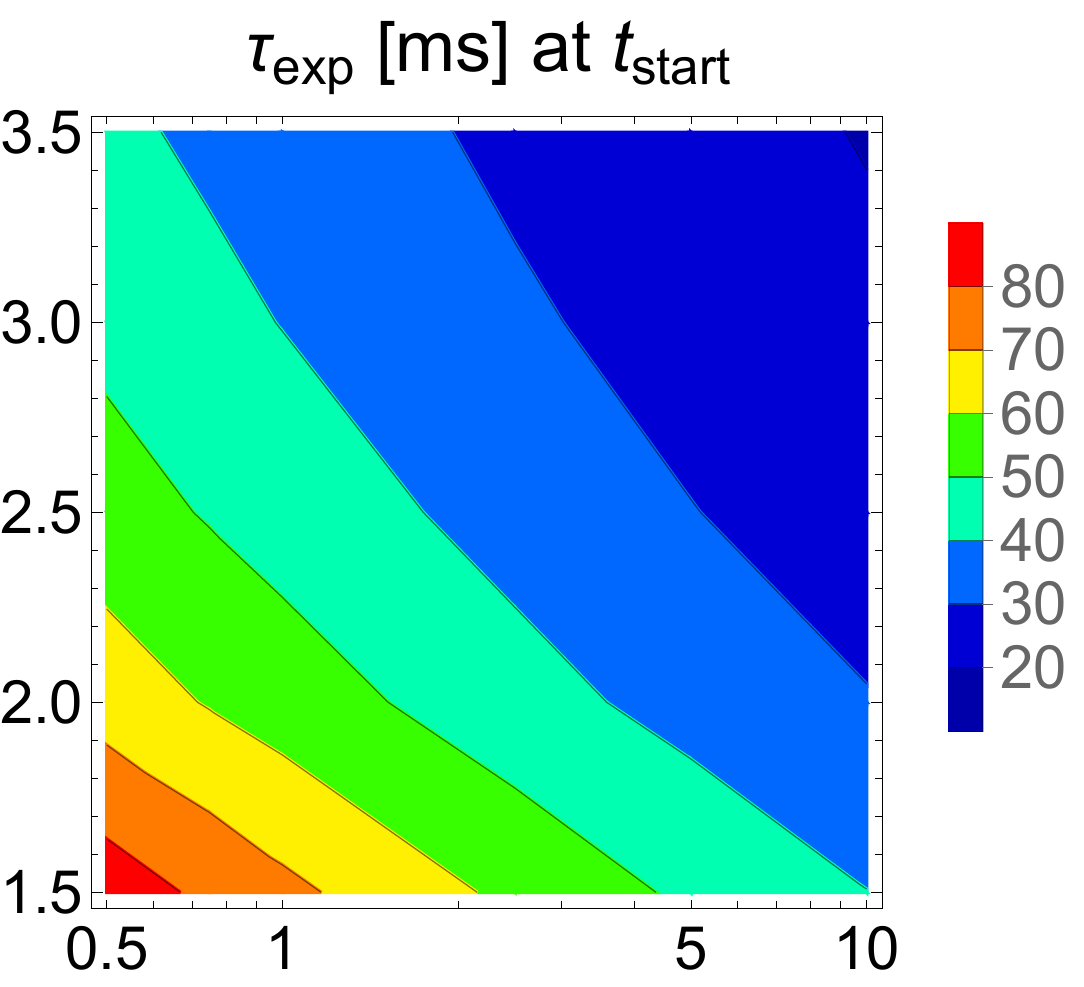}};
\node[left=67pt, node distance=0cm, rotate=90, anchor=center,font=\color{white}] {$~~|$};
\end{tikzpicture}
\end{minipage}
\begin{minipage}{0.245\textwidth}
\begin{tikzpicture}
\node (img) {\includegraphics[width=0.95\linewidth]{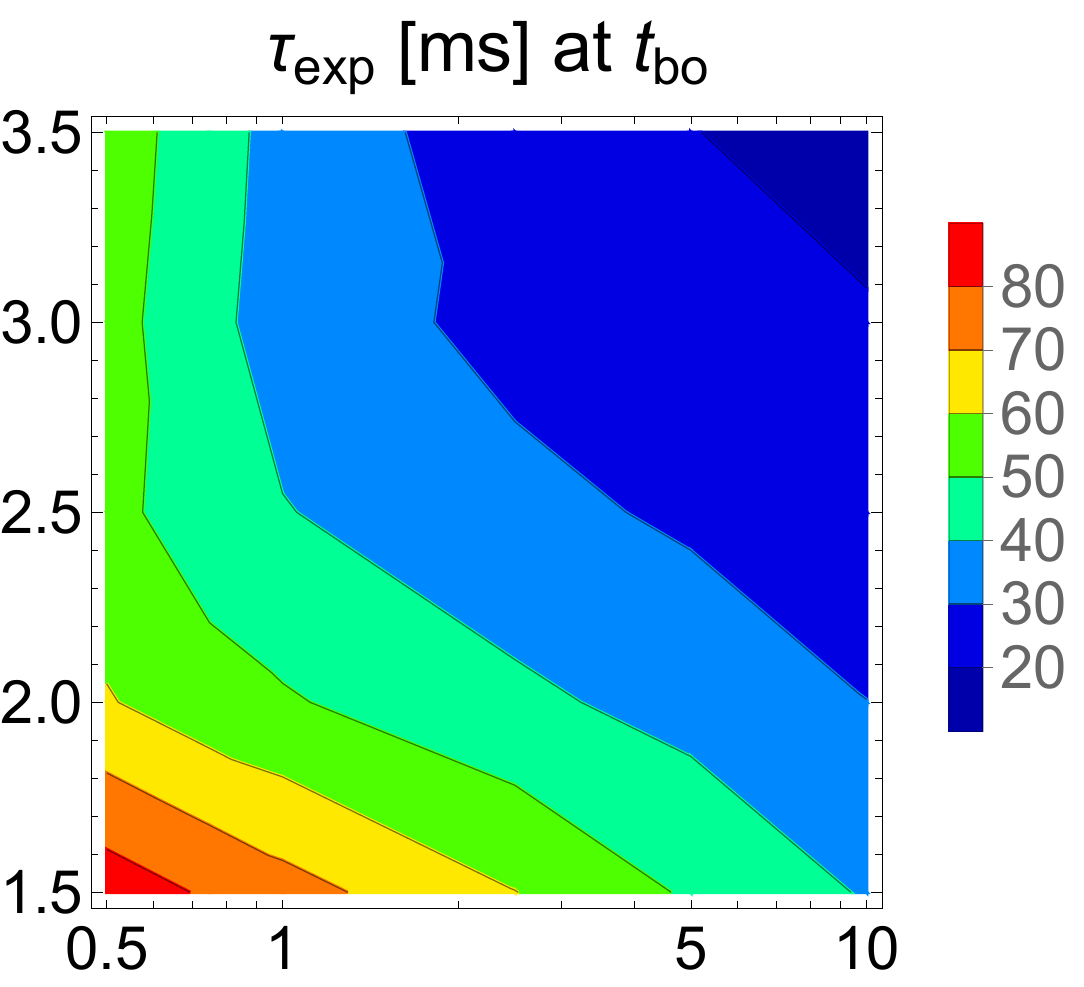}};
\node[left=67pt, node distance=0cm, rotate=90, anchor=center,font=\color{white}] {~~~$|$};
\end{tikzpicture}
\end{minipage}
\begin{minipage}{0.245\textwidth}
\begin{tikzpicture}
\node (img) {\includegraphics[width=0.95\linewidth]{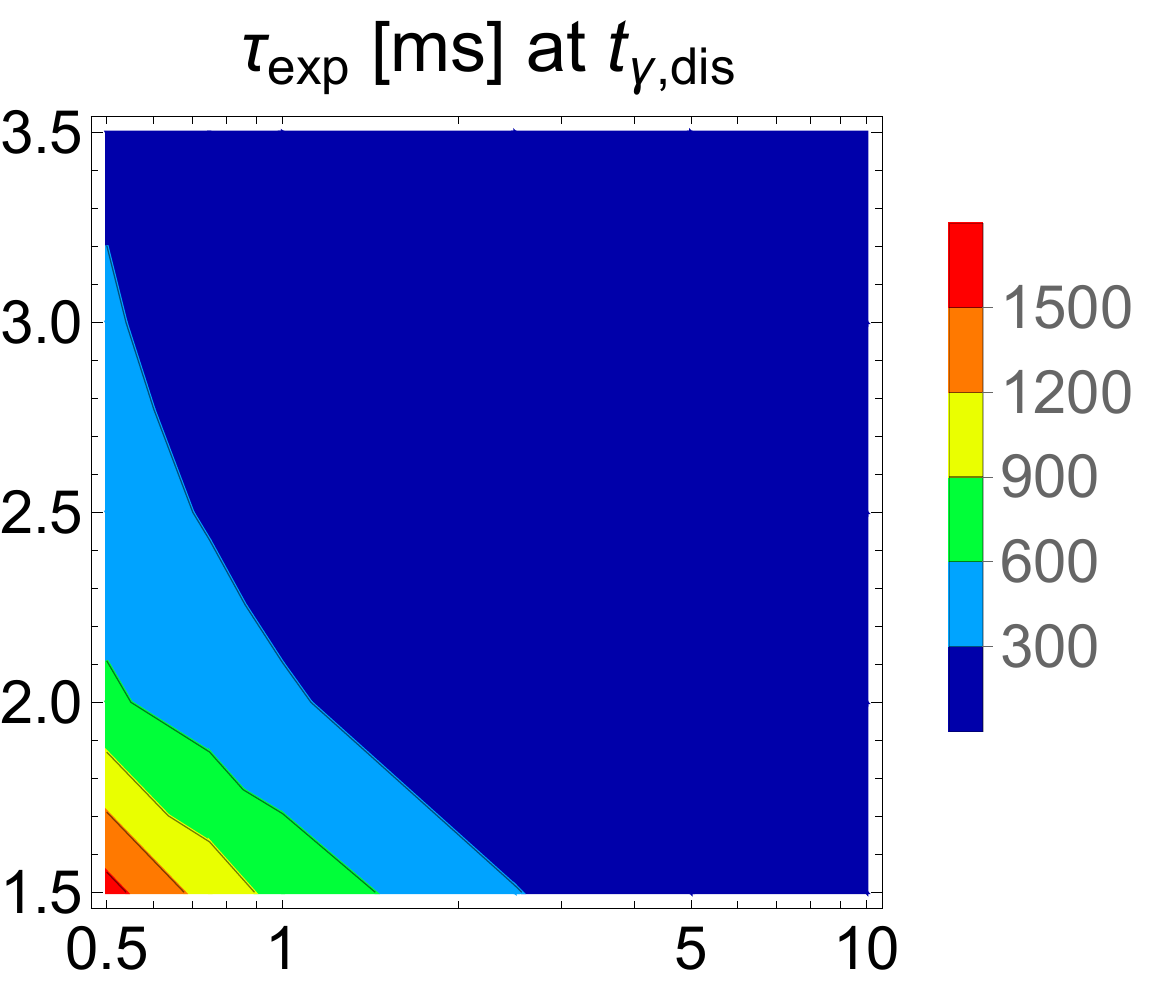}};
\node[left=67pt, node distance=0cm, rotate=90, anchor=center,font=\color{white}] {$~~~~|$};
\end{tikzpicture}
\end{minipage}
\begin{minipage}{0.245\textwidth}
\begin{tikzpicture}
\node (img) {\includegraphics[width=0.95\linewidth]{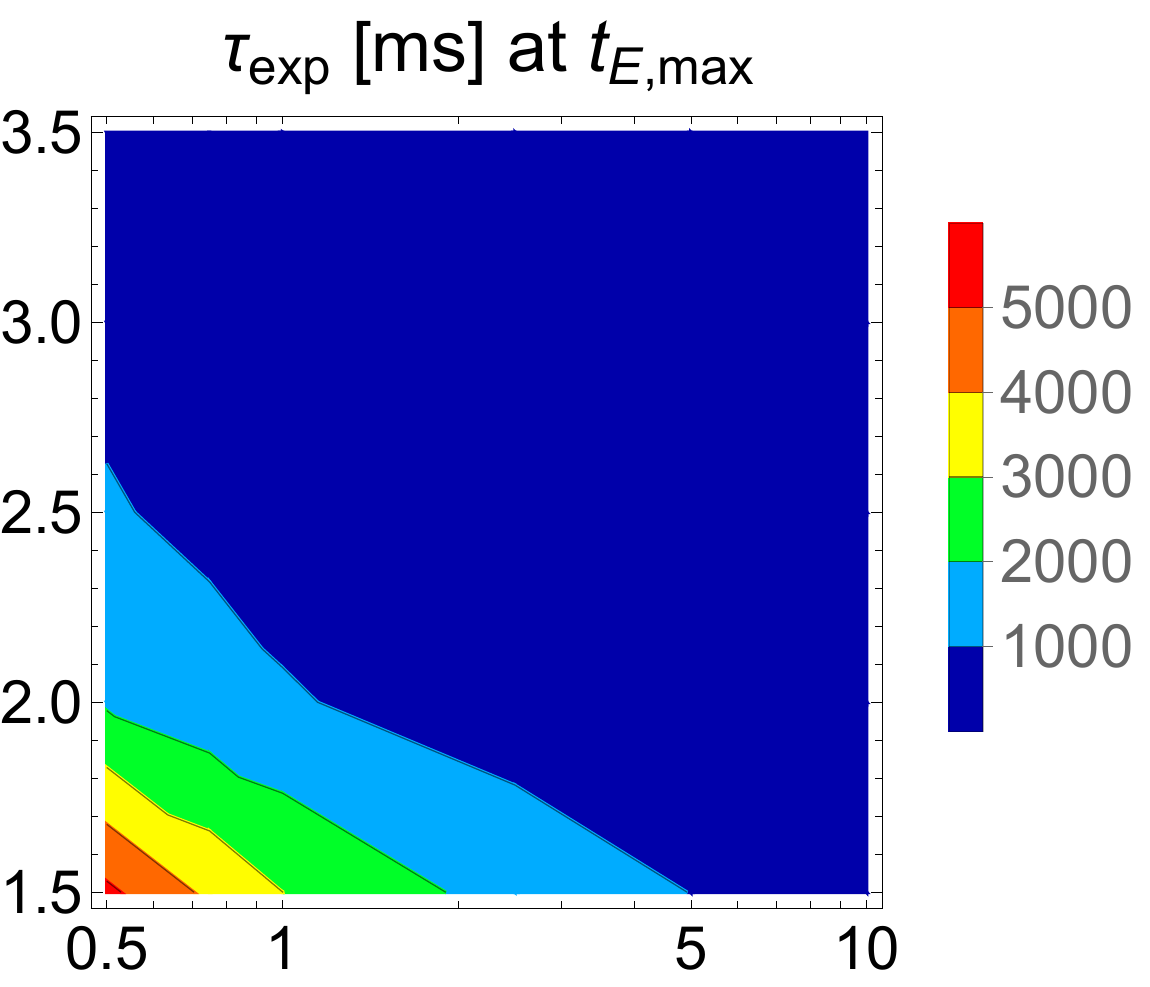}};
\node[left=67pt, node distance=0cm, rotate=90, anchor=center,font=\color{white}] {$~|$};
\end{tikzpicture}
\end{minipage}
\newline
\begin{minipage}{0.245\textwidth}
\begin{tikzpicture}
\node (img) {\includegraphics[width=0.95\linewidth]{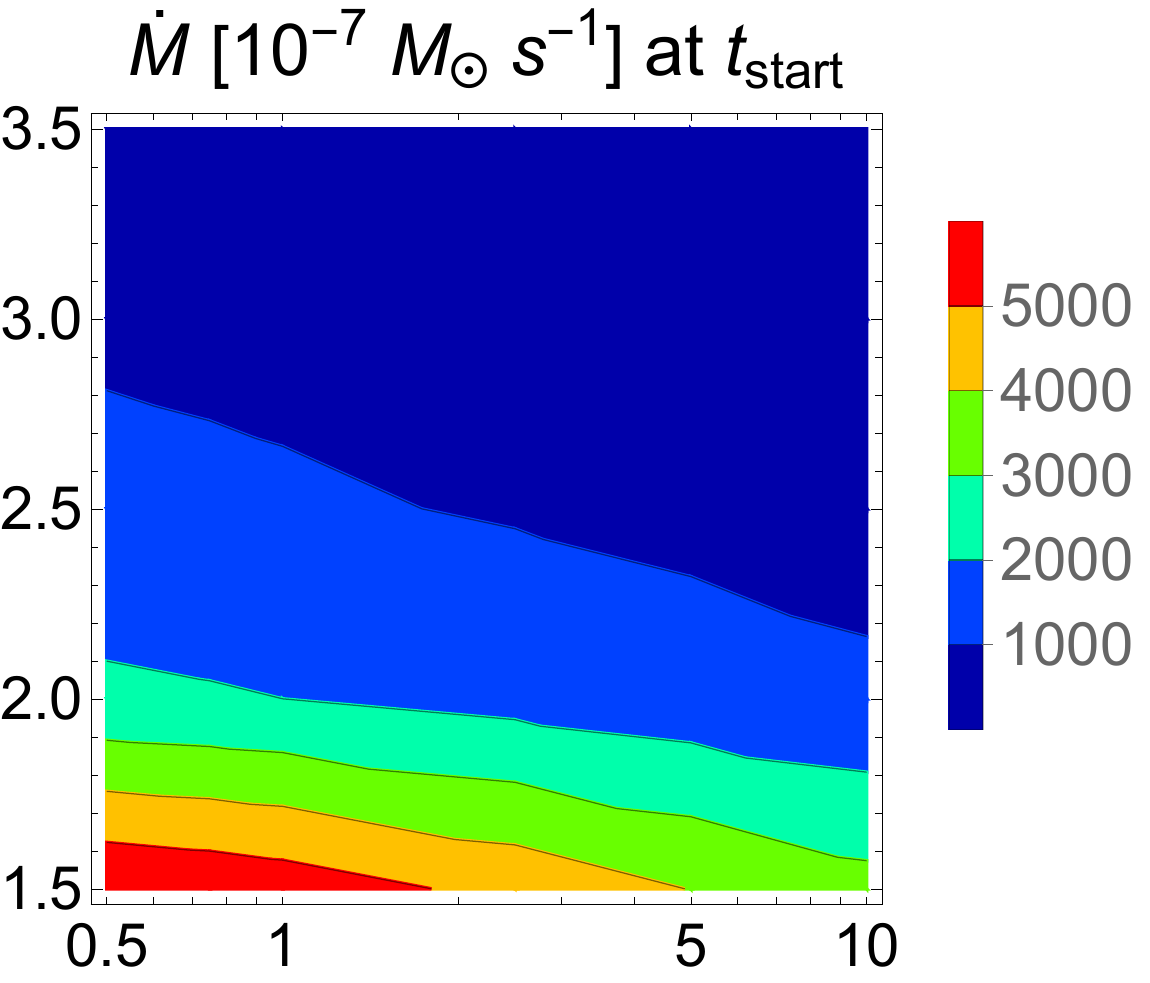}};
\node[below=55pt, node distance=0cm, font=\color{white}] {$|$};
\node[left=67pt, node distance=0cm, rotate=90, anchor=center,font=\color{white}] {$|$};
\end{tikzpicture}
\end{minipage}
\begin{minipage}{0.245\textwidth}
\begin{tikzpicture}
\node (img) {\includegraphics[width=0.95\linewidth]{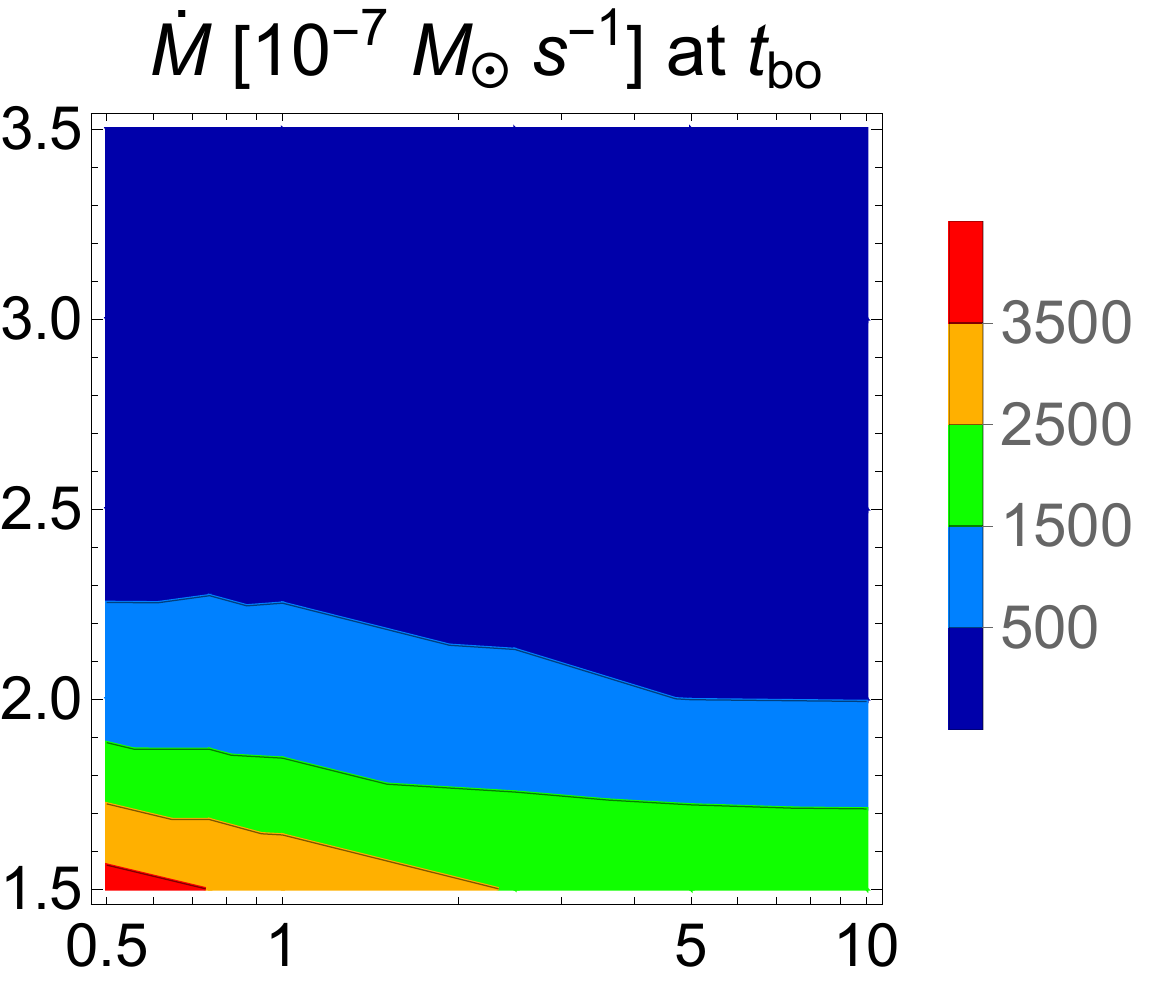}};
\node[below=54pt, node distance=0cm, xshift=2.1cm] {$B_{\textrm{dip}}$ [$10^{15}$ G]};
\node[left=67pt, node distance=0cm, rotate=90, anchor=center,font=\color{white}] {$~~~|$};
\end{tikzpicture}
\end{minipage}
\begin{minipage}{0.245\textwidth}
\begin{tikzpicture}
\node (img) {\includegraphics[width=0.95\linewidth]{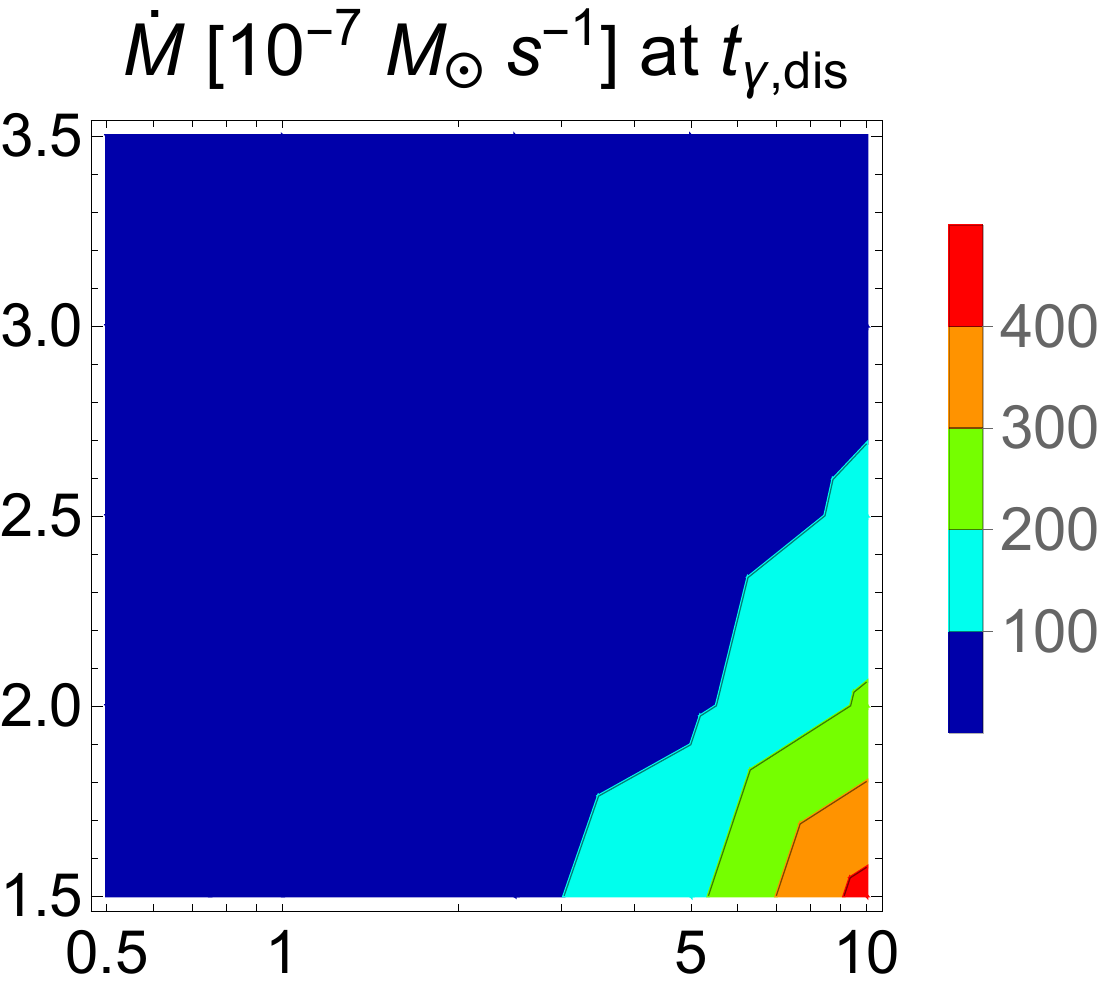}};
\node[below=55pt, node distance=0cm,font=\color{white}] {$|$};
\node[left=67pt, node distance=0cm, rotate=90, anchor=center,font=\color{white}] {$~~~|$};
\end{tikzpicture}
\end{minipage}
\begin{minipage}{0.245\textwidth}
\begin{tikzpicture}
\node (img) {\includegraphics[width=0.95\linewidth]{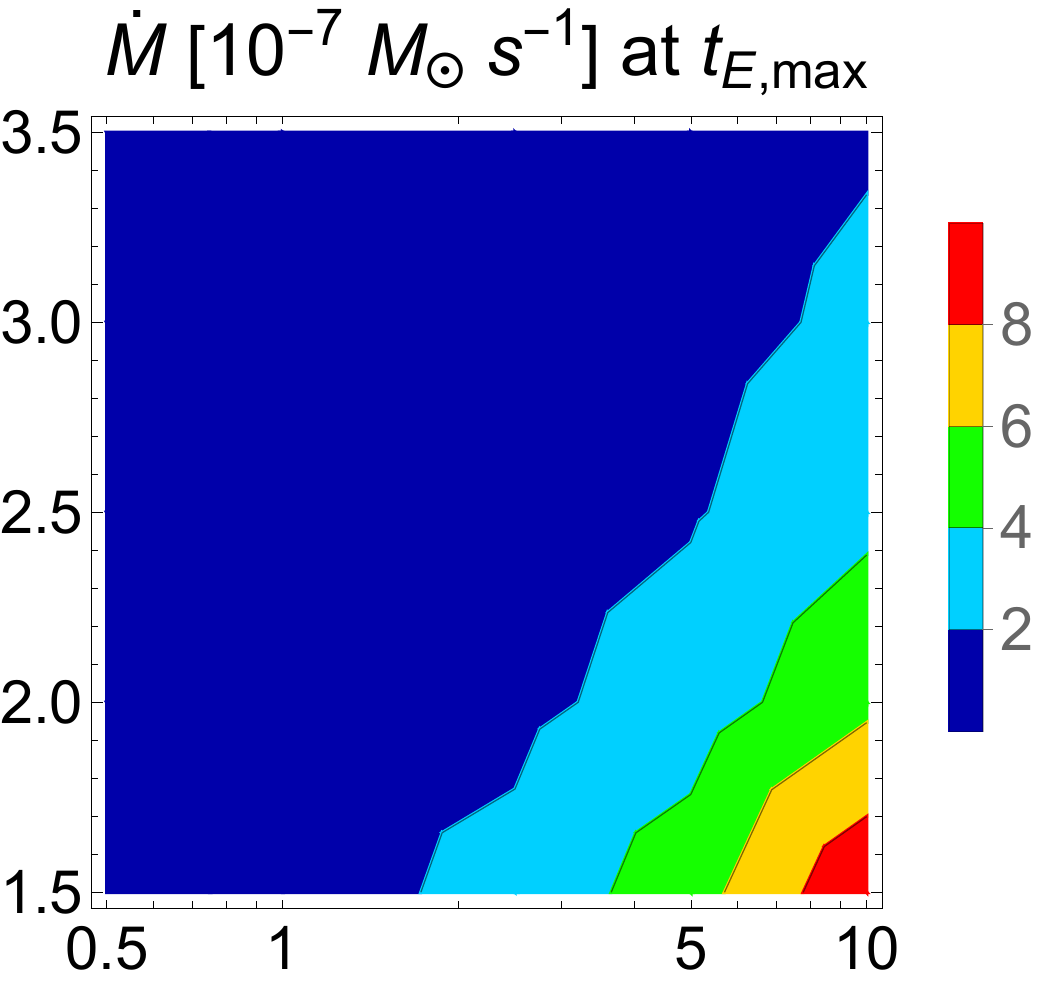}};
\node[below=55pt, node distance=0cm,font=\color{white}] {$|$};
\node[left=67pt, node distance=0cm, rotate=90, anchor=center,font=\color{white}] {~$|$};
\end{tikzpicture}
\end{minipage}
\newline
\vspace{-0.2cm}
\caption{\emph{Top-to-bottom:} The contours for the initial outflow conditions; density, temperature, entropy per baryon, expansion time-scale, and mass loss rate are shown. \emph{Left-to-right:} The figure panels are ordered by the IC time. Note the transition in $B_{\rm dip}$ trends for $\rho$, $T$, and $\dot{M}$, that is suggestive of the final trends in $\overline{A}$.} 
\label{fig:initialconditions}
\end{figure*}

The ICs for each numerical test ultimately determine the final abundance pattern. None of these ICs directly determine the final abundance pattern, but, from {\tt SkyNet}'s perspective, is rather a competition between the thermodynamic trajectories of density, temperature, and electron fraction over time. These trajectories are necessarily coupled to the entropy, expansion time-scale, mass loss rate, etc. So while a direct correlation of ICs do not tell the full story, the initial densities and temperatures are somewhat indicative of the final results.

\begin{figure*}
\begin{minipage}{0.32\textwidth}
\begin{tikzpicture}
\node (img) {\includegraphics[width=0.9\linewidth]{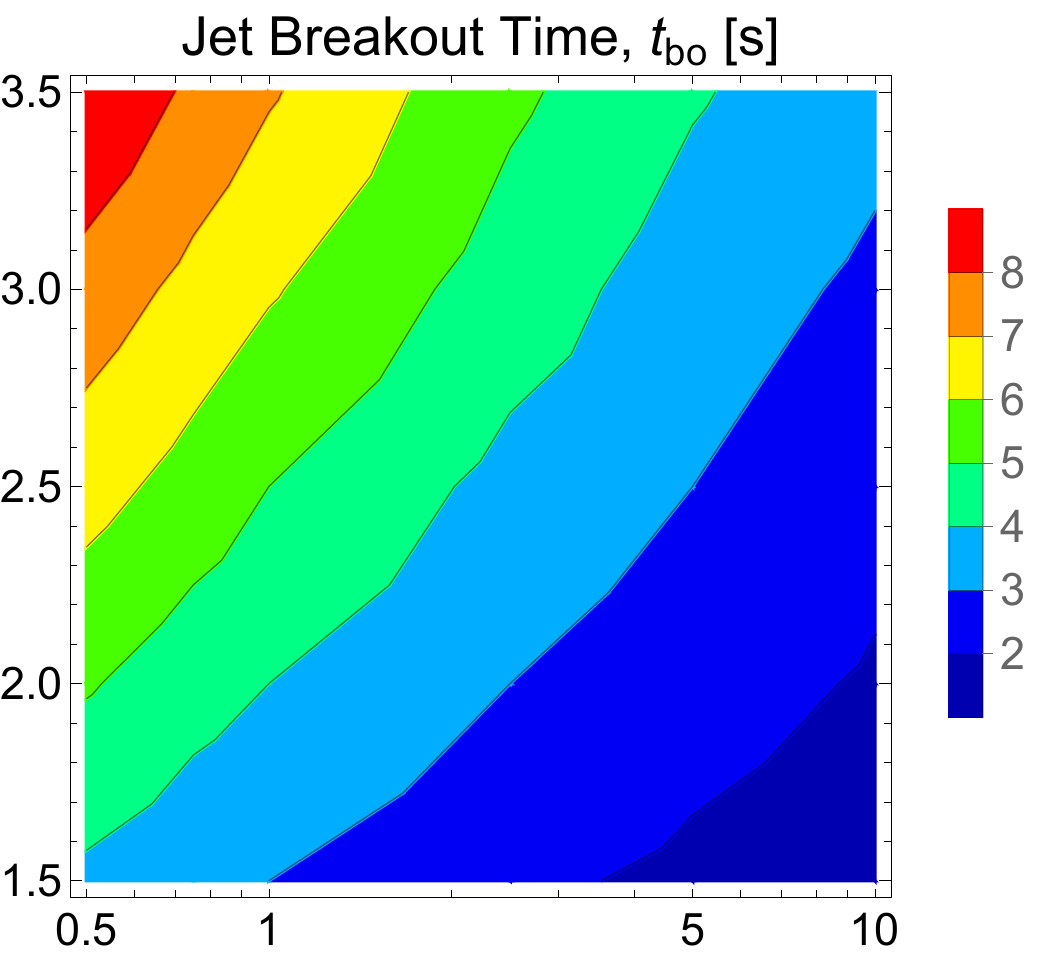}};
\node[below=70pt, node distance=0cm, font=\color{white}] {$B_{\textrm{dip}}$ [$10^{15}$ G]};
\node[left=85pt, node distance=0cm, rotate=90, anchor=center]{$P_0~$[ms]};
\end{tikzpicture}
\end{minipage}
\begin{minipage}{0.32\textwidth}
\begin{tikzpicture}
\node (img) {\includegraphics[width=0.9\linewidth]{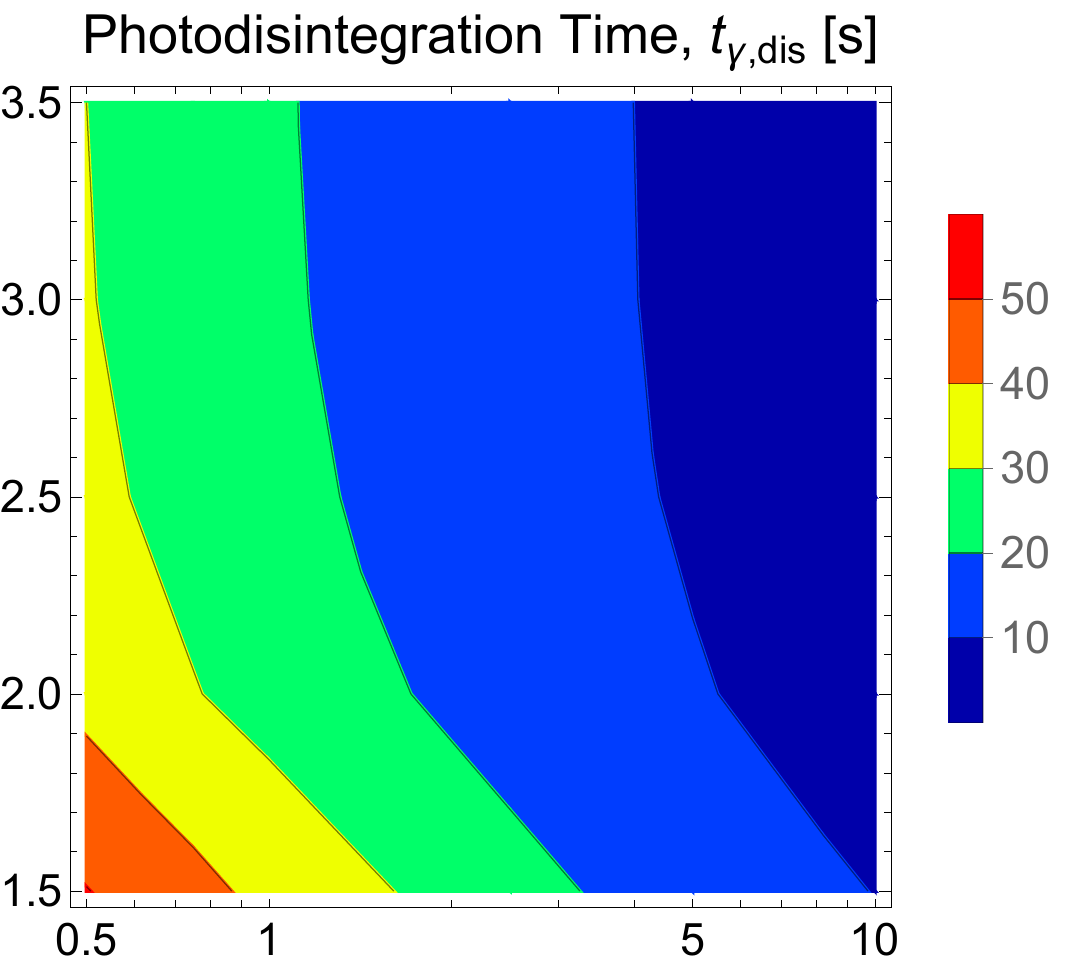}};
\node[below=70pt, node distance=0cm] {$B_{\textrm{dip}}$ [$10^{15}$ G]};
\node[left=85pt, node distance=0cm, rotate=90, anchor=center,font=\color{white}]{$~~|$};
\end{tikzpicture}
\end{minipage}
\begin{minipage}{0.32\textwidth}
\begin{tikzpicture}
\node (img) {\includegraphics[width=0.9\linewidth]{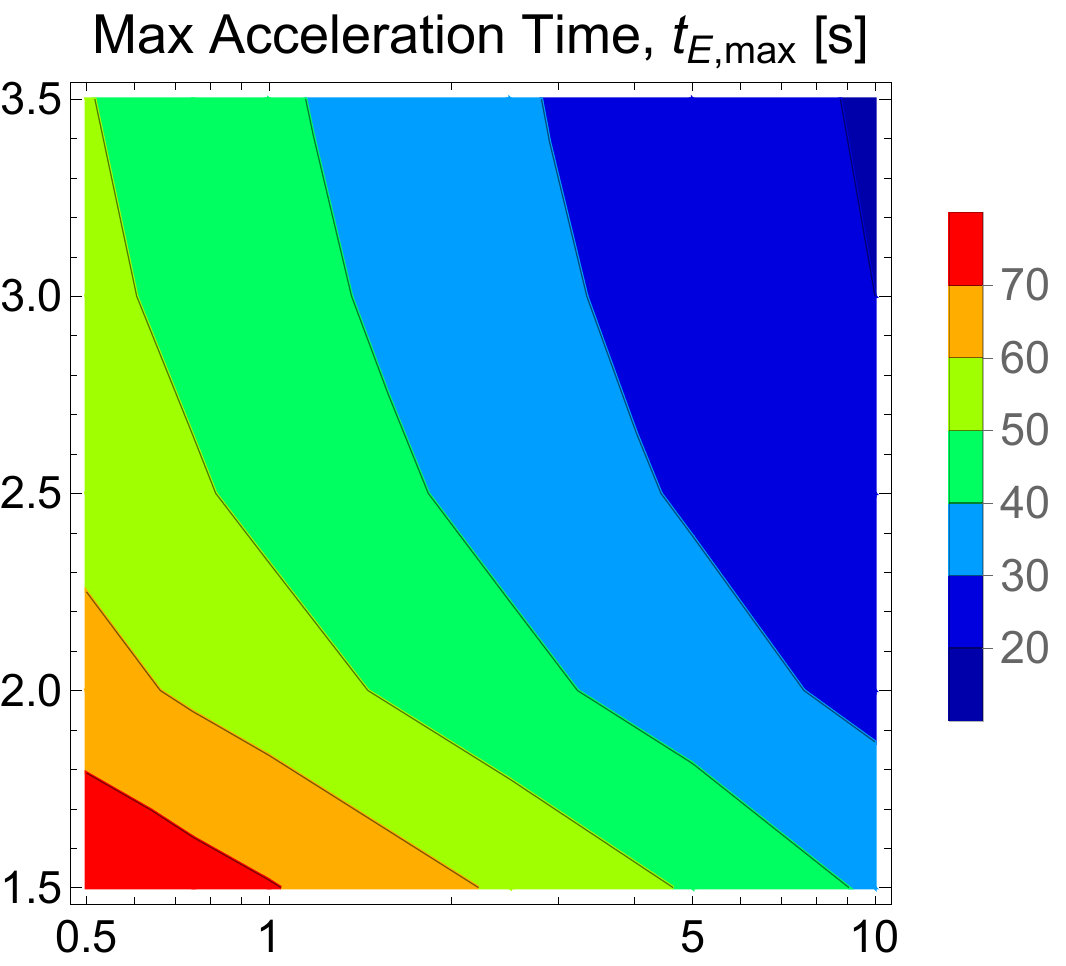}};
\node[below=70pt, node distance=0cm, font=\color{white}] {$A_{blank}$};
\node[left=85pt, node distance=0cm, rotate=90, anchor=center,font=\color{white}]{$~|$};
\end{tikzpicture}
\end{minipage}
\caption{
\emph{Left panel:} Breakout time as a function of $B_{\textrm{dip}}$ and $P_0$.
\emph{Middle panel:} Photodisintegration time, i.e.,  when $\tau_{\gamma-N}\sim$ 1 as a function of $B_{\textrm{dip}}$ and $P_0$.
\emph{Right panel:} Max acceleration IC time, i.e.,  the time when particles can no longer be accelerated above $E_{\rm max}=10^{20}\, {\rm eV}$ as a function of $B_{\textrm{dip}}$ and $P_0$.}
\label{fig:times}
\end{figure*}

Here we summarize some features and general trends from the ICs:
\begin{itemize}[leftmargin=*]
    \item Initial density, temperature, and mass loss rate show a transition in $B_{\textrm{dip}}$ preference between $t_{\rm bo}$ and $t_{\rm \gamma,dis}$.
    \item $\rho$ and $T$ have regions that aren't purely monotonic ($\rho$ at $t_{\rm \gamma,dis}$ and $t_{\rm E,max}$ and $T$ at $t_{\rm bo}$).
    \item Entropy is a function of spin period, but not directly a function of field strength. The field strength dependence at later IC times arises from the IC dependence on field strength and spin period.
    \item The $B_{\rm dip}$ and $P_0$ trends in $\tau_{\rm exp}$ and $\dot{M}$ are reflective of the magnetic field and spin period trends of the $f_{\rm open}$ and $f_{\rm cent}$ correction factors.
    \item Overall, $\rho$ and $T$ (and, to a lesser degree, $\Dot{M}$) may be indicative of the $B_{\rm dip}$ and IC time nucleosynthesis trends: generally, higher densities suggest higher values of $\overline{A}$ and higher temperatures suggest lower values of $\overline{A}$, but do not directly predict the results.
\end{itemize}

\section{Analytic vs. Numeric \texorpdfstring{$X_h$}{Xh}}\label{sec:appendXh}

\citet{Roberts_2010} estimates $X_h$ as~(\ref{xhanalytic}) as a function of expansion time-scale, entropy, and electron fraction:

\begin{equation}\label{xhanalytic}
\begin{split}
& X_h \approx \\
&
\begin{cases} 
      1-\exp\left[-8\times10^5Y_e^3\left(\frac{\tau_{\textrm{exp}}}{\textrm{ms}}\right)\left(\frac{S}{\textrm{k}_{\textrm{B}}\textrm{nuc}^{-1}}\right)^{-3}\right], \hfill Y_e<0.5\\
      1-\left[1+140(1-Y_e^2)\left(\frac{\tau_{\textrm{exp}}}{\textrm{ms}}\right)\left(\frac{S}{\textrm{k}_{\textrm{B}}\textrm{nuc}^{-1}}\right)^{-2}\right]^{-1/2}, \hfill Y_e \geq 0.5.
\end{cases}
\end{split}
\end{equation}

\begin{figure}
\centering
\includegraphics[width=\linewidth]{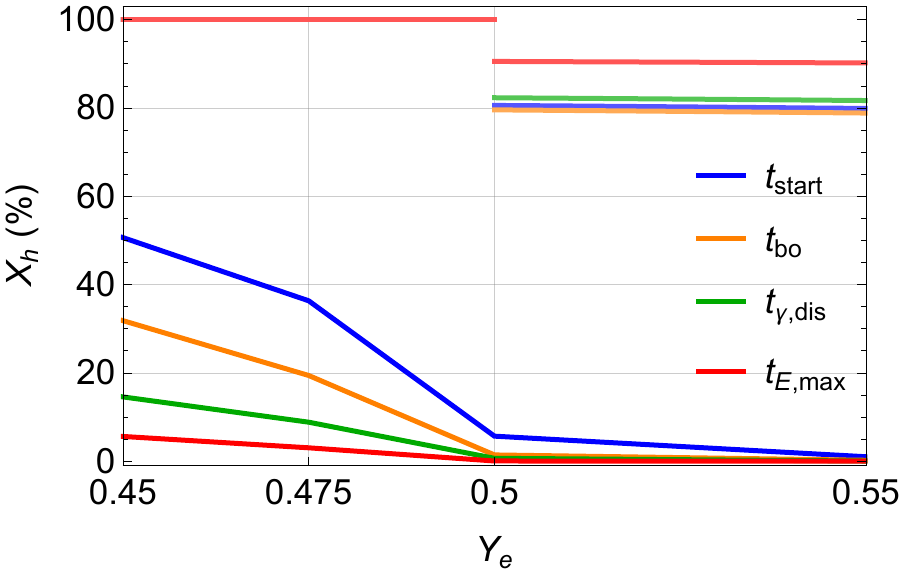}
\vspace{-0.5cm}
\caption{Comparison of analytic (equation (\ref{xh})) and numeric (this work) $X_h$ for $B_{\textrm{dip}} = 5\times10^{15}\, {\rm G}$, $P_0 = 2\, {\rm ms}$ model. Analytic is given by the translucent lines (all $\sim80-100\%$) while numeric is given by opaque colors.}
\label{fig:xhcomp}
\end{figure}

In Fig.~\ref{fig:xhcomp} we show that the analytic expression predicts an $X_h$ that is $\sim50-90\%$ higher than what is produced with {\tt SkyNet} numerically. When nucleosynthesis is most conducive for this model, there is a $\sim50\%$ difference that increases significantly at later IC times and for greater $Y_e$. Numerically, it also is much more sensitive to the electron fraction for this range. In proton-rich conditions, we find it very difficult to synthesize large fractions of heavy elements. These differences do depend on the particular configuration of $B_{\rm dip}$ and $P_0$, however. Finally, as IC time increases, $X_h$ decreases in the numerical implementation (and, in the limiting late-IC-time case, goes to zero). With the analytical expression, we see no time dependence in neutron-rich conditions and an inverse time dependence in proton-rich conditions.

These differences likely arise from a few key assumptions made in the analytic expression and may not apply well to the model studied in this work. Firstly, \citet{Roberts_2010} does not consider a magnetically dominated outflow; the rapid rotation from misaligned magnetars leads to an overall entropy decrease, although this decrease is only by a factor of $\sim$ 2. This work also uses a limited 19-isotope nuclear network; a larger network with updated reactions may result in different results. Unlike in the numerical calculations of {\tt SkyNet}, the analytical estimate assumes dominant reaction channels. For neutron-rich conditions, the reaction sequence is $^4\textrm{He}(\alpha\textrm{n},\gamma)^9\textrm{Be}(\alpha,\textrm{n})^{12}$C and for proton-rich conditions, $\alpha~$particles recombine into $^{12}$C via the triple-$\alpha$ reaction. The effective $\tau_{\textrm{exp}}$ and $S$ dependence of $X_h$ may not be described by the same polynomial as in the analytical treatment. 

In the analytic expression, the only temporal variance arises in proton-rich conditions, since for $Y_e \lesssim 0.5$, $X_h \approx 1$ throughout. In that case, $X_h$ rises with increasing time post-CC (with the ICs derived in this work). However, $X_h$ numerically trends with $\rho$ and $T$. These ICs depend on $\tau_{\textrm{exp}}$ and $S$, but to different order, and also are a function of mass loss rate and include the dynamical evolution of NS radius. When including these, the ICs (and thus, heavy element nucleosynthesis) decrease with time. In the limiting case, heavy elements are not produced at all.

\label{lastpage}

\end{document}